\documentclass[fleqn,usenatbib]{mnras}

\usepackage{newtxtext,newtxmath}
\usepackage[T1]{fontenc}

\DeclareRobustCommand{\VAN}[3]{#2}
\let\VANthebibliography\thebibliography
\def\thebibliography{\DeclareRobustCommand{\VAN}[3]{##3}\VANthebibliography}

\usepackage{graphicx}	% Including figure files
\usepackage{amsmath}	% Advanced maths commands
\usepackage{caption}
\usepackage{threeparttable}
\title[Stellar Population of Cluster UDGs and NUDGEs]{Diffuse Dwarf Galaxies in Galaxy Clusters: I. Stellar Populations and Radial Gradients}

\author[Levitskiy et al.]{
Arsen Levitskiy,$^{1}$\thanks{E-mail: arsenlv.1115@gmail.com}
Duncan A. Forbes,$^{1}$
Jonah S. Gannon,$^{1}$
Yimeng Tang,$^{2}$
Anna Ferr\'e-Mateu,$^{3,4,1}$
\newauthor{\,\,Aaron J. Romanowsky,$^{2,5}$
Jean P. Brodie,$^{1,5}$
Warrick J. Couch$^{1}$}
\\
$^{1}$\,Center for Astrophysics and Supercomputing, Swinburne University, John Street, Hawthorn VIC 3122, Australia\\
$^{2}$\,Department of Physics and Astronomy, San Jos\'e State University, One Washington Square, San Jose, CA 95192, USA\\
$^{3}$\,Instituto de Astrof\'isica de Canarias, Av. Via Lactea s/n, E38205 La Laguna, Spain\\
$^{4}$\,Departamento de Astrof\'isica, Universidad de La Laguna, E-38200, La Laguna, Tenerife, Spain\\
$^{5}$\,Department of Astronomy and Astrophysics, University of California Santa Cruz, 1156 High Street, Santa Cruz, CA 95064, USA \\}
\date{Accepted XXX. Received YYY; in original form ZZZ}
\pubyear{\the\year{}}

\begin{document}
\label{firstpage}
\pagerange{\pageref{firstpage}--\pageref{lastpage}}
\maketitle

\begin{abstract}

We use Keck/KCWI spectroscopy to study one ultra-diffuse galaxy (UDG) and five Nearly-UDGs (NUDGEs) in the Perseus cluster, together with an additional UDG in the Coma cluster. As the first paper in a series, we focus on the global and radial stellar population properties of our sample. We find that these galaxies host intermediate-to-old stellar populations, with typical ages of $\sim$7\,Gyr, low metallicities ([M/H]$\simeq -$0.9\,dex), and enhanced [Mg/Fe] abundances ($\sim$0.3\,dex), consistent with previous studies. Six galaxies lie within the scatter of the present-day mass–metallicity relation (MZR), whereas the Coma UDG (DF11) is more consistent with the MZR of high$-z$ galaxies ($z \sim 2$). We find no strong correlation between global stellar population properties and cluster infall parameters, suggesting that any environmental impact is not easily traceable through integrated stellar populations. We go one step further and measure radial gradients for three galaxies. Two show flat age and mildly negative metallicity gradients, similar to classical dwarfs, while one shows a rising metallicity profile as recently found in other UDGs. Comparing with classical dwarfs, we find a continuous correlation between metallicity gradient and globular cluster (GC) richness, where more GC-rich systems tend to show rising profiles. We propose that preferential tidal disruption of GCs in the inner regions of galaxies naturally produces rising metallicity profiles, unlike GC-poor classical dwarfs. This mechanism, potentially coupled with strong stellar feedback from early concentrated star formation, may explain the unusual rising metallicity profiles observed in GC-rich UDGs/NUDGEs.

\end{abstract}


%%%%%%%%%%%%%%%%%%%%%%%%%%%%%%%%%%%%%%%%%%%%%%%%%%

%%%%%%%%%%%%%%%%% BODY OF PAPER %%%%%%%%%%%%%%%%%%

\section{Introduction}

It has been more than a decade since low surface brightness, ultra diffuse galaxies (UDGs), sparked the renewed interest of many in the scientific community. The discovery by \citet{VanDokkum2015} of numerous systems in the Coma Cluster defined this population as galaxies with low central surface brightnesses of $\mu_{0,g} \ge 24$ mag arcsec$^{-2}$ and extended sizes, with effective radii $R_{\mathrm{e}} \ge 1.5$ kpc. This definition has since been widely adopted as the criterion for identifying UDGs. Subsequent works dramatically expanded the known UDG population, finding these systems in a broad range of environments, including clusters \citep[e.g.,][]{VanDokkum2015, Yagi2016, Janssens2017, Wittmann2017, Janssens2024}, groups \citep[e.g.,][]{Forbes2019, Prole2019, Iodice2020, Lim2020, LaMarca2022}, and the field \citep[e.g.,][]{Leisman2017, Janowiecki2019, Zaritsky2021, Jones2023}.
Considerable effort has since also been devoted to investigating the formation pathways of UDGs. These include scenarios such as intrinsically high halo spin \citep[e.g.,][]{Amorisco2016, Rong2017, Benavides2021}, strong and cyclic stellar feedback \citep[e.g.,][]{Dicintio2017}, and environmental processing through mechanisms such as tidal heating and ram pressure stripping \citep[e.g.,][]{Yozin2015, Mistani2016, Carleton2019, Sales2020}, or even combinations of these processes \citep[e.g.,][]{Jiang2019}. For further detailed elaboration of many of the outlined scenarios in the literature, see the comprehensive review by \citet{Gannon2026}. 

Ultimately, a central question regarding UDGs has been whether they all share the same physical properties and formation mechanisms as classical dwarf galaxies, or whether they represent a distinct population. A similar consideration also applies to low surface brightness galaxies classified as Nearly-UDGs (NUDGEs) \citep{Forbes2024}. While NUDGEs depart slightly from the conventional \citet{VanDokkum2015} criteria, broad-band imaging based studies such as \citet{Buzzo2025} have demonstrated through unsupervised clustering that NUDGes and UDGs are statistically similar across a wide range of physical properties. The answers to these questions may also differ depending whether they are gas-rich, star-forming systems or old, passive UDGs. The first kind, which are most commonly found in field environments, are generally galaxies with vast reservoirs of HI gas, host few globular clusters (GCs) and exhibit ongoing star formation \citep[e.g.][]{Leisman2017, Janowiecki2019, Zaritsky2021, Jones2023}, although not entirely limited to that \citep[e.g.][]{Sandoval2025}.  

Meanwhile, some old and passive UDGs, which preferentially inhabit group and cluster environments, exhibit a distinctly different set of morphological and stellar population properties. Many cluster UDGs, as well as NUDGEs, show stellar population properties similar to those of classical cluster dwarf galaxies, as demonstrated by \citet{FM2018, Chilingarian2019, FM2023}, and closely follow established galaxy scaling relations such as the present-day mass–metallicity relation  \citep[MZR;][]{Kirby2013, Simon2019}. Nonetheless, some cluster UDGs, particularly those with extraordinarily large GC populations ($\rm N_{GC}$ $\gtrsim$ 20), appear to display remarkably different stellar population properties. A fraction of these galaxies are often referred to as ``failed galaxies,” owing to their inferred dark matter halos being overly massive relative to their stellar masses \citep{VanDokkum2015, Peng2016, Danieli2022, Forbes2024}. Investigations of their stellar populations reveal predominantly old, significantly more metal poor, yet more $\alpha$-enhanced stars, with these galaxies following the MZR of galaxies at higher redshift, around $z$ $\sim$ 2 \citep{Buzzo2022, FM2023, Buzzo2025, Forbes2025}. This has led to the idea of very early stellar assembly, along with possible substantial contribution from stars originating in tidally disrupted GCs, given the large fractions of surviving GCs still observed in these galaxies today \citep[e.g.][]{Danieli2022}. Hence, given that UDGs may follow distinct formation histories driven by their environment \citep[e.g.][]{Jiang2019, FM2023} or their GC populations \citep[e.g.][]{Buzzo2025}, larger spectroscopic samples are essential to verify these patterns and elucidate the broader trends of UDG formation. The present work contributes to this effort with the study of seven new cluster UDGs and NUDGEs.

%particularly regarding whether these systems share the same properties as UDGs and differ from classical dwarf galaxies. 

\begin{figure}
\centering
\includegraphics[width=7.5cm]{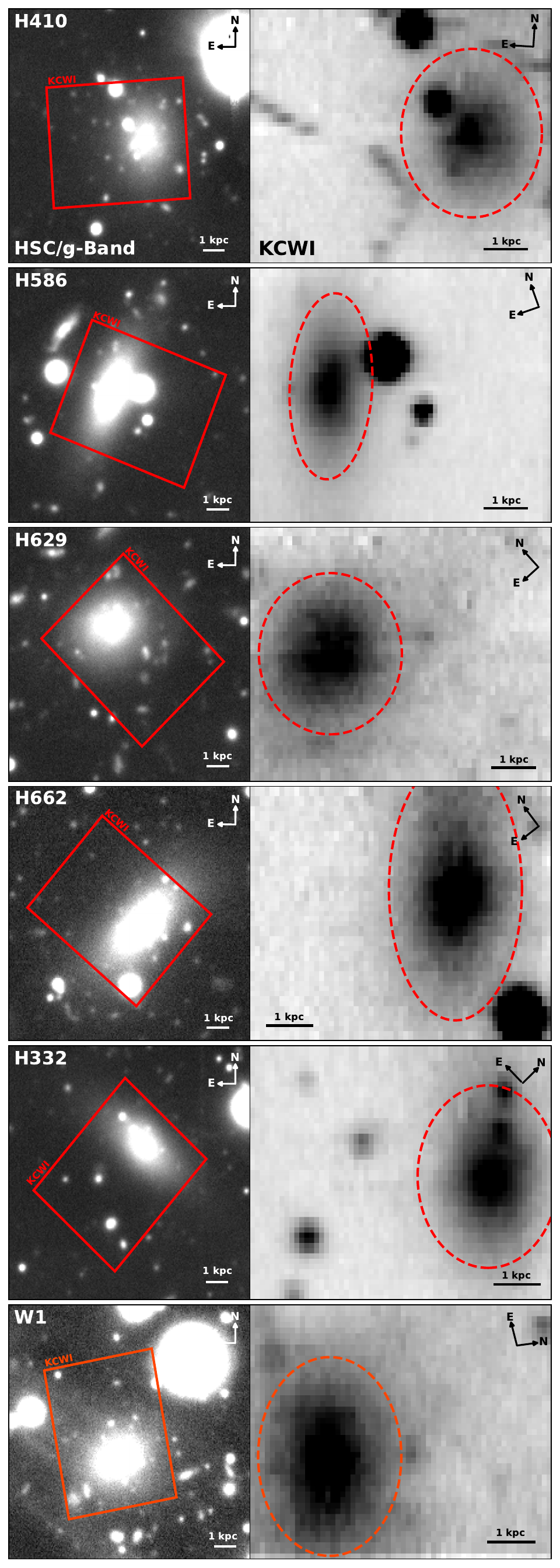}
\caption{Optical broadband imaging of the sample of Perseus cluster galaxies together with the corresponding KCWI white-light images. The left panels display Subaru Hyper Suprime-Cam (HSC) $g$-band images used in \citet{Tang2025}. The orange/red colour boxes overlaid on the UDGs/NUDGEs indicate the 16\arcsec $\times$ 20\arcsec \;KCWI Medium slicer field of view (FoV) used in this work. The corresponding KCWI white-light images from that FoV are shown in the right panels with orange/red dashed aperture indicating 1 $R_{\mathrm{e}}$ of the galaxy. }
\label{fig:hsc_hst_kcwi}
\end{figure}

%This paragraph beginning needs some edits In terms of their characteristics, most appear to align well with classical field dwarf galaxies \citep[e.g.,][]{Buzzo2025, Motiwala2025}, with their extended sizes likely resulting from a temporary or transitional phase in galaxy evolution.

With the advent of increasingly powerful IFU capabilities, studies of resolved stellar population properties have become another key aspect in understanding galaxy formation. 
%In classical dwarf galaxies such as dEs and dSphs, they are typically found to host flat to negative metallicity gradients \citep[e.g.][]{Koleva2011, Sybilska2017, Taibi2022, Bidaran2023, Li2026}, which are commonly interpreted within the framework of an outside in formation scenario \citep[e.g.,][]{Pipino2008, Benitez2013, Revaz2018}. 
In classical dwarf galaxies such as dwarf Elliptical (dEs) and Spheroidals (dSphs) galaxies, they are typically found to host flat to negative metallicity gradients \citep[e.g.][]{Koleva2011, Sybilska2017, Taibi2022, Bidaran2023, Li2026}, with younger and more metal-rich populations concentrated in their inner regions, consistent with the framework of an outside-in formation \citep[e.g.][]{Pipino2008, Benitez2013, Revaz2018}.  Specifically, recent works such as \citet{Bidaran2023} found negative [M/H], and flat age gradients in a sample of Virgo dEs, with further positive $[\alpha/Fe]$ gradients suggesting more extended star formation timescales in their inner regions, consistent with this picture. However, recent works focusing on UDGs have begun to reveal a different picture. While these observations are technically challenging and time consuming, often requiring substantial integration time on large telescopes, works including \citet{Villaume2022, FM2025, Buzzo2025a, AL2025} find relatively flat age gradients and, more puzzlingly, flat-to-rising metallicity gradients within the effective radius. These results are based on a limited number of UDGs/NUDGEs that are predominantly GC-rich. This apparent departure from the trends observed in classical dwarfs suggests that UDGs may have distinct evolutionary pathways, possibly linked to their GC populations. Determining whether these unusual metallicity gradients are unique to GC-rich UDGs or if they represent a universal UDG trait requires a larger and more diverse sample. Here, we build on previous works by adding 3 new GC poor UDGs and NUDGEs in dense cluster environments, thereby diversifying the available spectroscopic sample in terms of their GC populations.

%More recently, a new avenue in UDG studies has emerged that focuses on spatially resolved analyses tracing radial variations in stellar population ages and metallicities. Works including \citet{Villaume2022, FM2025, Buzzo2025a, AL2025} have begun to find relatively flat age gradients and, more puzzlingly, flat--to--rising metallicity gradients within the effective radius of predominantly cluster based, GC-rich UDGs/NUDGEs. This behavior appears to contrast with many studies of classical dEs or dSph galaxies, largely finding flat--to--negative metallicity gradients \citep[e.g.,][]{Koleva2011, Sybilska2017, Taibi2022}. These trends are commonly interpreted within the framework of the well known outside--in formation scenario for dwarf galaxies \citep[e.g.,][]{Pipino2008, Benitez2013, Revaz2018}. Thus, elucidating the spatially resolved stellar populations of UDGs can further shed light on their formation mechanisms and the evolution of stars within these systems, ultimately helping to explain what makes this subset of galaxies unique.

This work represents the first paper in a pair, with a forthcoming one focusing on the kinematic aspects of the sample. The paper is organized as follows. We begin with a brief introduction to the galaxy sample and the data acquisition in Section \ref{sec:obs}. This is followed by an investigation of the global stellar population properties in Sections \ref{subsec:global}--\ref{subsec:alpha}, including their placement on the MZR and any dependence on the cluster environment. We then present spatially resolved stellar population analyses for selected galaxies in Section \ref{subsec:gradient}. Finally, we present an interpretation and discussion of the results in Section \ref{sec:discussion}.

Throughout the manuscript, we adopt a $\Lambda$CDM cosmology with $H_0 = 70 \; \rm km  \; s^{\mathrm{-1}} \; Mpc^{\mathrm{-1}}$, a distance $d=75$\,Mpc to the Perseus cluster and $d=100$\,Mpc to the Coma cluster.

%--> Resolved studies becoming available, can point to studies of galaxy evolution\\
%--> Classical dwarfs interpretation, do UDGs follow that? Summarize Anna 2025 results slightly.\\
%--> Summarize the paper and define distances to Perseus and Coma and cosmology\\
%\renewcommand{\arraystretch}{0.8}%

\begin{table*}
    \hspace{-2.5cm}
	\caption{Galaxy sample and observation summary. Columns are: Galaxy name, target coordinates (deg), Effective radius, ellipticity, GC classification, KCWI gratings, central wavelength, program ID, observation date and exposure times used in this study}
	\label{tbl:kcwi_obs}
	\begin{tabular}{cccccccll} % four columns, alignment for each
		\hline
		Galaxy & RA & DEC & $R_{\mathrm{e}}$ & Ellipticity & GC & Central $\lambda$\, / Grating & Program ID & Exposures\\
     & (deg) & (deg) & (kpc) &  & richness  &  & (Obs Date) & \\
		\hline
        H410 & 50.13390 & 41.19850 & 1.88 & 0.18 & rich & 5110\,\AA\, / BH3 & U245 (2024 Oct 29) &  9 $\times$ 1620s\\
          & & & & & & 6750\,\AA\, / RH2 & W283 (2024 Nov 3) &  28 $\times$ 580s \\
        \hline
        H586 & 50.61892 & 41.53714 & 2.03 & 0.57 & poor & 5110\,\AA\, / BH3 & U245 (2024 Oct 29) &  6 $\times$ 1800s\\
          & & & & & & 6750\,\AA\, / RH2 &  & 18 $\times$ 580s\\
        \hline
        H629 & 50.74297 & 41.53384 & 1.72 & 0.16 & poor & 5110\,\AA\, / BH3 & W285 (2024 Nov 6) &  6 $\times$ 1800s\\
          & & & & & & 6750\,\AA\, / RH2 &  & 18 $\times$ 585s\\
        \hline
        H662 & 50.92158 & 41.57217 & 2.52 & 0.50 & rich & 5110\,\AA\, / BH3 & U245 (2024 Oct 28, 29) &  5 $\times$ 1800s\\
          & & & & & & 6750\,\AA\, / RH2 & W285 (2024 Nov 6) & 7 $\times$ 580s + 9 $\times$ 585s + 5 $\times$ 590s \\
        \hline
        H332 & 49.95743 & 41.66925 & 1.86 & 0.48 & poor & 4550\,\AA\, / BL & W283 (2024 Nov 3) &  8 $\times$ 1800s \\
          & & & & & & 6750\,\AA\, / RH2 & &  24 $\times$ 585s \\
        \hline
        W1 & 49.25156 & 41.32243 & 2.33 & 0.17 & poor & 4550\,\AA / BL & W285 (2024 Nov 7) &  5 $\times$ 1620s\\
          & & & & & & 6750\,\AA\, / RH2 &  & 13 $\times$ 600s\\
        \hline
        DF11 & 195.60625 & 28.23278 & 2.10 & 0.02 & poor$^{*}$ & 4550\,\AA\, / BL & U234 (2025 Apr 30) & 11 $\times$ 1800s\\
          & & & & & & 6750\,\AA\, / RH2 &  & 33 $\times$ 575s\\
        
        \hline
	\end{tabular}
    \begin{tablenotes}
    \item{\textbf{*}} DF11 is measured to have $9.4 \pm 12.7$ from the HST/ACS based analysis of \citet{Lim2018}, which nominally places it in the GC-poor regime, although a GC-rich classification remains possible given the measurement uncertainties.
    \end{tablenotes}
\end{table*}

\section{Observation}\label{sec:obs}

\begin{figure}
\centering
\includegraphics[width=8.3cm]{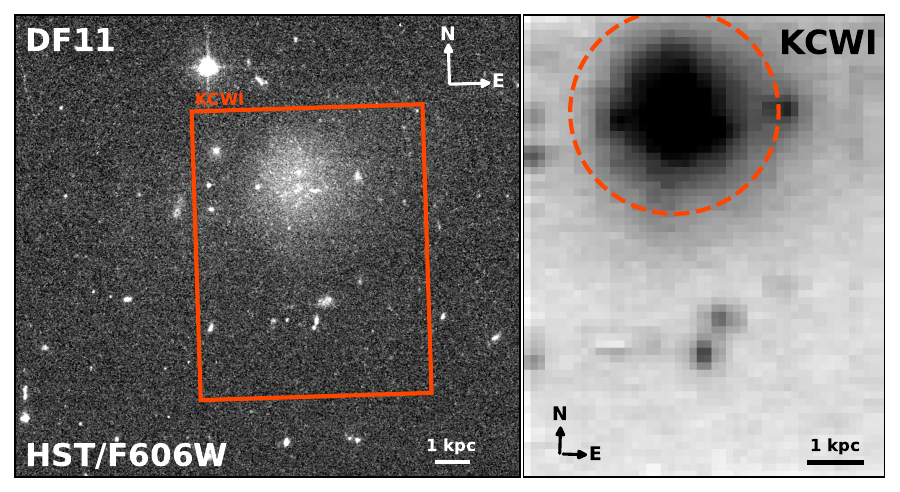}
\caption{Same as described previously in Figure\,\ref{fig:hsc_hst_kcwi}, but using HST/F606W image for DF11 in the Coma cluster, originating from \citet{Lim2018}, with the similar KCWI Medium slicer FoV box overlaid on top in orange. The corresponding KCWI white-light image is shown in the right panel, where the 1 $R_{\mathrm{e}}$ of the galaxies are indicated by a orange dashed aperture.}
\label{fig:hsc_hst_kcwi_2}
\end{figure}

\subsection{Data Sample}\label{subsec:sample}

The sample consists of seven galaxies, six of which are located in the Perseus cluster, namely H332, H410, H586, H629, H662, and W1. In addition, we include an UDG from the Coma cluster, DF11, as a candidate for a resolved stellar population study. This galaxy serves as a comparison to both our primary Perseus sample and existing literature on Coma UDGs \citep{Villaume2022, FM2025}. The Perseus cluster targets were selected based on previous observations from deep imaging data obtained with the Hyper Suprime Cam on the Subaru Telescope \citep[as described in][]{Tang2025}, as well as the HST study presented in \citet{Janssens2024}, with no prior spectroscopic studies available. The Coma cluster object was also only observed previously with HST, as reported in \citet{Lim2018}. We show cutout images of these objects in Figures \ref{fig:hsc_hst_kcwi} and \ref{fig:hsc_hst_kcwi_2}, including broadband imaging and their corresponding KCWI white-light images.

\begin{figure}
\centering
\includegraphics[width=\columnwidth]{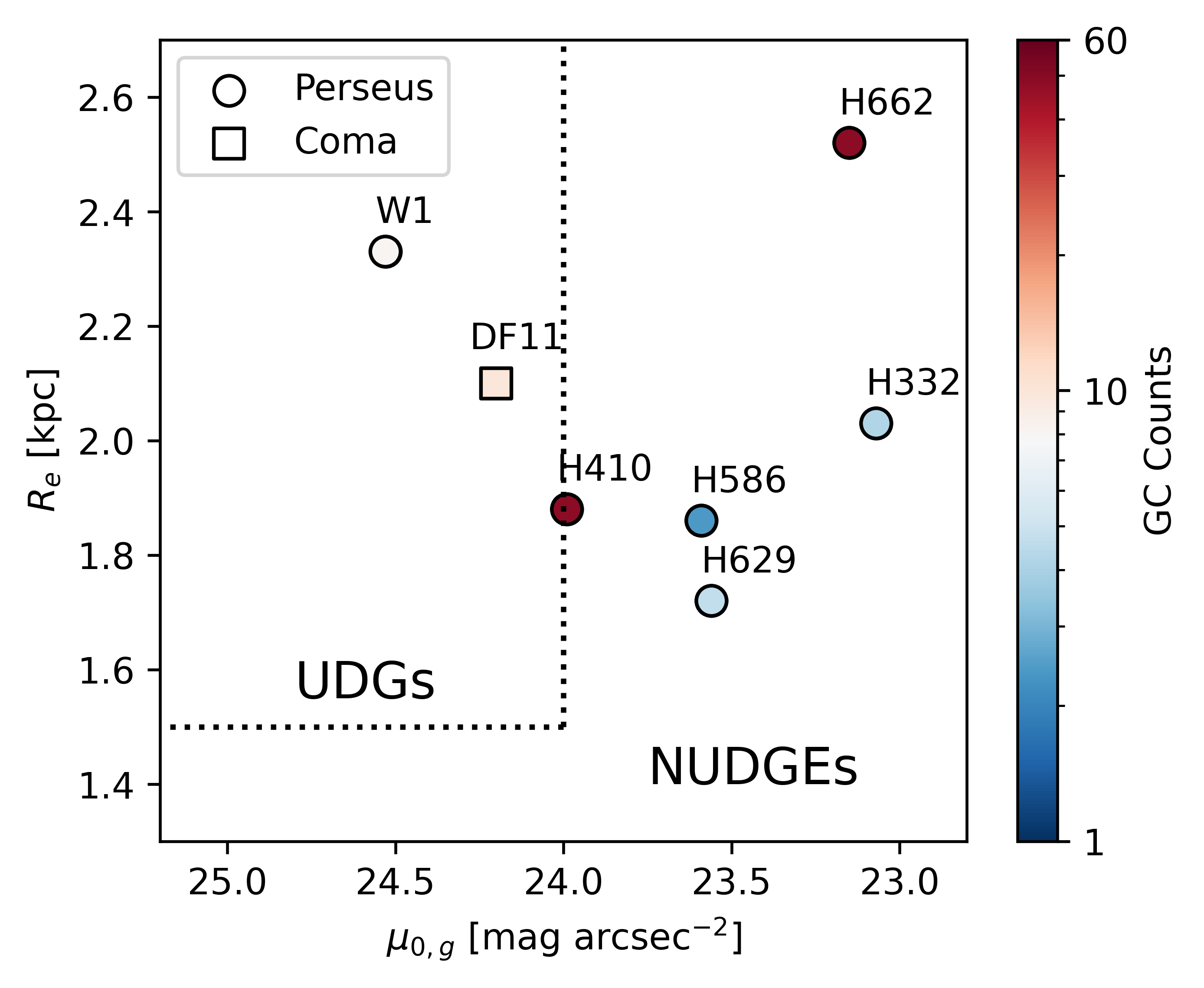}
\caption{Sample of Perseus and Coma UDGs/NUDGEs from this work in $\mu_{0,g}$ vs. $R_{\mathrm{e}}$ space, where point colors represent GC-richness. We consider galaxies with $\rm N_{GC}$ $\gtrsim$ 20 to be GC-rich. We All 7 galaxies have UDG-like $R_{\mathrm{e}}$, but two satisfy the \citet{VanDokkum2015} criteria, while the remaining five exhibit slightly $\mu_{0,g}$ and are therefore classified as NUDGEs.}
\label{fig:classification}
\end{figure}

Within the sample, two galaxies, W1 and DF11, meet the stringent definition of ``bona fide" UDGs, satisfying the criteria for central surface brightness in the $g$ band of $\mu_{0,g} \ge 24$ mag arcsec$^{-2}$ and effective radii of $R_{\mathrm{e}} \ge 1.5$ kpc. 
%The remaining five galaxies meet the effective radius criterion but are slightly brighter than typical UDGs, and can therefore be classified as NUDGEs, as defined in \citet{Forbes2024}.
The remaining five galaxies meet the effective radius criterion but are slightly brighter than typical UDGs, and are therefore considered NUDGEs. As described by \citet{Forbes2024} and \citet{Buzzo2024}, this term is used to describe galaxies that lie slightly outside of the \citet{VanDokkum2015} definition, typically having $\mu_{0,g}$ between 23 and 24 mag arcsec$^{-2}$ or an $R_\mathrm{e}$ slightly smaller than 1.5 kpc. In Figure\,\ref{fig:classification}, we show our sample in the $\mu_{0,g}$ -- $R_{\mathrm{e}}$ space, colour coded by their GC-richness based on the measurements of \citet{Janssens2024, Li2025} and \citet{Tang2025} for the Perseus cluster objects and \citet{Lim2018} for DF11. For the GC-classification, two of the galaxies (H410 and H662) are considered GC-rich based on the $\rm N_{GC}$ $\gtrsim$ 20 condition, while the rest are GC-poor as summarized in the Table\,\ref{tbl:kcwi_obs}.

\begin{figure}
\hspace*{-0.9cm}\includegraphics[width=9.8cm]{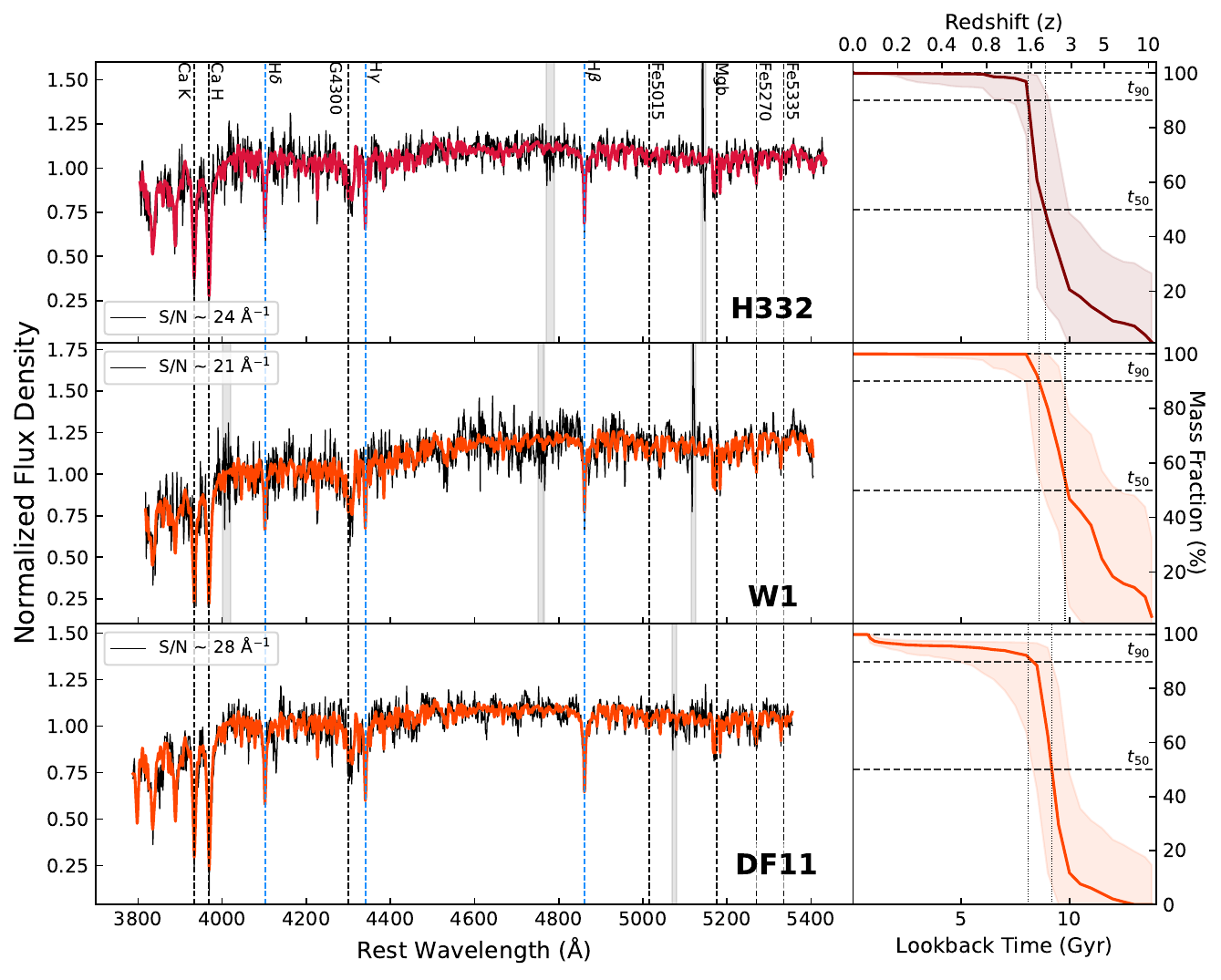}
\hspace*{-0.9cm}\includegraphics[width=9.8cm]{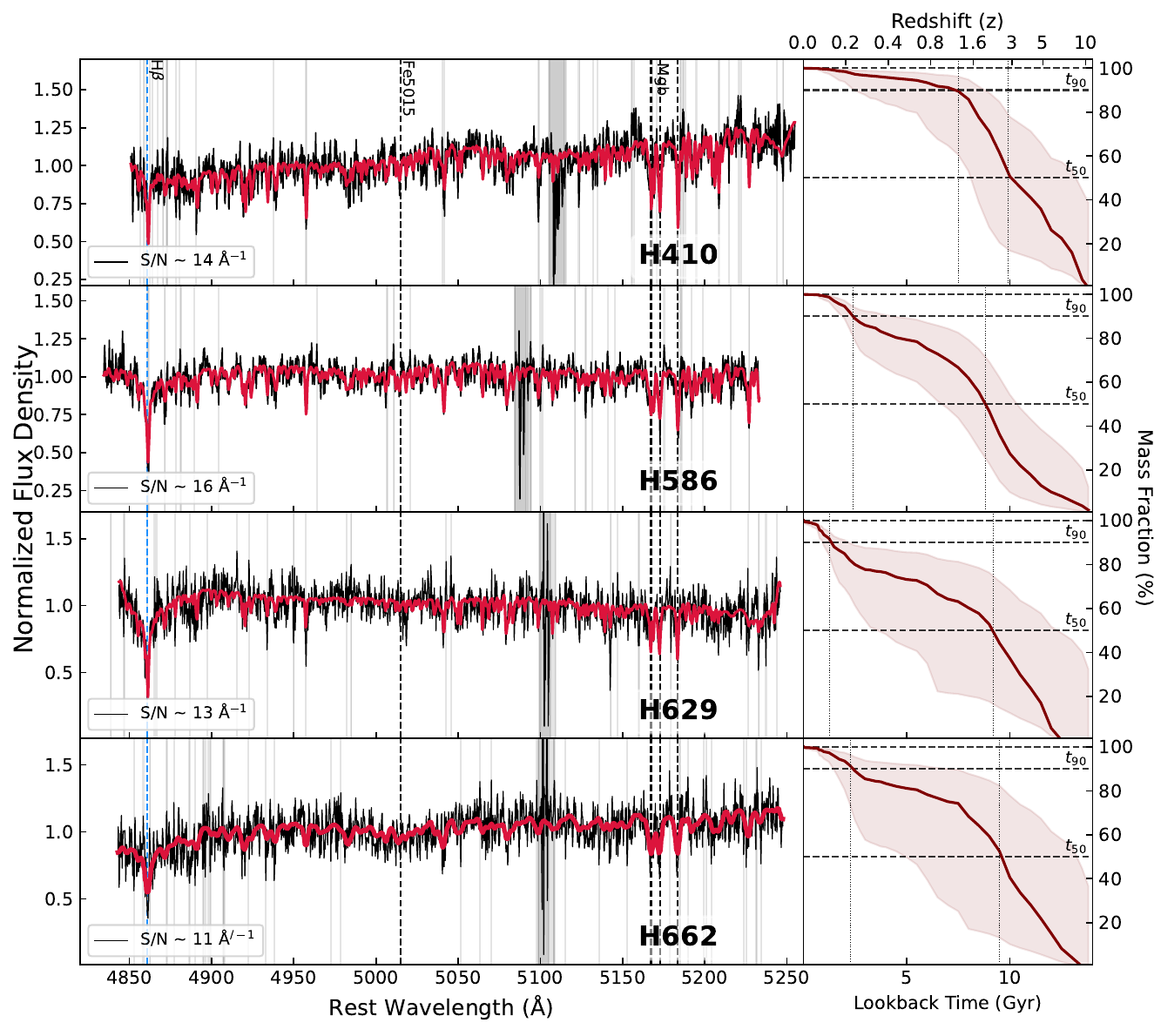}
\caption{{\tt pPXF} fits to the extracted global spectra of our sample galaxies. The sample is divided based on two instrumental configurations. The top three spectra (H332, W1, and DF11) were obtained with the BL grating, which provides a longer wavelength baseline and lower spectral resolution, while the remaining four galaxies (H410, H586, H629, and H662) were observed with the BH3 grating, having a shorter wavelength baseline and higher spectral resolution. The best-fitting {\tt pPXF} models are overlaid in orange/red colour for UDGs/NUDGEs. Prominent metal absorption features (G band, Mg$_{\rm b}$, Fe5015, Fe5270, and Fe5335) are indicated with black dotted lines, while Balmer absorption lines (H$\beta$, H$\delta$, and H$\gamma$) are shown with blue dotted lines. The corresponding cumulative mass fraction plots are shown in the right panels of each spectral fit, illustrating the median derived star formation histories with associated uncertainties indicated by shaded regions. The epochs at which 50\% and 90\% of the stellar mass are formed are marked by horizontal gray black lines, and their intersection with the galaxy's SFH indicated by vertical dotted lines. The first four galaxies show an earlier and more rapid stellar buildup, while the latter three shows a more gradual and continuous star formation over the epoch. }
\label{fig:global_fits}
\end{figure}

\subsection{Data acquisition and reduction}\label{subsec:reduction}

This work utilizes newly obtained spectroscopic data from the Keck Cosmic Web Imager (KCWI) \citep{Morrissey2018} with observations conducted on the nights of 2024 October 28, 29 (Program: U245, PI: Brodie), 2024 November 3 (Program: W283, PI: Forbes), 6, 7 (Program: W285, PI: Gannon), and 2025 April 30 (Program: U234, PI: Brodie). We used a Medium slicer with either the BL grating or the BH3 grating. The BL grating was set with a central wavelength of 4550\,\AA\,, while the BH3 grating was set with the central wavelength of 5110\,\AA\,. For all observations on the red arm, we made use of the RH2 grating, with the central wavelength set to 6750\,\AA\,. The detailed summary of the gratings used along with the number of exposures with their respective exposure times for each object used in this study is summarized in Table\,\ref{tbl:kcwi_obs}. The BL grating has an instrumental resolution  $\sigma_{\rm inst}=67$\,km\,s$^{\mathrm{-1}}$, while the BH3 grating has $\sigma_{\rm inst}=13$ km\,s$^{\mathrm{-1}}$. Across five of the six nights, the observations were mostly conducted in clear and dark skies with seeing conditions ranging from 1.2$\arcsec$ to 1.5$\arcsec$. On the sixth night, the night of October 28, we had only a few periods of cloudless conditions. For the work presented here, we only make use of the blue side datasets, and those obtained in cloudless conditions, while discarding frames affected by clouds. 

The main data reduction of IFU cubes was performed with the KCWI pipeline, except for the sky subtraction procedures, where a manual approach was preferred instead. We further performed cube trimming and additional flat-fielding procedures to remove residual instrumental gradient across the field as described in \citet{Gannon2020}. The cleaned cubes were then mosaic stacked using $\rm Montage$ \citep{Jacob2010}. As the Medium slicer was chosen for observation, the FOV encapsulates the bulk of the target galaxies as shown in Figure\,\ref{fig:hsc_hst_kcwi} with the 1\,$R_\mathrm{e}$ indicated in red ellipses. For the manual sky subtraction procedure, we extracted an on-chip global sky spectrum, which was taken from remaining empty sky regions within the slicer. 

%area of 16\arcsec$\,\times$\,20\arcsec (30\,$\times$\,44 spaxels). We then extracted on-chip global sky spectrum taken from an empty sky regions within the slicer. As a medium slicer is chosen for observation, all the targets are within the FOV area of 16\arcsec$\,\times$\,20\arcsec (30\,$\times$\,44 spaxels).

\section{Stellar Population Analysis}\label{sec:stellarpop}

To probe the stellar population properties of our sample of UDGs/NUDGEs, we made use of the full spectrum fitting code {\tt pPXF} \citep{Cappellari2017} to obtain properties such as mass-weighted ages and metallicities, along with the corresponding star formation histories. We additionally employ a classical line index approach to investigate the $\alpha$-abundances of these galaxies, using measurements of magnesium and iron sensitive absorption features. In all cases, we adopt E-MILES single stellar population models \citep{Vazdekis2015}, using the BASTI isochrone with a universal Kroupa initial mass function \citep{Kroupa2001} and scaled solar models abundances. The spectral resolution of the models is 2.52\,\AA, spanning the ages between 0.03 to 14\,Gyr and metallicities between [M/H]=$-$2.42 and +0.4\;dex. For each of the {\tt pPXF} fits, we incorporate only an optimal number of multiplicative polynomials, while avoiding the use of any additive polynomials in the derivation of ages and metallicities. As noted in \citet{AL2025}, we find irregular spectral continuum shapes that are likely due to imperfections in the spectral response calibration for low surface brightness targets, thereby necessitating a higher order multiplicative polynomial for the BL grating spectra (see also \citealp{FM2023}). So, for the multiplicative polynomials, we restrict the order to five for the shorter baseline BH3 grating spectra, and to seventeen for the BL grating spectra. The kinematic results, in the form of recessional velocities and velocity dispersion, attained in Paper II were also applied as additional constraints to the fits. Following the initial fit, we perform 1000 bootstrap fitting iterations for each target, adopting the median non-regularized solution as the final result, with uncertainties defined by the 16th and 84th percentiles.

%mention about the 3 sigma clipping for BH3 galaxies

\subsection{Global Ages, Metallicities and Star Formation Histories}\label{subsec:global}

%CHANGE THE METALLICITY AND ALPHA TO TWO DECIMAL PLACE??
For each galaxy target, we extracted a global galaxy spectrum by collapsing spaxels within one effective radius, $1\,R_{\mathrm{e}}$, while masking out any contaminating sources within the vicinity of the $1\,R_{\mathrm{e}}$ elliptical aperture. The results of {\tt pPXF} full spectrum fits are shown in Figure\,\ref{fig:global_fits}, with the median output ages and metallicities summarized in Table \ref{tbl:results_list}. %From the {\tt pPXF} full spectrum fits, we obtained mass-weighted stellar population ages, $t_{\rm M}$, and metallicities, [M/H]

The mean mass-weighted age of our sample is 7.1 $\pm$ 2.0 Gyr, spanning a range from $\sim$3.5 to 10.0 Gyr. Three of the NUDGEs exhibit young to intermediate ages between $\sim$3.0 and 6.0 Gyr, while the remaining systems have significantly older ages between $\sim$8 to 10 Gyr. We find that the uncertainties for the four NUDGEs observed using the BH3 grating tend to be elevated, owing to the lower S/N of the data and the much shorter wavelength baseline of the spectra, which is also reflected in the resultant star formation histories. We also caution that this is a known caveat, as such low S/N short baseline spectra can potentially bias inferred ages toward younger, intermediate values \citep{Forbes2022, Webb2022, FM2023}. All UDGs/NUDGEs in the sample are also found to have sub-solar metallicities, with a mass-weighted average of [M/H] = $-$0.9 $\pm$ 0.2\,dex. 

%We find that the uncertainties for the four NUDGEs observed using the BH3 grating tend to be likely elevated, owing to the lower S/N of the data and the much shorter wavelength baseline of the spectra. We also caution the reader about known caveats regarding a bias toward young-intermediate inferred ages when using a low S/N spectra shorter baseline \citep{Forbes2022, Web2022, FM2023}.

%Nonetheless, the three younger galaxies display high ellipticities, which may be indicative of a more disky morphologies, and it would therefore not be surprising to find younger stellar populations in these systems. The remaining galaxy observed with the BH3 grating, H410, which is the most GC rich system in the sample, instead shows an older age of 8.2 Gyr, consistent with expectations from previous studies of GC rich UDGs \citep[e.g.,][]{FM2018, FM2023}. 

%This is consistent with expectations from the MZR \citep{Kirby2013, Simon2019} in the dwarf galaxy stellar mass regime of $\sim 10^{8.5}\,\mathrm{M_{\odot}}$. The biggest outlier is DF11, which has the lowest metallicity in the sample, with [M/H] = -1.29 dex, placing it well below the relation. 

We additionally recovered star formation histories (SFHs), illustrated as the cumulative mass fraction as a function of lookback time, shown on the right panels for each spectrum in Figure\,\ref{fig:global_fits}. From this, we also calculated t$_{50}$ and t$_{90}$, corresponding to the time when 50\% and 90\% of the stellar mass of each galaxy has been formed. One can consider t$_{90}$ as the approximate quenching time of the galaxy, when most star formation has quenched \citep{FM2018}. Two of the galaxies with the youngest stellar ages consequently also exhibit more extended star formation histories, experiencing more gradual quenching that terminated only within the past $\sim$3 Gyr. H662 also shows a more extended star formation history, but with an earlier quenching timescale at intermediate epochs of $\sim$4 to 5 Gyr, while the remaining galaxies exhibit much shorter star formation histories, quenching well before $\sim$7 Gyr. All of the results on ages and metallicities for each target are summarized in Table \ref{tbl:results_list}.

\subsection{$\alpha$-abundances}\label{subsec:alpha}

To calculate the $\alpha$-abundances, in form of [Mg/Fe], for our sample of galaxies, we used the classical line indices fitting methodologies. Specifically, two different approaches are employed, both of which have been implemented and largely discussed in \citet{FM2018, FM2023, AL2025} and \citet{Doll2025}. The first approach involves direct comparisons between $\alpha$-sensitive absorption features such as Mg$_{\rm b}$ and iron sensitive features (e.g. Fe5015 or $\rm \langle Fe \rangle$\footnote{$\rm \langle Fe \rangle$ = 0.72 $\times$ Fe5270 + 0.28 $\times$ Fe5335}), using model index--index grids based on single stellar population models with varying levels of $\alpha$-enhancement. In this case, we make use of the new sMILES single stellar population models \citep{Knowles2023}, which provide an extended range of $\alpha$-enhancement from $-$0.2 to +0.6 dex, and with a finer step size of 0.2\,dex. 
Alternatively, we also compute [Mg/Fe] directly from the derivations of Z$_{\rm Mgb}$ and Z$_{\rm Fe}$, where [Mg/Fe] = Z$_{\rm Mgb}$--Z$_{\rm Fe}$. One can then convert between the two measures using the empirical relation [$\alpha$/Fe] = 0.02 + 0.56 $\times$ [Mg/Fe] \citep{Vazdekis2015}. In both methodologies, the ages of the single stellar population models are fixed to the {\tt pPXF} light-weighted ages obtained from the full spectrum fitting. To measure the line indices, we use the publicly available package indexf\footnote{https://indexf.readthedocs.io/en/latest/}\citep{Cardiel2010}.

As the BH3 spectra have a much finer wavelength resolution, we applied further smoothing, invoking the KCWI instrumental resolution ($\sigma_{\rm KCWI}$), the stellar velocity dispersion ($\sigma_{*}$) and templates resolution ($\sigma_{*}$), in the form of $\rm \sigma_{final}$ = $\rm\sqrt{\sigma^{2}_{template} - (\sigma^{2}_{KCWI}+\sigma^{2}_{*})}$ to match model spectra templates. As noted previously, the $\sigma_{*}$ used here was attained from Paper II. Given the shorter baseline of BH3, we could only make use of the Fe5015 line index to probe the iron abundance, whereas for the longer BL grating baseline spectra, we use $\rm \langle Fe \rangle$ (generally finding Fe5270 to be the strongest of the three aforementioned iron absorption features).

\begin{figure}
\centering
\includegraphics[width=7cm]{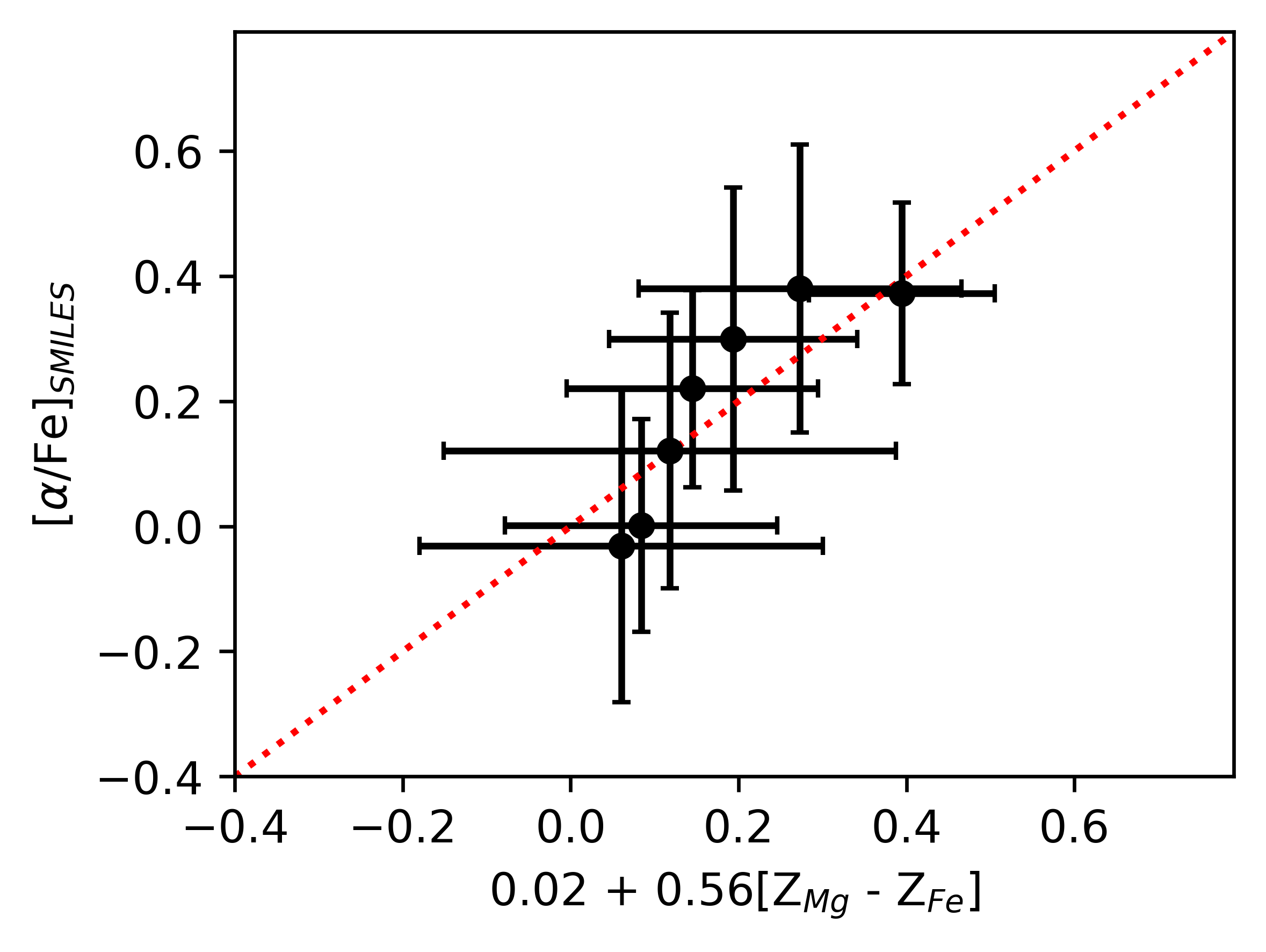}
\caption{Comparison of two line-index methodologies for deriving $\alpha$-abundances. The vertical axis shows values obtained using the S–MILES SSP grids to directly interpolate $\alpha$-enhancements, while the horizontal axis shows values derived from the \citet{Vazdekis2015} empirical relation between Z${\rm _{Mgb}}$ and Z${\rm _{Fe}}$. Overall, the two methods show good consistency, with measurements falling well within the uncertainties.}
\label{fig:index_check}
\end{figure}

Among the 7 galaxies, we find good consistency between the two methodologies, within the uncertainties, as shown in Figure\,\ref{fig:index_check}. For the final values reported in Table \ref{tbl:results_list}, we quote the average $\alpha$-abundance converted to [Mg/Fe], measured using both methodologies.

Overall, we find a wide range of [Mg/Fe] abundance patterns, with five galaxies exhibiting slightly elevated values between 0.0 and 0.3\,dex. The remaining two galaxies show much higher [Mg/Fe] $>$ 0.4\,dex, with DF11 reaching $\sim$0.5\,dex. This results are fully compatible with that of \citet{FM2018, FM2023} and \citet{Doll2025}, who report a similar spread in $\alpha$-abundance patterns for UDGs in cluster environments.

\renewcommand{\arraystretch}{1.2}%
\begin{table*}
    \centering
	\caption{A summary stellar population properties of our UDG/NUDGE sample}
	\label{tbl:results_list}
	\begin{tabular}{lcccccc@{\hspace{0.5em}\vline\hspace{0.5em}}cc} % four columns, alignment for each
		\hline
		Galaxy & Age ($\rm t_{M}$) & $\rm t_{50}$ & $\rm t_{90}$ & [M/H] & [Mg/Fe] & M$_{*}$ & $\nabla$log(Age) & $\nabla$log([M/H])\\
        & [Gyr] &  [Gyr] & [Gyr] & [dex] & [dex] & [10$^{8}\,\mathrm{M_{\odot}}$] & [log(Gyr)/log(R/R$\rm {_e}$)] & [dex/log(R/R$\rm{_e}$)]\\
		\hline
        H410 & 8.2 $\pm$ 2.7 & 9.9 $\pm$ 2.9 & 7.1 $\pm$ 3.6 & -0.85 $\pm$ 0.32 & 0.45 $\pm$ 0.32 & 7.1 & - & - \\
        H586 & 3.5 $\pm$ 2.3 & 8.8 $\pm$ 3.1 & 2.3 $\pm$ 2.8 & -0.69 $\pm$ 0.17 & 0.31 $\pm$ 0.24 & 8.2 & - & - \\
        H629 & 4.9 $\pm$ 2.9 & 8.2 $\pm$ 4.8 & 1.3 $\pm$ 3.6 & -0.87 $\pm$ 0.38 & 0.07 $\pm$ 0.32 & 2.7 & - & - \\
        H662 & 6.1 $\pm$ 3.1 & 9.5 $\pm$ 3.9 & 2.4 $\pm$ 3.7 & -0.66 $\pm$ 0.27 & 0.22 $\pm$ 0.23 & 12.8 & - & -  \\
        H332 & 8.7 $\pm$ 1.2 & 8.8 $\pm$ 1.4 & 8.1 $\pm$ 1.1 & -1.14 $\pm$ 0.11 & 0.18 $\pm$ 0.20 & 5.6 & -0.09 $\pm$ 0.07 & -0.33 $\pm$ 0.21 \\
        \hline
        W1 & 9.9 $\pm$ 1.3 & 9.8 $\pm$ 1.5 & 8.6 $\pm$ 1.4 & -0.93 $\pm$ 0.11 & 0.12 $\pm$ 0.22 & 6.7 & -0.07 $\pm$ 0.06 & -0.21 $\pm$ 0.14 \\
        DF11 & 8.1 $\pm$ 1.2 & 9.2 $\pm$ 1.1 & 8.1 $\pm$ 2.2 & -1.29 $\pm$ 0.08 & 0.49 $\pm$ 0.18 & 3.4 & 0.01 $\pm$ 0.05 & 0.34 $\pm$ 0.07\\
        \hline
	\end{tabular}
    \begin{tablenotes}
        \item{\textbf{Notes.}} Galaxy names (column 1). The global mass-weighted age from full spectrum fitting (column 2). The corresponding $\mathrm{t_{50}}$ and $\mathrm{t_{90}}$ are the ages when 50 and 90 percent of stellar mass was formed (columns 3 and 4). The global mass-weighted metallicity, [M/H], is listed in columns 5. [Mg/Fe] measure (column 6) based on a line index approach (see Section\,\ref{subsec:alpha}). The total stellar mass (column 7), computed based on the UDG luminosities in $g$ band and corresponding $M_{*}/L_{g}$ based on the aforementioned derived ages and metallicities. The age and metallicity gradients computed within 1\,R$\rm_{e}$ available for only three of the galaxies are listed (in columns 8 and 9). The sample of 5 NUDGEs and 2 UDGs are seperated by the horizontal line.  
    \end{tablenotes}
\end{table*}

%In comparison with previous literature, our results are consistent with the findings of \citet{FM2023}, who report a similar spread in $\alpha$-abundance patterns for UDGs in cluster environments.

%--> Compare method 1 vs method 2 plot?
%--> Further Full spectrum fitting method of deriving $\alpha$-enhancement is reserved for future studies
\begin{figure}
\hspace{-0.3cm}
\includegraphics[width=\columnwidth]{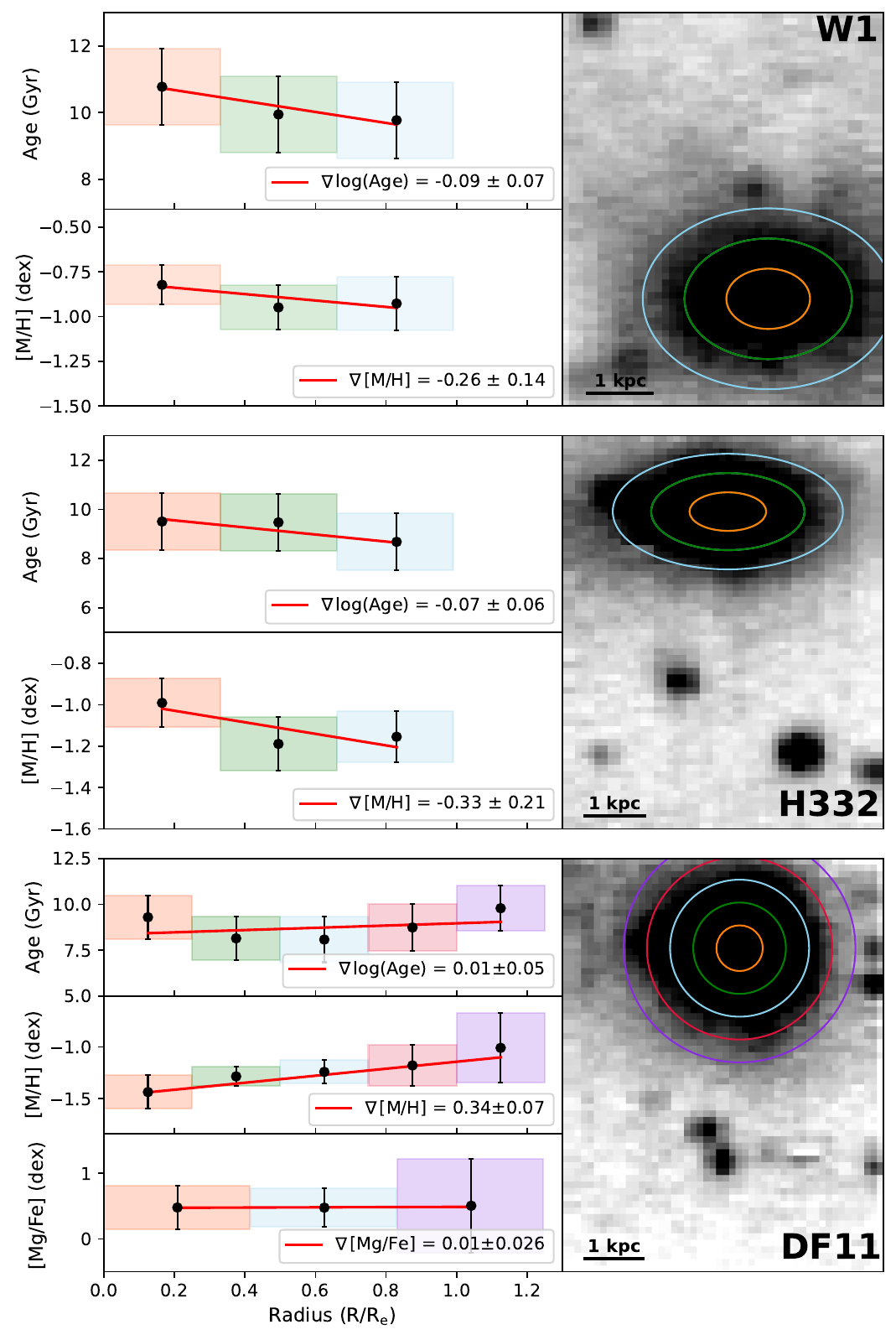}
\caption{Stellar population gradients for our three UDGs and NUDGEs with sufficient S/N. Right panels show white-light KCWI images with overlaid 1 kpc scale bars and annular regions. Left panels display derived age, metallicity, and [Mg/Fe] (for DF11) gradients, with shaded colour regions indicating uncertainties. Red lines represent best fit linear relations, with the respective slopes indicated. While W1 and H332 show flat age and flat to negative metallicity gradients, DF11 exhibits a positive metallicity gradient, unexpected in classical outside in dwarf formation.}
\label{fig:grad_all}
\end{figure}

\subsection{Stellar Population Gradients}\label{subsec:gradient}

\begin{figure*}
\hspace{-0.2cm}
\includegraphics[width=\textwidth]{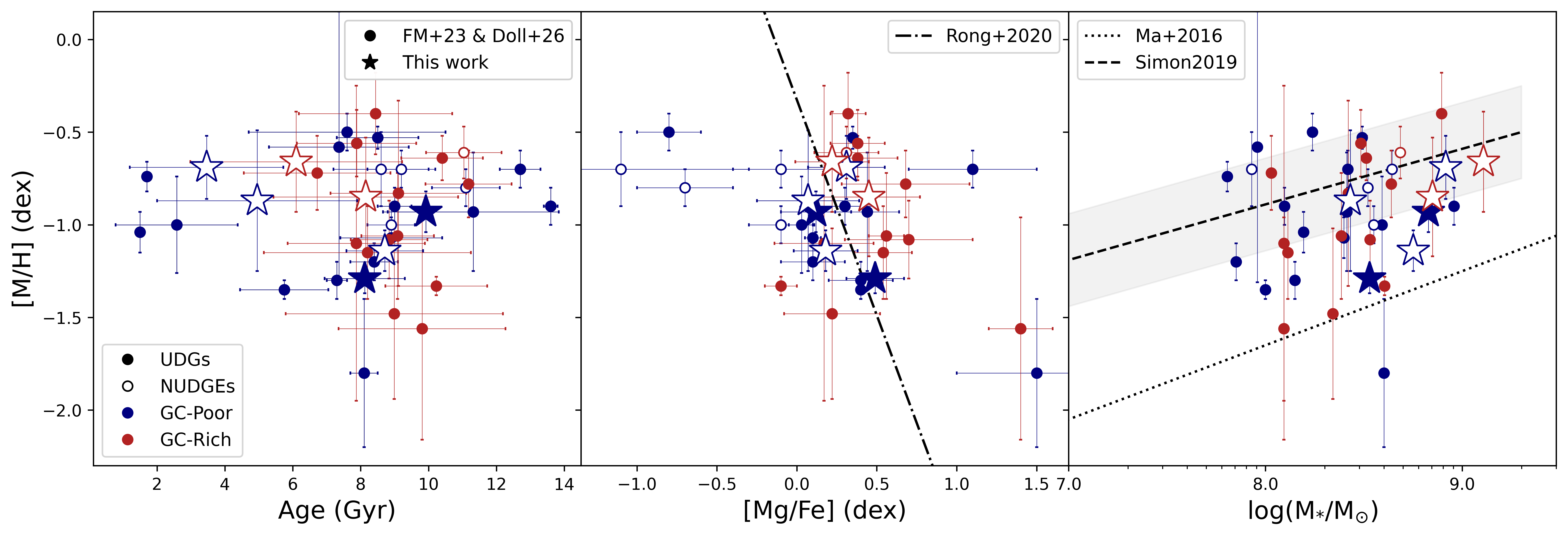}
\caption{Comparison of spectroscopically studied literature cluster-based UDGs/NUDGEs (filled/open circle symbols) with the sample investigated in this work (filled/open star symbols). The data points are also separated based on GC-richness classification, with galaxies that have $\rm N_{GC}$ $\gtrsim$ 20 considered GC-rich (red), while below that is characterized as GC-poor (blue). \textit{Left:} Distribution of cluster UDGs/NUDGEs from \citet{FM2023} and \citet{Doll2025} in age--[M/H] phase space. \textit{Middle:} Corresponding [M/H]--[Mg/Fe] distribution of literature and our sample. The dashed line represents the \citet{Rong2020} relation for field UDGs. \textit{Right:} Stellar mass--[M/H] relation for the sample, with the present-day MZR for low-mass galaxies \citep{Simon2019} including intrinsic scatter (dashed line with shading) and the MZR at $z \sim 2$ \citep{Ma2016} (dotted line). In all cases, our sample galaxies lie within the expected distribution in age, [M/H], and [Mg/Fe] found in previous works. In the mass–metallicity plane, the NUDGEs show slightly elevated stellar masses but still follow the present MZR. The most deviant point, DF11, lies closer to the higher-redshift MZR, further reinforcing its resemblance to a GC-rich UDG, and a failed galaxy candidate.}
\label{fig:MZ_plot}
\end{figure*}

With the increased efficiency of the BL grating compared to the BH3, we were able to achieve much higher signal-to-noise spectra for H332, W1, and DF11. We thus conducted a spatially resolved stellar population study, probing radial variations in stellar ages and metallicities. Similar to \citet{FM2025, Buzzo2025a} and \citet{AL2025}, we extracted spectra from annular elliptical regions around the centre of each galaxy, with the semi-major axis radius stepping outward at equal intervals, and having S/N of at least $\sim$15 \AA$^{-1}$. Three concentric annular regions were used, extending to one effective radius ($\rm 1\,R_{\mathrm{e}}$) for W1 and H332, while for DF11 we employed five regions extending to 1.25 $R_{\mathrm{e}}$ given the higher overall S/N of the data. We then applied the same stellar population analysis routine used for the global stellar population measurements to the spectra of each individual annular region. The annular regions and the derived age and metallicity gradients for all three galaxies are shown in Figure \ref{fig:grad_all} and also tabulated in Table\,\ref{tbl:results_list}. In addition to age and metallicity, we also obtained [Mg/Fe] measurements as a function of radius for DF11, making it only the second UDG after DF44 to have a measured [Mg/Fe] gradient. We use a similar line index approach as described in Section \ref{subsec:alpha}. For the [Mg/Fe] measurements, fewer annular regions were used in order to maximize the S/N of the individual spectra, restricting the analysis to three annular regions.

Among the three aforementioned galaxies, W1 and H332 show nearly flat age gradients, with $\nabla \log(\mathrm{Age})$ measured to be $-0.07 \pm 0.06$\,log(Gyr/$R$/$R_\mathrm{e}$) and $-0.09 \pm 0.07$\,Gyr/log($R$/$R_\mathrm{e}$), respectively. The metallicity gradients, $\nabla$[M/H], are mildly negative for both galaxies, with values of $-0.26 \pm 0.14$\,dex/log($R$/$R_\mathrm{e}$) and $-0.33 \pm 0.21$\,dex/log($R$/$R_\mathrm{e}$). These measurements are consistent with expectations from the literature on dwarf galaxies, which generally show similar trends in both age and metallicity gradients \citep[e.g.,][]{Koleva2011, Sybilska2017}. For DF11, we also find a nearly flat age gradient, with a value of $0.01 \pm 0.05$\,log(Gyr/$R$/$R_\mathrm{e}$), but a positive metallicity gradient of $+0.34 \pm 0.07$\,dex/log($R$/$R_\mathrm{e}$). This behavior contrasts with expectations for classical dwarf galaxies, but instead parallels the findings of \citet{FM2025}, who reported flat-to-positive metallicity gradients for samples of UDGs/NUDGEs. In addition, we derive $\nabla \log(\mathrm{Mg/Fe})$, which is also flat, with a slope of $0.01 \pm 0.03$\,dex/log10($R$/$R_\mathrm{e}$). The only other measured $\nabla$[Mg/Fe] of an UDG is DF44, having a negative gradient of $-$0.2 $\pm$0.18\,dex\,kpc$^{-1}$ within the inner $\sim$0.5 $R_\mathrm{e}$ \citep{Villaume2022}. 

%The natural interpretation would be the suggested inside-out formation model, while in the case here, much of star assembly must have occurred in very short duration for a flatter [Mg/Fe] gradient. 

%--> DF11, W1 and H332, 5 and 3 bins each --> Figure showing bins and coloured by age and metallicity? 

\section{Discussion}\label{sec:discussion}

\subsection{Global stellar population trends}\label{subsec:global_trends}

%\subsubsection{Literature UDG and NUDGEs comparison}\label{subsubsec:lit_global}
%1) Global properties, --> GC based differentiation ?

%2) Infall based diagnostics
% quick section to show no strong correlation here...

We begin by placing our sample of galaxies in the context of the growing spectroscopic sample of UDGs/NUDGEs in cluster environments \citep[e.g.,][]{FM2018, Chilingarian2019, FM2023, Gannon2024, Doll2025}. In Figure \ref{fig:MZ_plot}, we present different UDG scaling relations characterising galaxies in the parameter space of stellar age, metallicity, $\alpha$-abundance in the form of [Mg/Fe], and stellar mass, putting additional emphasis on the GC-richness and UDGs/NUDGEs classifications.

In the left panel of Figure \ref{fig:MZ_plot}, we show the age versus [M/H] parameter space. Three galaxies from this work display relatively young mean stellar ages with slightly higher metallicities. %These galaxies also display higher ellipticities of the sample, which may be indicative of a more disky morphologies, and it would therefore not be surprising to find younger stellar populations in these systems. 
These galaxies display the highest ellipticities in our sample, potentially consistent with the predictions of \citet{Pfeffer2024} that GC-poor systems are systematically younger and flatter, although a detailed investigation into their rotational support is deferred to Paper II. The remaining four systems show older ages and with slightly lower metallicities. When comparing to the overall available UDGs/NUDGEs population, we find that our sample is consistent with previous measurements. Looking at the population of UDGs and NUDGEs separately, we also do not see any clear separation between the two samples, echoing the findings of \citet{Buzzo2025} with photometric SED analysis. When distinguishing between the GC-richness, we note of a larger scatter in ages for the GC-poor population, whereas GC-rich UDGs generally occupy the older stellar ages phase space signifying an earlier stellar assembly and quenching. The metallicities distribution remains scattered for both GC-rich and GC-poor population of UDGs/NUDGEs in clusters \citep{FM2023, Buzzo2024}. 
 %This overall behavior broadly follows expectations from classical galaxy evolution, where younger stellar populations are typically associated with higher chemical enrichment. Consequently, as noted in \citet{FM2023}, the majority of UDGs/NUDGEs, including those presented here, are largely comparable to the classical dwarf population. 

A similar picture emerges when considering [Mg/Fe]. The middle panel of Figure \ref{fig:MZ_plot} presents the [M/H] versus [Mg/Fe] distribution, where the galaxies in our sample are consistent with previous literature measurements of UDGs/NUDGEs. For comparison, we also include the [Mg/Fe] versus metallicity relation from \citet{Rong2020}, derived for isolated star forming UDGs. While some of the galaxies in our sample scatter around this relation, the overall UDGs/NUDGEs population exhibits a relatively flat trend, as in \citet{FM2023}, and displays substantial scatter in metallicity at fixed $\alpha$-abundance. Given the uncertainties associated with individual [Mg/Fe] measurements, these trends should nevertheless be interpreted with appropriate caution. We also do not see a distinction among the UDGs/NUDGEs, but a preference for high [Mg/Fe] for GC-rich population is noted, confirming the results reported in \citet{FM2023}. 

However, one of the key relations to study in UDGs has been the mass–metallicity distribution shown in the right panel of Figure \ref{fig:MZ_plot}, which provides a useful constraint on the diverse evolutionary histories proposed for these systems. We observe that our sample of galaxies largely lies within the bulk of the literature sample, with a slight bias toward the higher stellar mass end. This, however, is expected given that some galaxies in our sample are classified as NUDGEs and are therefore slightly brighter than ``bona fide" UDGs, while having a similar physical size. We find that six galaxies in our sample lie within the scatter of the present-day MZR for dwarf galaxies \citep{Simon2019}, similar to the literature UDG/NUDGE sample, reflecting their close resemblance to classical dwarf galaxies. The remaining galaxy, DF11, lays slightly below the present MZR with a [M/H] = -1.29 dex, thus trending toward the theoretical relation at $z \sim 2$ \citep{Ma2016} instead. This behavior aligns more with the failed galaxy scenario, in which galaxies experience early and rapid star formation at high redshift, followed by fast quenching that limits further metal enrichment \citep[e.g.,][]{Forbes2024}. As a result, such systems are expected to host older stellar populations with very low metallicities, resembling the properties of their rich GC populations \citep[see][]{Buzzo2022, FM2023}. While DF11 is classified as a GC-poor UDG, with a reported GC count of $9.4 \pm 12.7$ \citep{Lim2018}, it still remains consistent within the uncertainties with also being GC-rich as well. DF11 also lies closest to the $z \sim 2$ MZR in our sample deviating only by $\sim$0.05 dex, thus offering further support for its interpretation as a possible failed galaxy candidate.

Overall, the sample in general also exhibits significant scatter across the MZR for both UDG and NUDGE populations, regardless of globular cluster richness as seen in previous works. While some GC-rich systems follow the local relation, others align with the high-redshift MZR; a similar dispersion is observed among GC-poor galaxies. The lack of clear segregation based on GC richness suggests that cluster UDGs represent a composite population formed through diverse evolutionary pathways \citep[e.g.,][]{Jiang2019}. %This population likely comprises a mixture of failed galaxies and dwarfs expanded by internal processes such as stellar feedback \citep[e.g.][]{Dicintio2017} and high halo spin \citep{Amorisco2016}, alongside those shaped by their external environment through processes like tidal heating or ram-pressure stripping \citep[][]{Jiang2019, Pfeffer2024}. 

% more internal process as feedback, or environment depended UDGs formation, associating with tidal heating or ram-pressure stripping.

%Nonetheless, the three younger galaxies display high ellipticities, which may be indicative of a more disky morphologies, and it would therefore not be surprising to find younger stellar populations in these systems. The remaining galaxy observed with the BH3 grating, H410, which is the most GC rich system in the sample, instead shows an older age of 8.2 Gyr, consistent with expectations from previous studies of GC rich UDGs \citep[e.g.,][]{FM2018, FM2023}. 

%This is consistent with expectations from the MZR \citep{Kirby2013, Simon2019} in the dwarf galaxy stellar mass regime of $\sim 10^{8.5}\,\mathrm{M_{\odot}}$. The biggest outlier is DF11, which has the lowest metallicity in the sample, with [M/H] = -1.29 dex, placing it well below the relation. 

%--> mass metallicity relation, age-metallicity --> oldest metal poor track GC population rather than dEs
%--> brief discussion on GC differentiation

\subsubsection{Cluster environment dependence}\label{subsubsec:cluster_dep}

\begin{figure}
\hspace{-0.5cm}
\includegraphics[width=9cm]{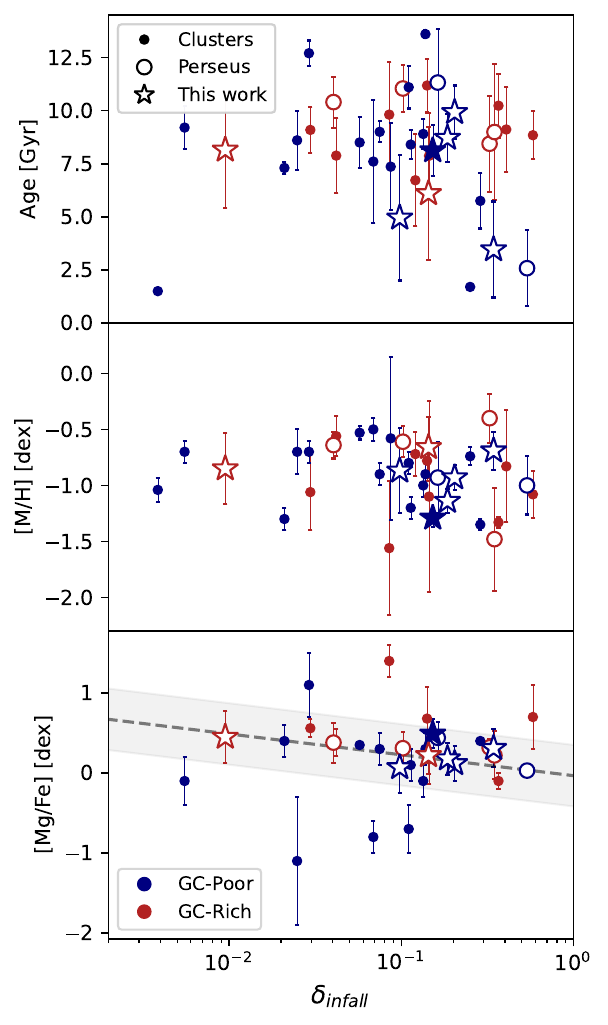}
\caption{UDG/NUDGE stellar population dependence on the infall parameter. \textit{Top:} Mass-weighted ages of UDGs/NUDGEs Perseus cluster objects \citep{FM2023, AL2025} are shown in open symbols; in other clusters (filled symbols) from \citet{FM2023} and \citet{Doll2025}; galaxies from this work are star symbols (filled for Perseus targets and open for Coma target). As in previous figure, the colours represent the GC-richness classification. The infall parameter is calculated following the prescription as done in \citet{Tang2025} (See text). \textit{Middle:} Same as the top panel, but showing mass-weighted metallicities. \textit{Bottom:} Comparison of the infall parameter with $\alpha$-abundance in the form of [Mg/Fe]. The grey dotted line indicates the best-fitted linear relation, with a minor slope of -0.17. Overall, no strong correlation is observed between stellar ages or metallicities and cluster infall, while a weak trend of increasing $\alpha$-abundance for earlier infall galaxies is seen, consistent with the conclusions of \citet{FM2023}.}
\label{fig:infall_plot}
\end{figure}

The cluster environment certainly plays a role in shaping UDGs/NUDGEs with many notable observational and simulation based works exploring this idea \citep[e.g.][]{Yozin2015, Carleton2019, Jiang2019, Sales2020, Grishin2021, Juanis2022, Forbes2023, Tang2025}. Yet whether such environmental effects leave a clear observational imprint on the present-day stellar population properties of these galaxies remains an open question. Previous works, such as \citet{FM2023, Buttitta2025, Doll2025}, have investigated UDGs/NUDGEs properties as a function of projected cluster-centric radius, and further probed their locations in phase space diagrams.

\begin{figure}
\hspace{-0.2cm}
\includegraphics[width=8.5cm]{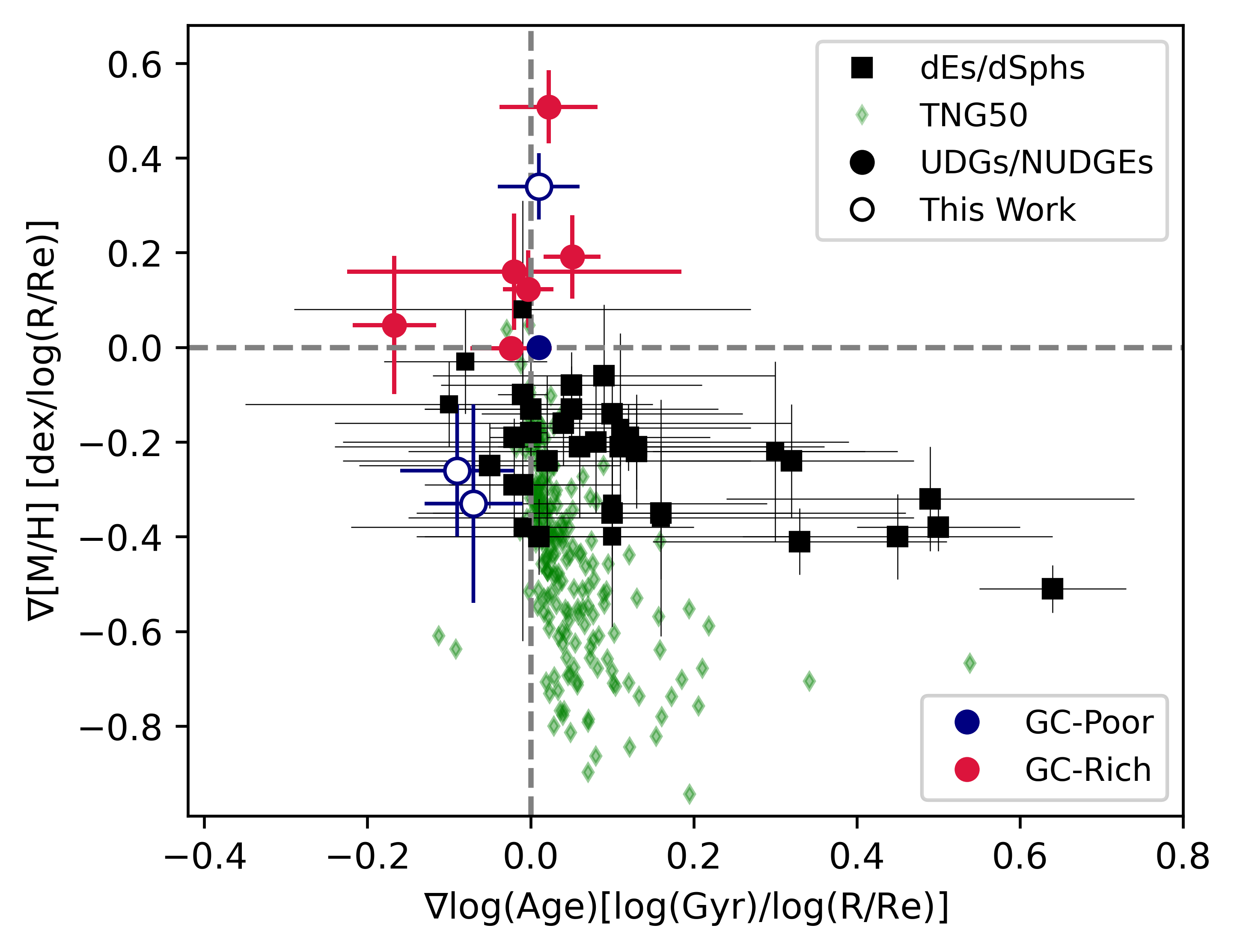}
\caption{Age and metallicity gradients within 1 $R_\mathrm{e}$ for UDGs/NUDGEs (filled circles for literature sample and open circles for this work sample), cluster-based classical dwarf galaxies (black rectangles), and simulated dwarf galaxies (green diamonds), similar to the figure from \citet{FM2025}. The color scheme of the UDGs/NUDGEs is the same as in Figure\,\ref{fig:infall_plot}, separated by the GC-richness. Both classical dwarf galaxies and simulated dwarfs show flat-to-negative metallicity gradients and flat-to-positive age gradients. Some UDGs follow this trend, while others, including DF11 of this work, exhibiting a clear flat-to-positive metallicity gradients, differing from the bulk of the remaining sample.}
\label{fig:grad_age_met}
\end{figure}

Here, we adopt the infall metric as similarly done in \citet{Tang2025}, which incorporates both the velocity and positional phase space of galaxies (see also \citealp{Haines2012, Alabi2018, FM2018}). In summary, the infall parameter, $\delta_{\rm infall}$, uses the projected cluster-centric radius of the galaxy, $R$; the measured radial velocity $V$; the systemic velocity and velocity dispersion of the cluster, $V_{0}$ and $\sigma$, respectively, similar to the projected phase-space analysis of \citet{Rhee2017}, arriving at the same conclusions. The equational form of $\delta_{\rm infall}$ is defined as:
\begin{equation}
\delta_{\rm infall} \equiv R/R_{200} \times |V - V_{0}|/\sigma, 
\end{equation}
Smaller values of $\delta_{\rm infall}$ indicate a higher likelihood of a galaxy being an early cluster infaller. In Figure\,\ref{fig:infall_plot}, we examine the dependence of UDGs/NUDGEs stellar population properties on the $\delta_{\rm infall}$ parameter.

In the cases of mass-weighted ages (top panel) and [M/H] (middle panels), no obvious correlation is seen in Figure\,\ref{fig:infall_plot}. Intuitively one would expect that older galaxies preferentially exhibit smaller $\delta_{\rm infall}$ values, which would align with the notion that low-mass galaxies that are predominantly old and $\alpha$-enhanced have resided in cluster environments for longer periods \citep{Liu2016, Pasquali2019, Gallazzi2021, Bidaran2022}. However, no strong dependence can be established here, with large scatter of ages across different $\delta_{\rm infall}$ as in \citet{FM2023}, consistent with \citet{RomeroGomez2024} who similarly report no strong dependence of age or metallicity on cluster environment for classical Fornax dwarfs. In terms of the Perseus cluster sample specifically, it is worth noting that \citet{Tang2025} found no correlation between the infall parameter and the GC-richness across a large sample of dwarf galaxies within the cluster. This was attributed to the potential unique merger history of the cluster. Although, \citet{Forbes2023} also similarly found no correlation between GC-richness and galaxy infall in a combined sample of UDGs in four clusters (Coma, Perseus, Hydra I and Virgo). In this context, the lack of observed trends in ages and metallicities within our sample may also be a reflection of the unique evolution history of each cluster. Additionally, the lack of a significant age trend may point toward the importance of pre-infall quenching for a subset of these systems, highlighting the different formation pathways of UDGs found in clusters as pointed out in e.g. \citet{Jiang2019} and \citet{Pfeffer2024}. The quenching then could be associated with either strong stellar feedback scenarios \citep[e.g.][]{Dicintio2017} or a failed galaxy origin \citep[e.g.][]{FM2018, Buzzo2025}. Nonetheless, a slight correlation can be observed in the [Mg/Fe]–$\delta_{\rm infall}$ phase space. We measure a moderate negative correlation ($p$-value of 0.02 and Spearman correlation coefficient of -0.43) using the Spearman correlation test, with the best-fit weakly negative linear relation with a slope of -0.17$\pm$0.05  plotted in the bottom panel of Figure\,\ref{fig:infall_plot}. This is consistent with the findings of \citet{FM2023}, where earlier infall UDGs presented more elevated $\alpha$-abundance measurements, as well as \citet{Liu2016}, who found a similar correlation between [$\alpha$/Fe] and low-mass early-type galaxies in the Virgo cluster. It is well known that $\alpha$-abundances act as effective chemical clocks of galaxy formation, due to the increasing contribution of iron from Type Ia supernovae over time \citep[e.g.][]{Worthey1992}, and such a trend with cluster infall is therefore expected.

We also note the ambiguity of phase-space tracers arising from the assumption of two-dimensional projected distances, which inevitably introduces interlopers and uncertainties in characterizing the infall times of galaxies within the sample (e.g. see Ivleva et al. in prep). The presence of backsplash galaxies further complicates this interpretation, as galaxies that have already completed a first passage through the cluster core can still exhibit large $\delta_{\rm infall}$ values \citep[e.g.][]{More2015}, thus all contributing to the potential lack of apparent trends observed here. Given the statistical nature of phase-space analysis, a much larger sample of UDGs is required to draw firmer conclusions regarding environment-based correlations with galaxy stellar population properties.

%--> Anna 2023 results on UDG properties -- no effect
%--> What we see here --> How does it compare
%--> Extention of Yi Meng paper for Perseus objects
%--> How reliable are these infall things --> Anna I. reference

%basic differentiation with age metallicity and morhpology, consistency with previous studies by Anna, Luisa etc...

\subsection{Stellar population gradients trends}\label{subsec:gradient_trends}

%\subsubsection{Literature Comparison}\label{subsubsec:lit_comp}

Recent investigations into spatially resolved studies of UDG stellar population properties have yielded intriguing results, with \citet{FM2025} showing that some UDGs exhibit flat-to-positive metallicity gradients (see also \citealp{Villaume2022}). This appears contrary to the bulk of measured metallicity gradients in classical dwarf galaxies in cluster-based environments. As an example, in Figure\,\ref{fig:grad_age_met} we reproduce the $\nabla$log(Age) versus $\nabla$[M/H] diagram following \citet{FM2025}, but with a more extended sample of UDGs/NUDGEs compiled from \citet{FM2025, Buzzo2025} and \citet{AL2025}, together with the three additional galaxies from this work. The comparison sample of classical dwarfs is drawn from \citet{Koleva2011, Sybilska2017} and \citet{Bidaran2023}, focusing primarily on Virgo and Fornax cluster dwarf galaxies as the closest available counterparts to our UDG/NUDGE sample. Complementary to this, we also include age and metallicity gradient measurements within 1 $R_{\mathrm{e}}$ from the TNG50 simulation of quenched UDGs \citep{Benavides2024}, in low-mass cluster environment setting, for further comparison. While our UDGs/NUDGEs sample is drawn from high-mass clusters, like Coma and Perseus clusters, where tidal effects are more pronounced than in simulations, we still consider this a valuable comparison, but noting that simulated gradients remain subject to uncertainties in the treatment of supernova feedback and the resulting stellar distributions in dwarfs \citep[e.g.][]{Zhang2025}.

The cluster-based classical dwarf galaxies and simulated UDGs broadly show consistency in their age and metallicity gradients, with flat-to-negative metallicity gradients and flat-to-positive age gradients. We note that variation in age gradients is more truncated for the simulated UDGs, but would be understandable given the use of only quenched UDGs sample. For the sample of GC-poor UDGs/NUDGEs in this work, similar results are found for two systems, with W1 and H332 showing mildly negative to flat metallicity gradients. Many of the UDGs/NUDGEs, however, appear to be disjoint from bulk of the cluster-based classical dwarfs and simulated UDGs, differing in their metallicity gradients, exhibiting flatter or even positive gradients. This includes one galaxy in this work, DF11, which shows the puzzling rising metallicity gradient, as in the cases of other GC-rich UDGs like DF44, DFX1 and PUDG-R27 \citep{FM2025}. 

\begin{figure}
\hspace{-0.3cm}
\includegraphics[width=\columnwidth]{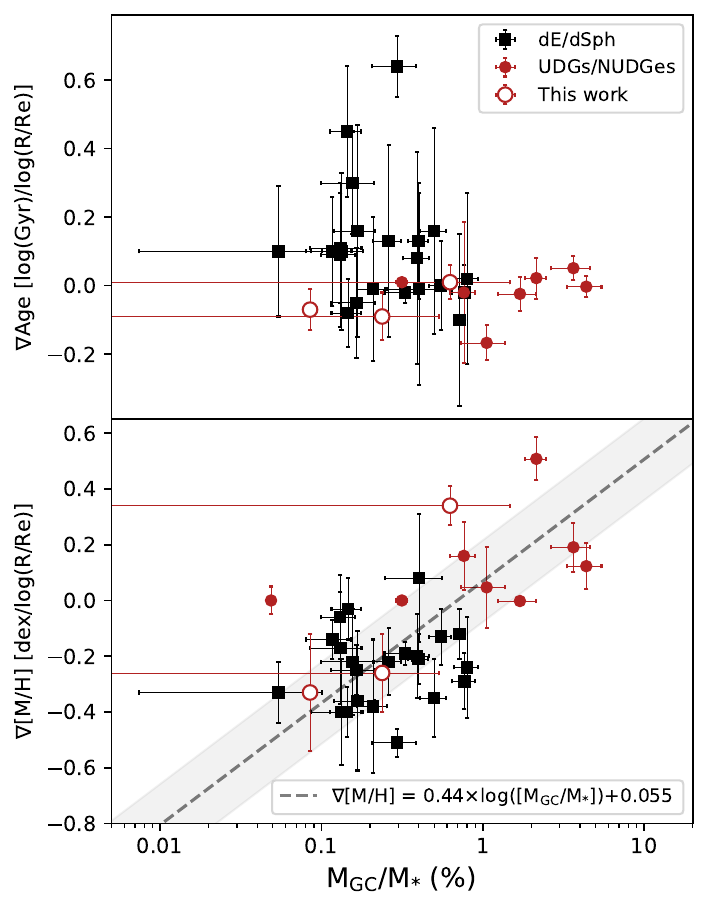}
\caption{Age (top panel) and metallicity (bottom panel) gradients versus GC to galaxy stellar mass ratio ($\rm M_{GC}/M_{*}$) for a sample of UDGs/NUDGEs and classical dwarfs (dEs and dSphs) in cluster environments. The classical dwarf sample (black squares) is the same as in Figure\,\ref{fig:grad_age_met}, with the GC and stellar masses from \citet[][for Virgo cluster galaxies]{Peng2008} and \citet[][for Fornax cluster galaxies]{Liu2019} as part of the HST/ACS surveys. The UDG/NUDGE sample includes gradients from \citep{FM2025, Buzzo2025, AL2025} (red open circles) and three galaxies from this work (red filled circles). Age gradients show no clear correlation with $\rm M_{GC}/M_{*}$, with most UDGs/NUDGEs having flat gradients similar to the bulk of classical dwarfs. However, a stronger relation is observed for metallicity gradients, with UDGs/NUDGEs having the highest $\rm M_{GC}/M_{*}$ showing more positive metallicity gradients. We find this correlation to be statistically significant, with a linear-fit indicated by the black dashed line and its associated uncertainty shown as grey bands. Overall, we find GC-richness to be correlated with metallicity gradients but not with age gradients}
\label{fig:grad_MGC}
\end{figure}

In the conventional picture, galaxies without significant tidal interactions or mergers are expected to develop negative metallicity gradients, as gas becomes progressively self-enriched over time and flows toward the central regions, accompanied by continued star formation that produces younger and more metal-rich stellar populations \citep[e.g.][]{Pipino2008, Benitez2013, Revaz2018}. This behavior follows the standard outside-in formation scenario. Some low-mass dwarf galaxies, however, exhibit flatter stellar metallicity gradients compared to higher-mass systems \citep[e.g.][]{Taibi2022}. Several mechanisms have been proposed to explain this behavior, including feedback-driven redistribution of stellar orbits \citep{Koleva2011, Mercado2021}. This effect is especially pronounced in simulations where radial movements of stars are highly sensitive to the subgrid treatment of supernova feedback \citep{Zhang2025}, which can lead to drastic changes in the predicted gradients. Other mechanisms include late-time gas accretion followed by star formation in galaxy outskirts \citep{Mercado2021}, external effects such as significant tidal stripping \citep{Benavides2024}, mergers that incorporate old, metal-poor populations while redistributing existing stellar orbits \citep{Mercado2021}, and the outward migration of star-forming regions in rotationally supported galaxies \citep{Cardona2023}. The latter is further explored in Paper II, investigating rotation properties of the galaxies. Therefore, each of these processes can soften or partially erase pre-existing stellar population gradients. In the case here, W1 and H332 are entirely consistent with such outside-in formation models, thus being similar to classical dwarfs. Meanwhile, DF11, like the other investigated GC-rich UDGs in literature, appears to form its positive metallicity gradient in a different manner than the mechanisms discussed above.

%For the sample of GC-poor UDGs/NUDGEs in this work, similar results are found for two systems, with W1 and H332 showing mildly negative to flat metallicity gradients, consistent with the outside-in formation picture. 

%--> Caveat on Fe/H gradients 

%--> Talk about NIHAO simulation results briefly, but are field dwarfs

%--> Segway into looking at the difference of UDGs vs dwarfs --> are all of those simply tail end?

%--> How does the ages and metallicity gradients compare to other dwarfs? 

\subsubsection{Relation with GC-richness}\label{subsubsec:GC_relation}

%Why do the current samples of UDGs/NUDGEs generally show a tendency toward flat-to-positive metallicity gradients, unlike many classical dwarf galaxies? This may be linked to a key factor that distinguishes some UDGs, namely their unusually large number of globular clusters relative to their stellar mass. To test this idea, 
Given that GC richness is one of the most striking features distinguishing some UDGs from classical dwarfs, it is natural to ask whether it bears any relation to their stellar population gradients. Here we plot the age and metallicity gradients as a function of the ratio between stellar mass and total GC stellar mass in Figure\,\ref{fig:grad_MGC}, analogous to the approach 
of \citet{Forbes2025} for global stellar population properties. We adopted the previously mentioned UDGS/NUDGEs sample, but now restricting the analysis to galaxies with only available GC number counts. The GC counts for the classical dwarf sample are taken from \citet{Liu2019} for the Fornax cluster and \citet{Peng2008} for the Virgo cluster, both based on HST imaging. For the remaining UDGs/NUDGEs, we adopt the values reported in the catalog of \citet{Gannon2024}, if they were available. Since GC number counts naturally increase with increasing galaxy stellar mass, much of the underlying trend may be obscured given that the stellar masses of the dwarf galaxies with available GC counts span range of $\sim$8 to 9.5 in $\log(M_{*})$. Therefore, in Figure\,\ref{fig:grad_MGC} we instead adopt the commonly used metric of the $\rm M_{GC}/M_{*}$ ratio, assuming a typical $M_\mathrm{GC}$ $\sim$ 2$\times$10$^{5}\,M_{\odot}$ \citep{Forbes2025}. In regards to the dependence of gradients on stellar mass of galaxies, previous studies have shown correlations for higher-mass early-type galaxies \citep[e.g.][]{Spolaor2009}, but lack of an apparent one in the lower-mass regime \citep{Taibi2022, Tau2025}.
%As before, we adopt the UDG/NUDGE sample from \citet{FM2025, Buzzo2025, AL2025} and the classical dwarf sample from \citet{Koleva2011, Sybilska2017} and \citet{Taibi2022}, restricting the analysis to galaxies with available GC number counts.

As shown in Figure\,\ref{fig:grad_MGC}, we find no significant correlation between $M_\mathrm{GC}/M_{*}$ and the age gradients, which is not surprising given the old stellar ages and overall similarity in age gradients between UDGs/NUDGEs and classical dwarf galaxies.  In contrast,  the metallicity gradients display a more apparent correlation with $M_\mathrm{GC}/M_{*}$ once UDGs/NUDGEs are included. Specifically, we observe a general trend toward more positive metallicity gradients with increasing $M_\mathrm{GC}/M_{*}$, with UDGs/NUDGEs occupying the region of higher $M_\mathrm{GC}/M_{*}$ values ($\gtrsim$ 1\%). Conversely, classical dwarf galaxies, together with the two UDGs in this work that have relatively low $M_\mathrm{GC}/M_{*}$, occupy the region characterized by flat-to-negative metallicity gradients and smaller $M_\mathrm{GC}/M_{*}$. The overall correlation is moderately strong and statistically significant as indicated by Spearman correlation with a coefficient of 0.55 and $p$-value of 0.02, with the best-fitting linear slope of 0.44$\pm$0.12. Considering the UDG/NUDGE sample alone, and although limited in sample size, we still find a similar positive trend with a Spearman correlation coefficient of 0.58, albeit with marginal significance.

\subsection{What drives the positive metallicity gradients?}\label{subsec:positive_grad}

The positive metallicity gradients in UDGs/NUDGEs are difficult to reconcile with the standard evolutionary pathways invoked for classical dwarf galaxies that generally produce flat-to-negative gradients. In massive galaxies, for instance, most of the flattening occurs via major mergers during ex-situ formation pathways, which dynamically mix and redistribute stars through violent relaxation, erasing pre-existing gradients \citep[e.g.][]{Kobayashi2004}. This type of ex-situ formation, however, is typically absent in dwarf galaxy populations. For dwarf galaxies, one obvious explanation that has been proposed relates back to environmental processing, such as tidal stripping, which can preferentially remove metal-poor stars from the outskirts of galaxies and thus flatten, and in extreme cases, invert pre-existing negative metallicity gradients \citep{Benavides2024}. This idea could be attractive to explain cluster UDGs/NUDGEs, however, none of our sample ones exhibit clear signatures of strong tidal disturbance, or any signs of tidal stripping. While the tidal disturbance evidence might have faded over time, we still find the metallicities of our galaxies to be well within or below present MZR, which differs from the ones generally found in tidal stripped galaxies \citep[e.g.][]{Williamson2016}. The explanation also doesn't address the observed relation between $M_\mathrm{GC}/M_{*}$ and metallicity gradients as shown in Figure\,\ref{fig:grad_MGC}. So, the tidal stripping scenario alone would likely not be the dominant mechanism responsible for the observed positive gradients. 

A plausible mechanism to generate positive gradients was explored in \citet{Mori1997}, invoking the idea of a supersonic spherical outflow from an initial starburst concentrated at the galaxy center. The shock wave propagates outwards and collides with infalling accreted gas, forming high density super shell. With further cooling and accretion of gas, it is enough to spark new star formation. The propagating shock front, further accelerated by subsequent SN feedback, simultaneously also enriching the surrounding gas, slowly drives new star formation with increasing radius. This swiftly builds up the stellar contents of the dwarf galaxies, resulting in a minor age differentiation as function of radius, while still naturally generates a positive metallicity gradients due to continuous metal enrichment of gas throughout the process. 

To incorporate GCs into this idealized simulation scenario, high enough natal pressure and gas densities would be required to trigger such strong burst of star formation \citep[e.g.][]{Kruijssen2014, Pfeffer2024}. This would be aligned with the initial conditions of violent centrally concentrated star formation proposed in the scenario above. The subsequent feedback would eventually expel most of the gas from the galaxy, potentially quenching the galaxy, and altering its gravitational potential to transition to a more cored dark matter profile, as we see in many of the cluster UDGs \citep[e.g.][]{Gannon2022, Forbes2024}. Within this framework, galaxies with the most efficient GC formation would generate the strongest feedback, which may result in the steepest positive gradients. Whether equivalent GC-rich dE galaxies follow the same mechanism remains uncertain. The majority of literature dEs presented here tend to be relatively GC-poor and generally show more conventional negative metallicity gradients. This could be related to the strength of the initial shock or their higher stellar masses, which enable greater gas retention and facilitate more effective rejuvenation of star formation. That said, while this feedback model offers a plausible explanation to the observed positive gradients, the continuous distribution of metallicity gradients with GC richness also suggests that the GC populations themselves may play a more direct role.

%From observation, we find a variation of metallicity gradients among UDGs, which seemingly favors positive metallicity gradients only in GC-rich UDGs. %which subsequently triggers the formation of stars in the expanding, outward-propagating, metal-enriched shell, thereby producing a positive metallicity gradient. In this simulation, 

We therefore alternatively propose an explanation that naturally links back to the GC populations of these galaxies. Given that GCs consist of sightly older and more metal-poor stellar population than bulk of their host galaxy stars, \citet{Danieli2022} and \citet{Forbes2025} explored the idea that disrupted GCs could contribute to the field star population of UDGs, particularly in failed galaxy candidates, inferring very high GC formation efficiencies together with modest rates of GC destruction in GC-rich UDGs. Building on this model, we consider the same process could also contribute to the observed metallicity gradient trends presented here. 

\begin{figure}
\hspace{-0.7cm}
\includegraphics[width=9.5cm]{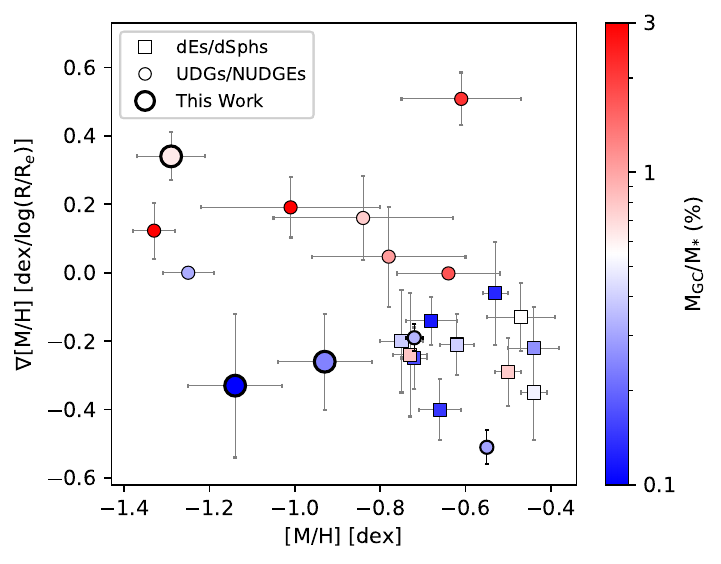}
\caption{Galaxy metallicity versus metallicity gradient for classical dwarfs and a sample of UDGs/NUDGEs coloured by the $\rm M_{GC}/M_{}$. The sample of classical dwarf galaxies in rectangular symbols are limited to those with measured [M/H] (see text). Small circular symbols represent literature UDGs/NUDGEs, while galaxies from this work are highlighted with a thick black outline. From the sample here, UDGs/NUDGEs, being generally more metal-poor than classical dwarfs, tend to exhibit more positive metallicity gradients, while also having high $\rm M_{GC}/M_{}$ ratios.}
\label{fig:grad_Z}
\end{figure}

The idea revolves around the in-spiral of some GCs due to dynamical friction and their continuous tidal stripping during migration toward the galaxy centers, ultimately tidally dissolving the cluster or contributing to the formation of a nuclear star cluster. Such a mechanism of GCs destruction have been explored in \citet{Meng2022} and \citet{Rodriguez2023}, simulating Milky Way type galaxy GCs, finding more GCs to be destroyed in the galaxy inner regions owing to the increasing strength of the tidal fields. In this scenario, the old and more metal-poor GCs in dwarf galaxies under the tidal field would deposit their stars into the central region, naturally causing the stellar metallicity of the inner galaxy region to be lower. Here we can adopt a simple model with an average GC metallicity of [M/H] = -1.5\,dex, while a uniform [M/H] of -1\,dex for the galaxy field stars, similar to the toy-model in \citet{Forbes2025}. Age is not considered here for simplicity, as we expect a fast stellar assembly for most of these galaxies and also find no corresponding trend in our age gradients. So, for instance, to raise the metallicity by 0.1\,dex within 1 $R_\mathrm{e}$, we would require up to 20\% of stellar mass contribution from disrupted GCs. For comparison, the contribution to the stellar mass of the Milky Way inner halo from disrupted GCs is estimated to be at least 25\% \citep{Schiavon2017}. While the required stellar mass contribution from disrupted globular clusters must be high especially in the cases of $\nabla$[M/H] of $\sim$0.34\,dex/log($R$/$R_\mathrm{e}$) as in DF11, the idea of significant early destruction combined with continued gradual dissolution over cosmic time may still make this a plausible mechanism for generating weak positive metallicity gradients in UDGs.

A natural consequence of this scenario may also be reflected in the global metallicity of the galaxy, as the disruption of GCs stars would lower the global metallicity as well. Following this, we plot the measured metallicity gradients against the galaxies global metallicity in Figure\,\ref{fig:grad_Z} with additional differentiation by GC counts in colour. As expected, we generally find UDGs/NUDGEs with flat-to-positive metallicity gradients have lower global metallicities comparing to the bulk of the sample classical dwarf galaxies and a preference for high GC counts. While a similar GC contribution may occur in GC-rich classical dwarfs, their higher central stellar densities likely dilute any such effect on the metallicity gradients. On the other hand, GC-poor UDGs/NUDGEs often have later and more extended stellar assemblies and lacked the high-pressure environments and dense gas conditions necessary to facilitate large-scale GCs formation. Hence, the stellar contribution from disrupted GC populations in these systems is reduced, being consistent with the continuous trend observed between metallicity gradients and GC richness.

This trend is also consistent with the findings of \citet{Forbes2025}, in which GC-rich UDGs preferentially show lower metallicities, likely due to the substantial contribution of the tidally disrupted GCs to the total stellar mass of the host galaxy. And with more of these GCs dissolving at smaller radii, it subsequently would lead to a flatter or perhaps even positive metallicity gradient. There also exist the possibility of a combination of stellar feedback and GC disruption acting complementarily, producing the steepest positive metallicity gradients in the most GC-rich UDGs. Nonetheless, further investigation with an expanded observational sample will be essential, particularly for extreme systems like NGC5846\_UDG1 \citep[e.g.][]{FM2023} and DGSAT I \citep[e.g.][]{MartinNavarro2019} that exhibit very low metallicities and very high $\rm M_{GC}/M_{}$ ratios. Detailed simulations of GC evolution and disruption will also be critical for evaluating whether these mechanisms can reproduce the full range of positive metallicity gradients observed in GC-rich UDGs.

%The concentration of GC stars toward the inner regions of the galaxy may result from the in-spiral of GCs due to dynamical friction, followed by their tidal dissolution in the central regions of the host galaxy over the cosmic time. 

\section{Conclusion}\label{sec:conclusion}

We present newly obtained spectroscopic data for 7 UDGs/NUDGEs in the Perseus (6) and Coma clusters (1), obtained with KCWI on the Keck II Telescope. Two of the galaxies satisfy the ``bona fide" UDG definition, while the rest, slightly brighter, are considered NUDGEs. Notably, five of these systems are GC-poor and two are GC-rich, thereby diversifying the available spectroscopic sample of galaxies across a broad range of GC-richness. For this work:
\begin{itemize}
    \item We performed full-spectrum fitting using {\tt pPXF} to derive global ages and metallicities, and employed a classical line-index approach to estimate $\alpha$-abundances. Overall, we find a mean age of 7.1$\pm$2.0\,Gyr, mean [M/H] of $-$0.9 $\pm$0.2\,dex, and a mean [Mg/Fe] of 0.3$\pm$0.2\,dex.
    \item All galaxies lie within the general spread of ages and metallicities of previously studied UDGs/NUDGEs. Overall, we also find UDGs and NUDGEs share similar stellar population properties. Six of our galaxies follow the standard stellar mass–metallicity relation, indicating properties similar to classical dwarf galaxies. Only DF11 appears offset from the relation, exhibiting lower metallicities respectively.
    \item Probing environmental dependence reveals no strong trends within a combined sample of cluster UDGs/NUDGEs from this work and the literature. No clear correlation is observed between age or metallicity and galaxy infall parameter, while only a weak negative relation is found for $\alpha$-abundance, with more elevated [Mg/Fe] systems preferentially residing closer to cluster centers.
    \item We conducted a spatially resolved analysis of three galaxies (H332, W1, and DF11), enabled by the high S/N of the BL grating spectra. Two systems (H332 and W1) exhibit flat age profiles and slightly negative metallicity gradients, consistent with trends observed in classical dwarf galaxies and simulated TNG50 UDGs. In contrast, DF11 shows a flat age profile, but a positive metallicity gradient akin to other GC-rich UDGs from \citet{FM2025}.
    \item Comparing with a broader dwarf galaxy sample, we find a strong correlation between metallicity gradients and the GC–to–galaxy mass fraction, while lack of any significant trend for age gradients. UDGs/NUDGEs with higher $\rm M_{GC}/M_{*}$, tend to exhibit increasingly positive metallicity gradients. We also observe a potential trend in global stellar metallicity, with galaxies showing more positive [M/H] gradients also having lower galaxy global metallicity.
    \item Building on the findings from \citet{Forbes2025}, we explore the idea of the dissolution of GCs driving the positive metallicity gradients in UDGs/NUDGEs. Since GCs may have played an important role in the buildup of stellar mass in some UDGs \citep{Forbes2025}, stars released from these systems may not be expected to be distributed uniformly throughout the galaxy. Owing to stronger tidal fields toward galaxy centers, stars from disrupted GCs are more likely to be deposited at smaller galactocentric radii. As GC stars are typically more metal-poor than the galaxy stellar population, this process can potentially naturally give rise to a reversed (positive) metallicity gradient with inner radii being more metal-poor than outer parts.  

    Nevertheless, we also explored a feedback scenario resembling that of \citet{Mori1997}, invoking feedback driven processes that create positive metallicity gradients in some of these UDGs. An early centrally concentrated burst of intense star formation, responsible for forming many of the GCs, could have generated sufficient feedback to trigger cascading star formation. This process, occurring in continuously metal enriching gas, may have built up most of the stellar body of the UDG. So a potentially combination of both GC destruction and feedback driven processes, could therefore provide a plausible mechanism for producing positive metallicity gradients as well as the continuous relation with $M_\mathrm{GC}/M_{*}$.
\end{itemize}

While these results provide insight into the stellar population properties and assembly histories of UDGs and NUDGEs in dense cluster environments, larger samples, particularly extending to lower stellar masses, are crucial for a more robust study of age and metallicity gradients in dwarf galaxies, as well as environmental trends such as cluster infall dependencies. Such expanded samples, complemented with more detailed simulations of GC formation and disruption in UDGs, will be essential for disentangling and verifying the formation pathways and the physical drivers behind the observed diversity in stellar population gradients. 

\section*{Acknowledgments}

%We thank the anonymous referee for their constructive feedback and suggestions that helped improve this manuscript. We also thank L. Buzzo, Y. Tang and S. Janssens for their help with observations and R. Davis, A. Di Cintio, S. Cardona-Barrero for insightful discussions. 
We thank the R. Peletier for his constructive feedback and suggestions that helped improve this manuscript. We also thank L. Haacke for her help with observations and K. Bekki, the AGATE team members: D. Vaz, B. van Heumen, H. Christie and M. Monaci for the insightful discussions and support. AL acknowledges financial support received through a Swinburne University Postgraduate Research Award throughout the making of this work. DAF, JPB, WJC thank the ARC for financial support via DP220101863 and DP25010673. AFM has received support from RYC2021-031099-I and PID2024-162088NB-I00. AJR was supported by National Science Foundation grant AST-2308390. Some of the data presented herein were obtained at Keck Observatory.%, which is a private 501(c)3 non-profit organization operated as a scientific partnership among the California Institute of Technology, the University of California, and the National Aeronautics and Space Administration. The Observatory was made possible by the generous financial support of the W. M. Keck Foundation.
The authors wish to recognize and acknowledge the very significant cultural role and reverence that the summit of Maunakea has always had within the Native Hawaiian community. We are most fortunate to have the opportunity to conduct observations from this mountain. This research also employed observations made with the NASA/ESA Hubble Space Telescope and made use of archival data from the Hubble Legacy Archive, which is a collaboration between the Space Telescope Science Institute (STScI/NASA), the Space Telescope European Coordinating Facility (STECF/ESAC/ESA) and the Canadian Astronomy Data Centre (CADC/NRC/CSA). This research made use of Photutils, an Astropy package for detection and photometry of astronomical sources \citep{Bradley2024}. This research also made use of Montage. It is funded by the National Science Foundation under Grant Number ACI-1440620, and was previously funded by the National Aeronautics and Space Administration's Earth Science Technology Office, Computation Technologies Project, under Cooperative Agreement Number NCC5-626 between NASA and the California Institute of Technology. This works also makes use of the catalogue of UDG spectroscopic properties from \citep{Gannon2024}, including data from: \cite{mcconnachie2012, VanDokkum2015, Beasley2016, Martin2016, Yagi2016, MartinezDelgado2016, vanDokkum2016, vanDokkum2017, Karachentsev2017, vanDokkum2018, Toloba2018, Gu2018, Lim2018, RuizLara2018, Alabi2018, FM2018, Forbes2018, MartinNavarro2019, Chilingarian2019, Fensch2019, Danieli2019, vanDokkum2019b, torrealba2019, Iodice2020, Collins2020, Muller2020, Gannon2020, Lim2020, Muller2021, Forbes2021, Shen2021, Ji2021, Huang2021, Gannon2021, Gannon2022, Mihos2022, Danieli2022, Villaume2022, Webb2022, Saifollahi2022, Janssens2022, Gannon2023, FM2023, Toloba2023, Iodice2023, Shen2023, Janssens2024, Gannon2024, Buttitta2025}

%%%%%%%%%%%%%%%%%%%%%%%%%%%%%%%%%%%%%%%%%%%%%%%%%%
\section*{Data Availability}

The KCWI data presented are available via the Keck Observatory Archive (KOA): https://www2.keck.hawaii.edu/koa/public/koa.php 18 months after observations are taken.



\bibliographystyle{mnras}
\bibliography{example} % if your bibtex file is called example.bib

@ARTICLE{FM2023,
       author = {{Ferr{\'e}-Mateu}, Anna and {Gannon}, Jonah S. and {Forbes}, Duncan A. and {Buzzo}, Maria Luisa and {Romanowsky}, Aaron J. and {Brodie}, Jean P.},
        title = "{The star formation histories of quiescent ultra-diffuse galaxies and their dependence on environment and globular cluster richness}",
      journal = {\mnras},
         year = 2023,
        month = dec,
       volume = {526},
       number = {3},
        pages = {4735-4754},
          doi = {10.1093/mnras/stad3102},
archivePrefix = {arXiv},
       eprint = {2309.15148},
 primaryClass = {astro-ph.GA},
       adsurl = {https://ui.adsabs.harvard.edu/abs/2023MNRAS.526.4735F}
}

@ARTICLE{FM2018,
       author = {{Ferr{\'e}-Mateu}, Anna and {Alabi}, Adebusola and {Forbes}, Duncan A. and {Romanowsky}, Aaron J. and {Brodie}, Jean and {Pandya}, Viraj and {Mart{\'\i}n-Navarro}, Ignacio and {Bellstedt}, Sabine and {Wasserman}, Asher and {Stone}, Maria B. and {Okabe}, Nobuhiro},
        title = "{Origins of ultradiffuse galaxies in the Coma cluster - II. Constraints from their stellar populations}",
      journal = {\mnras},
         year = 2018,
        month = oct,
       volume = {479},
       number = {4},
        pages = {4891-4906},
          doi = {10.1093/mnras/sty1597},
archivePrefix = {arXiv},
       eprint = {1801.09695},
 primaryClass = {astro-ph.GA},
       adsurl = {https://ui.adsabs.harvard.edu/abs/2018MNRAS.479.4891F}
}

@ARTICLE{Gannon2020,
       author = {{Gannon}, Jonah S. and {Forbes}, Duncan A. and {Romanowsky}, Aaron J. and {Ferr{\'e}-Mateu}, Anna and {Couch}, Warrick J. and {Brodie}, Jean P.},
        title = "{On the stellar kinematics and mass of the Virgo ultradiffuse galaxy VCC 1287}",
      journal = {\mnras},
         year = 2020,
        month = jul,
       volume = {495},
       number = {3},
        pages = {2582-2598},
          doi = {10.1093/mnras/staa1282},
archivePrefix = {arXiv},
       eprint = {2005.03041},
 primaryClass = {astro-ph.GA},
       adsurl = {https://ui.adsabs.harvard.edu/abs/2020MNRAS.495.2582G}
}

@ARTICLE{Gannon2023,
       author = {{Gannon}, Jonah S. and {Forbes}, Duncan A. and {Brodie}, Jean P. and {Romanowsky}, Aaron J. and {Couch}, Warrick J. and {Ferr{\'e}-Mateu}, Anna},
        title = "{Keck spectroscopy of the coma cluster ultra-diffuse galaxy Y358: dynamical mass in a wider context}",
      journal = {\mnras},
         year = 2023,
        month = jan,
       volume = {518},
       number = {3},
        pages = {3653-3666},
          doi = {10.1093/mnras/stac3264},
archivePrefix = {arXiv},
       eprint = {2211.03915},
 primaryClass = {astro-ph.GA},
       adsurl = {https://ui.adsabs.harvard.edu/abs/2023MNRAS.518.3653G}
}

@ARTICLE{Buzzo2022,
       author = {{Buzzo}, Maria Luisa and {Forbes}, Duncan A. and {Brodie}, Jean P. and {Romanowsky}, Aaron J. and {Cluver}, Michelle E. and {Jarrett}, Thomas H. and {Laine}, Seppo and {Couch}, Warrick J. and {Gannon}, Jonah S. and {Ferr{\'e}-Mateu}, Anna and {Okabe}, Nobuhiro},
        title = "{The stellar populations of quiescent ultra-diffuse galaxies from optical to mid-infrared spectral energy distribution fitting}",
      journal = {\mnras},
         year = 2022,
        month = dec,
       volume = {517},
       number = {2},
        pages = {2231-2250},
          doi = {10.1093/mnras/stac2442},
archivePrefix = {arXiv},
       eprint = {2208.11819},
 primaryClass = {astro-ph.GA},
       adsurl = {https://ui.adsabs.harvard.edu/abs/2022MNRAS.517.2231B}
}

@ARTICLE{Cappellari2017,
       author = {{Cappellari}, Michele},
        title = "{Improving the full spectrum fitting method: accurate convolution with Gauss-Hermite functions}",
      journal = {\mnras},
         year = 2017,
        month = apr,
       volume = {466},
       number = {1},
        pages = {798-811},
          doi = {10.1093/mnras/stw3020},
archivePrefix = {arXiv},
       eprint = {1607.08538},
 primaryClass = {astro-ph.GA},
       adsurl = {https://ui.adsabs.harvard.edu/abs/2017MNRAS.466..798C}
}

@ARTICLE{Vazdekis2015,
       author = {{Vazdekis}, A. and {Coelho}, P. and {Cassisi}, S. and {Ricciardelli}, E. and {Falc{\'o}n-Barroso}, J. and {S{\'a}nchez-Bl{\'a}zquez}, P. and {La Barbera}, F. and {Beasley}, M.~A. and {Pietrinferni}, A.},
        title = "{Evolutionary stellar population synthesis with MILES - II. Scaled-solar and {\ensuremath{\alpha}}-enhanced models}",
      journal = {\mnras},
         year = 2015,
        month = may,
       volume = {449},
       number = {2},
        pages = {1177-1214},
          doi = {10.1093/mnras/stv151},
archivePrefix = {arXiv},
       eprint = {1504.08032},
 primaryClass = {astro-ph.GA},
       adsurl = {https://ui.adsabs.harvard.edu/abs/2015MNRAS.449.1177V}
}

@ARTICLE{Janssens2024,
       author = {{Janssens}, Steven R. and {Forbes}, Duncan A. and {Romanowsky}, Aaron J. and {Gannon}, Jonah and {Pfeffer}, Joel and {Couch}, Warrick J. and {Brodie}, Jean P. and {Harris}, William E. and {Durrell}, Patrick R. and {Bekki}, Kenji},
        title = "{The PIPER survey. II. The globular cluster systems of low surface brightness galaxies in the Perseus cluster}",
      journal = {\mnras},
         year = 2024,
        month = oct,
       volume = {534},
       number = {1},
        pages = {783-799},
          doi = {10.1093/mnras/stae2137},
archivePrefix = {arXiv},
       eprint = {2409.07518},
 primaryClass = {astro-ph.GA},
       adsurl = {https://ui.adsabs.harvard.edu/abs/2024MNRAS.534..783J},
}

@ARTICLE{Kroupa2001,
       author = {{Kroupa}, Pavel},
        title = "{On the variation of the initial mass function}",
      journal = {\mnras},
         year = 2001,
        month = apr,
       volume = {322},
       number = {2},
        pages = {231-246},
          doi = {10.1046/j.1365-8711.2001.04022.x},
archivePrefix = {arXiv},
       eprint = {astro-ph/0009005},
 primaryClass = {astro-ph},
       adsurl = {https://ui.adsabs.harvard.edu/abs/2001MNRAS.322..231K}
}

@ARTICLE{Gannon2022,
       author = {{Gannon}, Jonah S. and {Forbes}, Duncan A. and {Romanowsky}, Aaron J. and {Ferr{\'e}-Mateu}, Anna and {Couch}, Warrick J. and {Brodie}, Jean P. and {Huang}, Song and {Janssens}, Steven R. and {Okabe}, Nobuhiro},
        title = "{Ultra-diffuse galaxies in the perseus cluster: comparing galaxy properties with globular cluster system richness}",
      journal = {\mnras},
         year = 2022,
        month = feb,
       volume = {510},
       number = {1},
        pages = {946-958},
          doi = {10.1093/mnras/stab3297},
archivePrefix = {arXiv},
       eprint = {2111.06007},
 primaryClass = {astro-ph.GA},
       adsurl = {https://ui.adsabs.harvard.edu/abs/2022MNRAS.510..946G}
}

@ARTICLE{Gannon2021,
       author = {{Gannon}, Jonah S. and {Dullo}, Bililign T. and {Forbes}, Duncan A. and {Rich}, R. Michael and {Rom{\'a}n}, Javier and {Couch}, Warrick J. and {Brodie}, Jean P. and {Ferr{\'e}-Mateu}, Anna and {Alabi}, Adebusola and {Mould}, Jeremy},
        title = "{A photometric and kinematic analysis of UDG1137+16 (dw1137+16): Probing ultradiffuse galaxy formation in a group environment}",
      journal = {\mnras},
         year = 2021,
        month = apr,
       volume = {502},
       number = {3},
        pages = {3144-3157},
          doi = {10.1093/mnras/stab277},
archivePrefix = {arXiv},
       eprint = {2102.00598},
 primaryClass = {astro-ph.GA},
       adsurl = {https://ui.adsabs.harvard.edu/abs/2021MNRAS.502.3144G}
}

@ARTICLE{Villaume2022,
       author = {{Villaume}, Alexa and {Romanowsky}, Aaron J. and {Brodie}, Jean and {van Dokkum}, Pieter and {Conroy}, Charlie and {Forbes}, Duncan A. and {Danieli}, Shany and {Martin}, Christopher and {Matuszewski}, Matt},
        title = "{Spatially Resolved Stellar Spectroscopy of the Ultra-diffuse Galaxy Dragonfly 44. III. Evidence for an Unexpected Star Formation History under Conventional Galaxy Evolution Processes}",
      journal = {\apj},
         year = 2022,
        month = jan,
       volume = {924},
       number = {1},
          eid = {32},
        pages = {32},
          doi = {10.3847/1538-4357/ac341e},
archivePrefix = {arXiv},
       eprint = {2101.02220},
 primaryClass = {astro-ph.GA},
       adsurl = {https://ui.adsabs.harvard.edu/abs/2022ApJ...924...32V},

}

@ARTICLE{FM2025,
       author = {{Ferr{\'e}-Mateu}, A. and {Gannon}, J. and {Forbes}, D.~A. and {Romanowsky}, A.~J. and {Buzzo}, M.~L. and {Brodie}, J.~P.},
        title = "{Signs of 'Everything Everywhere All at Once' formation in low-surface-brightness globular-cluster-rich dwarf galaxies}",
      journal = {\aap},
         year = 2025,
        month = feb,
       volume = {694},
          eid = {L6},
        pages = {L6},
          doi = {10.1051/0004-6361/202453393},
archivePrefix = {arXiv},
       eprint = {2501.04088},
 primaryClass = {astro-ph.GA},
       adsurl = {https://ui.adsabs.harvard.edu/abs/2025A&A...694L...6F}
}

@ARTICLE{Benavides2024,
       author = {{Benavides}, Jos{\'e} A. and {Sales}, Laura V. and {Abadi}, Mario. G. and {Vogelsberger}, Mark and {Marinacci}, Federico and {Hernquist}, Lars},
        title = "{Large Dark Matter Content and Steep Metallicity Profile Predicted for Ultradiffuse Galaxies Formed in High-spin Halos}",
      journal = {\apj},
         year = 2024,
        month = dec,
       volume = {977},
       number = {2},
          eid = {169},
        pages = {169},
          doi = {10.3847/1538-4357/ad8de8},
archivePrefix = {arXiv},
       eprint = {2407.15938},
 primaryClass = {astro-ph.GA},
       adsurl = {https://ui.adsabs.harvard.edu/abs/2024ApJ...977..169B}
}

@ARTICLE{Dicintio2017,
       author = {{Di Cintio}, Arianna and {Brook}, Chris B. and {Dutton}, Aaron A. and {Macci{\`o}}, Andrea V. and {Obreja}, Aura and {Dekel}, Avishai},
        title = "{NIHAO - XI. Formation of ultra-diffuse galaxies by outflows}",
      journal = {\mnras},
         year = 2017,
        month = mar,
       volume = {466},
       number = {1},
        pages = {L1-L6},
          doi = {10.1093/mnrasl/slw210},
archivePrefix = {arXiv},
       eprint = {1608.01327},
 primaryClass = {astro-ph.GA},
       adsurl = {https://ui.adsabs.harvard.edu/abs/2017MNRAS.466L...1D}
}

@ARTICLE{Cardona2023,
       author = {{Cardona-Barrero}, S. and {Di Cintio}, A. and {Battaglia}, G. and {Macci{\`o}}, A.~V. and {Taibi}, S.},
        title = "{Metallicity profiles of ultradiffuse galaxies in NIHAO simulations}",
      journal = {\mnras},
         year = 2023,
        month = feb,
       volume = {519},
       number = {1},
        pages = {1545-1561},
          doi = {10.1093/mnras/stac3243},
archivePrefix = {arXiv},
       eprint = {2206.10481},
 primaryClass = {astro-ph.GA},
       adsurl = {https://ui.adsabs.harvard.edu/abs/2023MNRAS.519.1545C}
}

@ARTICLE{VanDokkum2015,
       author = {{van Dokkum}, Pieter G. and {Abraham}, Roberto and {Merritt}, Allison and {Zhang}, Jielai and {Geha}, Marla and {Conroy}, Charlie},
        title = "{Forty-seven Milky Way-sized, Extremely Diffuse Galaxies in the Coma Cluster}",
      journal = {\apjl},
         year = 2015,
        month = jan,
       volume = {798},
       number = {2},
          eid = {L45},
        pages = {L45},
          doi = {10.1088/2041-8205/798/2/L45},
archivePrefix = {arXiv},
       eprint = {1410.8141},
 primaryClass = {astro-ph.GA},
       adsurl = {https://ui.adsabs.harvard.edu/abs/2015ApJ...798L..45V}
}

@ARTICLE{Buzzo2025,
       author = {{Buzzo}, Maria Luisa and {Forbes}, Duncan A. and {Jarrett}, Thomas H. and {Marleau}, Francine R. and {Duc}, Pierre-Alain and {Brodie}, Jean P. and {Romanowsky}, Aaron J. and {Ferr{\'e}-Mateu}, Anna and {Hilker}, Michael and {Gannon}, Jonah S. and {Pfeffer}, Joel and {Haacke}, Lydia},
        title = "{The multiple classes of ultra-diffuse galaxies: can we tell them apart?<SUP></SUP>}",
      journal = {\mnras},
         year = 2025,
        month = jan,
       volume = {536},
       number = {3},
        pages = {2536-2557},
          doi = {10.1093/mnras/stae2700},
archivePrefix = {arXiv},
       eprint = {2412.01901},
 primaryClass = {astro-ph.GA},
       adsurl = {https://ui.adsabs.harvard.edu/abs/2025MNRAS.536.2536B}
}

@ARTICLE{Koleva2011,
       author = {{Koleva}, Mina and {Prugniel}, Philippe and {De Rijcke}, Sven and {Zeilinger}, Werner W.},
        title = "{Age and metallicity gradients in early-type galaxies: a dwarf-to-giant sequence}",
      journal = {\mnras},
         year = 2011,
        month = nov,
       volume = {417},
       number = {3},
        pages = {1643-1671},
          doi = {10.1111/j.1365-2966.2011.19057.x},
archivePrefix = {arXiv},
       eprint = {1105.4809},
 primaryClass = {astro-ph.CO},
       adsurl = {https://ui.adsabs.harvard.edu/abs/2011MNRAS.417.1643K}
}

@ARTICLE{Taibi2022,
       author = {{Taibi}, S. and {Battaglia}, G. and {Leaman}, R. and {Brooks}, A. and {Riggs}, C. and {Munshi}, F. and {Revaz}, Y. and {Jablonka}, P.},
        title = "{Stellar metallicity gradients of Local Group dwarf galaxies}",
      journal = {\aap},
         year = 2022,
        month = sep,
       volume = {665},
          eid = {A92},
        pages = {Af92},
          doi = {10.1051/0004-6361/202243508},
archivePrefix = {arXiv},
       eprint = {2206.08988},
 primaryClass = {astro-ph.GA},
       adsurl = {https://ui.adsabs.harvard.edu/abs/2022A&A...665A..92T}
}

@ARTICLE{Forbes2024,
       author = {{Forbes}, Duncan A. and {Gannon}, Jonah},
        title = "{Do ultra diffuse galaxies with rich globular clusters systems have overly massive haloes?}",
      journal = {\mnras},
         year = 2024,
        month = feb,
       volume = {528},
       number = {1},
        pages = {608-619},
          doi = {10.1093/mnras/stad4004},
archivePrefix = {arXiv},
       eprint = {2401.07388},
 primaryClass = {astro-ph.GA},
       adsurl = {https://ui.adsabs.harvard.edu/abs/2024MNRAS.528..608F}
}

@ARTICLE{Kirby2013,
       author = {{Kirby}, Evan N. and {Cohen}, Judith G. and {Guhathakurta}, Puragra and {Cheng}, Lucy and {Bullock}, James S. and {Gallazzi}, Anna},
        title = "{The Universal Stellar Mass-Stellar Metallicity Relation for Dwarf Galaxies}",
      journal = {\apj},
         year = 2013,
        month = dec,
       volume = {779},
       number = {2},
          eid = {102},
        pages = {102},
          doi = {10.1088/0004-637X/779/2/102},
archivePrefix = {arXiv},
       eprint = {1310.0814},
 primaryClass = {astro-ph.GA},
       adsurl = {https://ui.adsabs.harvard.edu/abs/2013ApJ...779..102K}
}

@ARTICLE{Sybilska2017,
       author = {{Sybilska}, A. and {Lisker}, T. and {Kuntschner}, H. and {Vazdekis}, A. and {van de Ven}, G. and {Peletier}, R. and {Falc{\'o}n-Barroso}, J. and {Vijayaraghavan}, R. and {Janz}, J.},
        title = "{The hELENa project - I. Stellar populations of early-type galaxies linked with local environment and galaxy mass}",
      journal = {\mnras},
         year = 2017,
        month = sep,
       volume = {470},
       number = {1},
        pages = {815-838},
          doi = {10.1093/mnras/stx1138},
archivePrefix = {arXiv},
       eprint = {1706.00014},
 primaryClass = {astro-ph.GA},
       adsurl = {https://ui.adsabs.harvard.edu/abs/2017MNRAS.470..815S}
}

@ARTICLE{Buzzo2025a,
       author = {{Buzzo}, Maria Luisa and {Forbes}, Duncan A. and {Romanowsky}, Aaron J. and {Haacke}, Lydia and {Gannon}, Jonah S. and {Tang}, Yimeng and {Hilker}, Michael and {Ferr{\'e}-Mateu}, Anna and {Janssens}, Steven R. and {Brodie}, Jean P. and {Valenzuela}, Lucas M.},
        title = "{A new class of dark matter-free dwarf galaxies?: I. Clues from FCC 224, NGC 1052-DF2, and NGC 1052-DF4}",
      journal = {\aap},
         year = 2025,
        month = mar,
       volume = {695},
          eid = {A124},
        pages = {A124},
          doi = {10.1051/0004-6361/202453522},
archivePrefix = {arXiv},
       eprint = {2502.05405},
 primaryClass = {astro-ph.GA},
       adsurl = {https://ui.adsabs.harvard.edu/abs/2025A&A...695A.124B}
}

@software{Jacob2010,
       author = {{Jacob}, Joseph C. and {Katz}, Daniel S. and {Berriman}, G. Bruce and {Good}, John and {Laity}, Anastasia C. and {Deelman}, Ewa and {Kesselman}, Carl and {Singh}, Gurmeet and {Su}, Mei-Hui and {Prince}, Thomas A. and {Williams}, Roy},
        title = "{Montage: An Astronomical Image Mosaicking Toolkit}",
 howpublished = {Astrophysics Source Code Library, record ascl:1010.036},
         year = 2010,
        month = oct,
          eid = {ascl:1010.036},
       adsurl = {https://ui.adsabs.harvard.edu/abs/2010ascl.soft10036J}
}

@ARTICLE{AL2025,
       author = {{Levitskiy}, Arsen and {Forbes}, Duncan A. and {Gannon}, Jonah S. and {Ferr{\'e}-Mateu}, Anna and {Romanowsky}, Aaron J. and {Brodie}, Jean P. and {Couch}, Warrick J. and {Haacke}, Lydia},
        title = "{A comprehensive look at PUDG-R21: stellar population and kinematics of a globular cluster-rich ultra-diffuse galaxy in the Perseus Cluster}",
      journal = {\mnras},
         year = 2025,
        month = aug,
       volume = {541},
       number = {3},
        pages = {2761-2772},
          doi = {10.1093/mnras/staf1140},
archivePrefix = {arXiv},
       eprint = {2507.05679},
 primaryClass = {astro-ph.GA},
       adsurl = {https://ui.adsabs.harvard.edu/abs/2025MNRAS.541.2761L}
}

@ARTICLE{Gannon2024,
       author = {{Gannon}, Jonah S. and {Ferr{\'e}-Mateu}, Anna and {Forbes}, Duncan A. and {Brodie}, Jean P. and {Buzzo}, Maria Luisa and {Romanowsky}, Aaron J.},
        title = "{A Catalogue and analysis of ultra-diffuse galaxy spectroscopic properties}",
      journal = {\mnras},
         year = 2024,
        month = jun,
       volume = {531},
       number = {1},
        pages = {1856-1869},
          doi = {10.1093/mnras/stae1287},
archivePrefix = {arXiv},
       eprint = {2405.09104},
 primaryClass = {astro-ph.GA},
       adsurl = {https://ui.adsabs.harvard.edu/abs/2024MNRAS.531.1856G}
}

@ARTICLE{Buttitta2025,
       author = {{Buttitta}, Chiara and {Iodice}, Enrichetta and {Doll}, Goran and {Hartke}, Johanna and {Hilker}, Michael and {Forbes}, Duncan A. and {Corsini}, Enrico M. and {Rossi}, Luca and {Arnaboldi}, Magda and {Cantiello}, Michele and {D'Ago}, Giuseppe and {Falc{\'o}n-Barroso}, Jesus and {Gullieuszik}, Marco and {La Marca}, Antonio and {Mieske}, Steffen and {Mirabile}, Marco and {Paolillo}, Maurizio and {Rejkuba}, Marina and {Spavone}, Marilena and {Spiniello}, Chiara and {Sarzi}, Marc},
        title = "{Looking into the faintEst WIth MUSE (LEWIS): Exploring the nature of ultra-diffuse galaxies in the Hydra-I cluster: II. Stellar kinematics and dynamical masses}",
      journal = {\aap},
         year = 2025,
        month = feb,
       volume = {694},
          eid = {A276},
        pages = {A276},
          doi = {10.1051/0004-6361/202452915},
archivePrefix = {arXiv},
       eprint = {2501.16190},
 primaryClass = {astro-ph.GA},
       adsurl = {https://ui.adsabs.harvard.edu/abs/2025A&A...694A.276B}
}

@ARTICLE{Shen2023,
       author = {{Shen}, Zili and {van Dokkum}, Pieter and {Danieli}, Shany},
        title = "{Confirmation of an Anomalously Low Dark Matter Content for the Galaxy NGC 1052-DF4 from Deep, High-resolution Continuum Spectroscopy}",
      journal = {\apj},
         year = 2023,
        month = nov,
       volume = {957},
       number = {1},
          eid = {6},
        pages = {6},
          doi = {10.3847/1538-4357/acfa70},
archivePrefix = {arXiv},
       eprint = {2309.08592},
 primaryClass = {astro-ph.GA},
       adsurl = {https://ui.adsabs.harvard.edu/abs/2023ApJ...957....6S}
}

@ARTICLE{Iodice2023,
       author = {{Iodice}, Enrichetta and {Hilker}, Michael and {Doll}, Goran and {Mirabile}, Marco and {Buttitta}, Chiara and {Hartke}, Johanna and {Mieske}, Steffen and {Cantiello}, Michele and {D'Ago}, Giuseppe and {Forbes}, Duncan A. and {Gullieuszik}, Marco and {Rejkuba}, Marina and {Spavone}, Marilena and {Spiniello}, Chiara and {Arnaboldi}, Magda and {Corsini}, Enrico M. and {Greggio}, Laura and {Falc{\'o}n-Barroso}, Jesus and {Fahrion}, Katja and {Fritz}, Jacopo and {La Marca}, Antonio and {Paolillo}, Maurizio and {Angela Raj}, Maria and {Rampazzo}, Roberto and {Sarzi}, Marc and {Capasso}, Giulio},
        title = "{Looking into the faintEst WIth MUSE (LEWIS): Exploring the nature of ultra-diffuse galaxies in the Hydra-I cluster. I. Project description and preliminary results}",
      journal = {\aap},
         year = 2023,
        month = nov,
       volume = {679},
          eid = {A69},
        pages = {A69},
          doi = {10.1051/0004-6361/202347129},
archivePrefix = {arXiv},
       eprint = {2308.11493},
 primaryClass = {astro-ph.GA},
       adsurl = {https://ui.adsabs.harvard.edu/abs/2023A&A...679A..69I}
}

@ARTICLE{Iodice2020,
       author = {{Iodice}, E. and {Cantiello}, M. and {Hilker}, M. and {Rejkuba}, M. and {Arnaboldi}, M. and {Spavone}, M. and {Greggio}, L. and {Forbes}, D.~A. and {D'Ago}, G. and {Mieske}, S. and {Spiniello}, C. and {La Marca}, A. and {Rampazzo}, R. and {Paolillo}, M. and {Capaccioli}, M. and {Schipani}, P.},
        title = "{The first detection of ultra-diffuse galaxies in the Hydra I cluster from the VEGAS survey}",
      journal = {A\&A},
         year = 2020,
        month = oct,
       volume = {642},
          eid = {A48},
        pages = {A48},
          doi = {10.1051/0004-6361/202038523},
archivePrefix = {arXiv},
       eprint = {2007.11533},
 primaryClass = {astro-ph.GA},
       adsurl = {https://ui.adsabs.harvard.edu/abs/2020A&A...642A..48I}
}

@ARTICLE{Leisman2017,
       author = {{Leisman}, Lukas and {Haynes}, Martha P. and {Janowiecki}, Steven and {Hallenbeck}, Gregory and {J{\'o}zsa}, Gyula and {Giovanelli}, Riccardo and {Adams}, Elizabeth A.~K. and {Bernal Neira}, David and {Cannon}, John M. and {Janesh}, William F. and {Rhode}, Katherine L. and {Salzer}, John J.},
        title = "{(Almost) Dark Galaxies in the ALFALFA Survey: Isolated H I-bearing Ultra-diffuse Galaxies}",
      journal = {\apj},
         year = 2017,
        month = jun,
       volume = {842},
       number = {2},
          eid = {133},
        pages = {133},
          doi = {10.3847/1538-4357/aa7575},
archivePrefix = {arXiv},
       eprint = {1703.05293},
 primaryClass = {astro-ph.GA},
       adsurl = {https://ui.adsabs.harvard.edu/abs/2017ApJ...842..133L}
}

@ARTICLE{Benavides2021,
       author = {{Benavides}, Jos{\'e} A. and {Sales}, Laura V. and {Abadi}, Mario. G. and {Pillepich}, Annalisa and {Nelson}, Dylan and {Marinacci}, Federico and {Cooper}, Michael and {Pakmor}, Ruediger and {Torrey}, Paul and {Vogelsberger}, Mark and {Hernquist}, Lars},
        title = "{Quiescent ultra-diffuse galaxies in the field originating from backsplash orbits}",
      journal = {Nature Astronomy},
         year = 2021,
        month = sep,
       volume = {5},
        pages = {1255-1260},
          doi = {10.1038/s41550-021-01458-1},
archivePrefix = {arXiv},
       eprint = {2109.01677},
 primaryClass = {astro-ph.GA},
       adsurl = {https://ui.adsabs.harvard.edu/abs/2021NatAs...5.1255B}
}

@ARTICLE{Lim2018,
       author = {{Lim}, Sungsoon and {Peng}, Eric W. and {C{\^o}t{\'e}}, Patrick and {Sales}, Laura V. and {den Brok}, Mark and {Blakeslee}, John P. and {Guhathakurta}, Puragra},
        title = "{The Globular Cluster Systems of Ultra-diffuse Galaxies in the Coma Cluster}",
      journal = {ApJ},
         year = 2018,
        month = jul,
       volume = {862},
       number = {1},
          eid = {82},
        pages = {82},
          doi = {10.3847/1538-4357/aacb81},
archivePrefix = {arXiv},
       eprint = {1806.05425},
 primaryClass = {astro-ph.GA},
       adsurl = {https://ui.adsabs.harvard.edu/abs/2018ApJ...862...82L}
}

@ARTICLE{Muller2020,
       author = {{M{\"u}ller}, Oliver and {Marleau}, Francine R. and {Duc}, Pierre-Alain and {Habas}, Rebecca and {Fensch}, J{\'e}r{\'e}my and {Emsellem}, Eric and {Poulain}, M{\'e}lina and {Lim}, Sungsoon and {Agnello}, Adriano and {Durrell}, Patrick and {Paudel}, Sanjaya and {S{\'a}nchez-Janssen}, Rub{\'e}n and {van der Burg}, Remco F.~J.},
        title = "{Spectroscopic study of MATLAS-2019 with MUSE: An ultra-diffuse galaxy with an excess of old globular clusters}",
      journal = {A\&A},
         year = 2020,
        month = aug,
       volume = {640},
          eid = {A106},
        pages = {A106},
          doi = {10.1051/0004-6361/202038351},
archivePrefix = {arXiv},
       eprint = {2006.04606},
 primaryClass = {astro-ph.GA},
       adsurl = {https://ui.adsabs.harvard.edu/abs/2020A&A...640A.106M}
}

@ARTICLE{Ji2021,
       author = {{Ji}, Alexander P. and {Koposov}, Sergey E. and {Li}, Ting S. and {Erkal}, Denis and {Pace}, Andrew B. and {Simon}, Joshua D. and {Belokurov}, Vasily and {Cullinane}, Lara R. and {Da Costa}, Gary S. and {Kuehn}, Kyler and {Lewis}, Geraint F. and {Mackey}, Dougal and {Shipp}, Nora and {Simpson}, Jeffrey D. and {Zucker}, Daniel B. and {Hansen}, Terese T. and {Bland-Hawthorn}, Joss and {S5 Collaboration}},
        title = "{Kinematics of Antlia 2 and Crater 2 from the Southern Stellar Stream Spectroscopic Survey (S$^{5}$)}",
      journal = {\apj},
         year = 2021,
        month = nov,
       volume = {921},
       number = {1},
          eid = {32},
        pages = {32},
          doi = {10.3847/1538-4357/ac1869},
archivePrefix = {arXiv},
       eprint = {2106.12656},
 primaryClass = {astro-ph.GA},
       adsurl = {https://ui.adsabs.harvard.edu/abs/2021ApJ...921...32J}
}

@ARTICLE{Huang2021,
       author = {{Huang}, Kuan-Wei and {Koposov}, Sergey E.},
        title = "{Search for globular clusters associated with the Milky Way dwarf galaxies using Gaia DR2}",
      journal = {\mnras},
         year = 2021,
        month = jan,
       volume = {500},
       number = {1},
        pages = {986-997},
          doi = {10.1093/mnras/staa3297},
archivePrefix = {arXiv},
       eprint = {2005.14014},
 primaryClass = {astro-ph.GA},
       adsurl = {https://ui.adsabs.harvard.edu/abs/2021MNRAS.500..986H}
}

@ARTICLE{Toloba2023,
       author = {{Toloba}, Elisa and {Sales}, Laura V. and {Lim}, Sungsoon and {Peng}, Eric W. and {Guhathakurta}, Puragra and {Roediger}, Joel and {Wang}, Kaixiang and {Mihos}, J. Christopher and {C{\^o}t{\'e}}, Patrick and {Durrell}, Patrick R. and {Ferrarese}, Laura},
        title = "{The Next Generation Virgo Cluster Survey (NGVS). XXXV. First Kinematical Clues of Overly Massive Dark Matter Halos in Several Ultradiffuse Galaxies in the Virgo Cluster}",
      journal = {\apj},
         year = 2023,
        month = jul,
       volume = {951},
       number = {1},
          eid = {77},
        pages = {77},
          doi = {10.3847/1538-4357/acd336},
archivePrefix = {arXiv},
       eprint = {2305.06369},
 primaryClass = {astro-ph.GA},
       adsurl = {https://ui.adsabs.harvard.edu/abs/2023ApJ...951...77T}
}

@ARTICLE{Yagi2016,
       author = {{Yagi}, Masafumi and {Koda}, Jin and {Komiyama}, Yutaka and {Yamanoi}, Hitomo},
        title = "{Catalog of Ultra-diffuse Galaxies in the Coma Clusters from Subaru Imaging Data}",
      journal = {ApJS},
         year = 2016,
        month = jul,
       volume = {225},
       number = {1},
          eid = {11},
        pages = {11},
          doi = {10.3847/0067-0049/225/1/11},
       adsurl = {https://ui.adsabs.harvard.edu/abs/2016ApJS..225...11Y}
}

@ARTICLE{Prole2019,
       author = {{Prole}, D.~J. and {van der Burg}, R.~F.~J. and {Hilker}, M. and {Davies}, J.~I.},
        title = "{Observational properties of ultra-diffuse galaxies in low-density environments: field UDGs are predominantly blue and star forming}",
      journal = {\mnras},
         year = 2019,
        month = sep,
       volume = {488},
       number = {2},
        pages = {2143-2157},
          doi = {10.1093/mnras/stz1843},
archivePrefix = {arXiv},
       eprint = {1907.01559},
 primaryClass = {astro-ph.GA},
       adsurl = {https://ui.adsabs.harvard.edu/abs/2019MNRAS.488.2143P}
}

@ARTICLE{Zaritsky2021,
       author = {{Zaritsky}, Dennis and {Donnerstein}, Richard and {Karunakaran}, Ananthan and {Barbosa}, C.~E. and {Dey}, Arjun and {Kadowaki}, Jennifer and {Spekkens}, Kristine and {Zhang}, Huanian},
        title = "{Systematically Measuring Ultra-diffuse Galaxies (SMUDGes). II. Expanded Survey Description and the Stripe 82 Catalog}",
      journal = {\apjs},
         year = 2021,
        month = dec,
       volume = {257},
       number = {2},
          eid = {60},
        pages = {60},
          doi = {10.3847/1538-4365/ac2607},
archivePrefix = {arXiv},
       eprint = {2109.03345},
 primaryClass = {astro-ph.GA},
       adsurl = {https://ui.adsabs.harvard.edu/abs/2021ApJS..257...60Z}
}

@ARTICLE{Carleton2019,
       author = {{Carleton}, Timothy and {Errani}, Rapha{\"e}l and {Cooper}, Michael and {Kaplinghat}, Manoj and {Pe{\~n}arrubia}, Jorge and {Guo}, Yicheng},
        title = "{The formation of ultra-diffuse galaxies in cored dark matter haloes through tidal stripping and heating}",
      journal = {\mnras},
         year = 2019,
        month = may,
       volume = {485},
       number = {1},
        pages = {382-395},
          doi = {10.1093/mnras/stz383},
archivePrefix = {arXiv},
       eprint = {1805.06896},
 primaryClass = {astro-ph.GA},
       adsurl = {https://ui.adsabs.harvard.edu/abs/2019MNRAS.485..382C}
}

@ARTICLE{Janssens2017,
       author = {{Janssens}, Steven and {Abraham}, Roberto and {Brodie}, Jean and {Forbes}, Duncan and {Romanowsky}, Aaron J. and {van Dokkum}, Pieter},
        title = "{Ultra-diffuse and Ultra-compact Galaxies in the Frontier Fields Cluster Abell 2744}",
      journal = {\apjl},
         year = 2017,
        month = apr,
       volume = {839},
       number = {1},
          eid = {L17},
        pages = {L17},
          doi = {10.3847/2041-8213/aa667d},
archivePrefix = {arXiv},
       eprint = {1701.00011},
 primaryClass = {astro-ph.GA},
       adsurl = {https://ui.adsabs.harvard.edu/abs/2017ApJ...839L..17J}
}

@ARTICLE{Sales2020,
       author = {{Sales}, Laura V. and {Navarro}, Julio F. and {Pe{\~n}afiel}, Louis and {Peng}, Eric W. and {Lim}, Sungsoon and {Hernquist}, Lars},
        title = "{The formation of ultradiffuse galaxies in clusters}",
      journal = {\mnras},
         year = 2020,
        month = may,
       volume = {494},
       number = {2},
        pages = {1848-1858},
          doi = {10.1093/mnras/staa854},
archivePrefix = {arXiv},
       eprint = {1909.01347},
 primaryClass = {astro-ph.CO},
       adsurl = {https://ui.adsabs.harvard.edu/abs/2020MNRAS.494.1848S}
}

@ARTICLE{Amorisco2016,
   author = {{Amorisco}, N.~C. and {Loeb}, A.},
    title = "{Ultradiffuse galaxies: the high-spin tail of the abundant dwarf galaxy population}",
  journal = {\mnras},
archivePrefix = "arXiv",
   eprint = {1603.00463},
     year = 2016,
    month = jun,
   volume = 459,
    pages = {L51-L55},
      doi = {10.1093/mnrasl/slw055},
   adsurl = {http://adsabs.harvard.edu/abs/2016MNRAS.459L..51A}
}

@ARTICLE{Rong2017,
       author = {{Rong}, Yu and {Guo}, Qi and {Gao}, Liang and {Liao}, Shihong and {Xie}, Lizhi and {Puzia}, Thomas H. and {Sun}, Shuangpeng and {Pan}, Jun},
        title = "{A Universe of ultradiffuse galaxies: theoretical predictions from {\ensuremath{\Lambda}}CDM simulations}",
      journal = {\mnras},
         year = 2017,
        month = oct,
       volume = {470},
       number = {4},
        pages = {4231-4240},
          doi = {10.1093/mnras/stx1440},
archivePrefix = {arXiv},
       eprint = {1703.06147},
 primaryClass = {astro-ph.GA},
       adsurl = {https://ui.adsabs.harvard.edu/abs/2017MNRAS.470.4231R}
}

@ARTICLE{Jiang2019,
       author = {{Jiang}, Fangzhou and {Dekel}, Avishai and {Freundlich}, Jonathan and {Romanowsky}, Aaron J. and {Dutton}, Aaron A. and {Macci{\`o}}, Andrea V. and {Di Cintio}, Arianna},
        title = "{Formation of ultra-diffuse galaxies in the field and in galaxy groups}",
      journal = {\mnras},
         year = 2019,
        month = aug,
       volume = {487},
       number = {4},
        pages = {5272-5290},
          doi = {10.1093/mnras/stz1499},
archivePrefix = {arXiv},
       eprint = {1811.10607},
 primaryClass = {astro-ph.GA},
       adsurl = {https://ui.adsabs.harvard.edu/abs/2019MNRAS.487.5272J}
}

@ARTICLE{Alabi2018,
       author = {{Alabi}, Adebusola and {Ferr{\'e}-Mateu}, Anna and {Romanowsky}, Aaron J. and {Brodie}, Jean and {Forbes}, Duncan A. and {Wasserman}, Asher and {Bellstedt}, Sabine and {Mart{\'\i}n-Navarro}, Ignacio and {Pandya}, Viraj and {Stone}, Maria B. and {Okabe}, Nobuhiro},
        title = "{Origins of ultradiffuse galaxies in the Coma cluster - I. Constraints from velocity phase space}",
      journal = {MNRAS},
         year = 2018,
        month = sep,
       volume = {479},
       number = {3},
        pages = {3308-3318},
          doi = {10.1093/mnras/sty1616},
archivePrefix = {arXiv},
       eprint = {1801.09686},
 primaryClass = {astro-ph.GA},
       adsurl = {https://ui.adsabs.harvard.edu/abs/2018MNRAS.479.3308A}
}

@ARTICLE{Toloba2018,
       author = {{Toloba}, Elisa and {Lim}, Sungsoon and {Peng}, Eric and {Sales}, Laura V. and {Guhathakurta}, Puragra and {Mihos}, J. Christopher and {C{\^o}t{\'e}}, Patrick and {Boselli}, Alessandro and {Cuillandre}, Jean-Charles and {Ferrarese}, Laura and {Gwyn}, Stephen and {Lan{\c{c}}on}, Ariane and {Mu{\~n}oz}, Roberto and {Puzia}, Thomas},
        title = "{Dark Matter in Ultra-diffuse Galaxies in the Virgo Cluster from Their Globular Cluster Populations}",
      journal = {ApJL},
         year = 2018,
        month = apr,
       volume = {856},
       number = {2},
          eid = {L31},
        pages = {L31},
          doi = {10.3847/2041-8213/aab603},
archivePrefix = {arXiv},
       eprint = {1803.09768},
 primaryClass = {astro-ph.GA},
       adsurl = {https://ui.adsabs.harvard.edu/abs/2018ApJ...856L..31T}
}

@ARTICLE{Janssens2022,
       author = {{Janssens}, Steven R. and {Romanowsky}, Aaron J. and {Abraham}, Roberto and {Brodie}, Jean P. and {Couch}, Warrick J. and {Forbes}, Duncan A. and {Laine}, Seppo and {Mart{\'\i}nez-Delgado}, David and {van Dokkum}, Pieter G.},
        title = "{The globular clusters and star formation history of the isolated, quiescent ultra-diffuse galaxy DGSAT I}",
      journal = {MNRAS},
         year = 2022,
        month = nov,
       volume = {517},
       number = {1},
        pages = {858-871},
          doi = {10.1093/mnras/stac2717},
archivePrefix = {arXiv},
       eprint = {2209.09910},
 primaryClass = {astro-ph.GA},
       adsurl = {https://ui.adsabs.harvard.edu/abs/2022MNRAS.517..858J}
}

@ARTICLE{Forbes2021,
       author = {{Forbes}, Duncan A. and {Gannon}, Jonah S. and {Romanowsky}, Aaron J. and {Alabi}, Adebusola and {Brodie}, Jean P. and {Couch}, Warrick J. and {Ferr{\'e}-Mateu}, Anna},
        title = "{Stellar velocity dispersion and dynamical mass of the ultra diffuse galaxy NGC 5846\_UDG1 from the keck cosmic web imager}",
      journal = {MNRAS},
         year = 2021,
        month = jan,
       volume = {500},
       number = {1},
        pages = {1279-1284},
          doi = {10.1093/mnras/staa3289},
archivePrefix = {arXiv},
       eprint = {2010.07313},
 primaryClass = {astro-ph.GA},
       adsurl = {https://ui.adsabs.harvard.edu/abs/2021MNRAS.500.1279F}
}

@ARTICLE{Muller2021,
       author = {{M{\"u}ller}, Oliver and {Durrell}, Patrick R. and {Marleau}, Francine R. and {Duc}, Pierre-Alain and {Lim}, Sungsoon and {Posti}, Lorenzo and {Agnello}, Adriano and {S{\'a}nchez-Janssen}, Rub{\'e}n and {Poulain}, M{\'e}lina and {Habas}, Rebecca and {Emsellem}, Eric and {Paudel}, Sanjaya and {van der Burg}, Remco F.~J. and {Fensch}, J{\'e}r{\'e}my},
        title = "{Dwarf Galaxies in the MATLAS Survey: Hubble Space Telescope Observations of the Globular Cluster System in the Ultra-diffuse Galaxy MATLAS-2019}",
      journal = {ApJ},
         year = 2021,
        month = dec,
       volume = {923},
       number = {1},
          eid = {9},
        pages = {9},
          doi = {10.3847/1538-4357/ac2831},
archivePrefix = {arXiv},
       eprint = {2101.10659},
 primaryClass = {astro-ph.GA},
       adsurl = {https://ui.adsabs.harvard.edu/abs/2021ApJ...923....9M}
}

@ARTICLE{MartinezDelgado2016,
       author = {{Mart{\'\i}nez-Delgado}, David and {L{\"a}sker}, Ronald and {Sharina}, Margarita and {Toloba}, Elisa and {Fliri}, J{\"u}rgen and {Beaton}, Rachael and {Valls-Gabaud}, David and {Karachentsev}, Igor D. and {Chonis}, Taylor S. and {Grebel}, Eva K. and {Forbes}, Duncan A. and {Romanowsky}, Aaron J. and {Gallego-Laborda}, J. and {Teuwen}, Karel and {G{\'o}mez-Flechoso}, M.~A. and {Wang}, Jie and {Guhathakurta}, Puragra and {Kaisin}, Serafim and {Ho}, Nhung},
        title = "{Discovery of an Ultra-diffuse Galaxy in the Pisces--Perseus Supercluster}",
      journal = {AJ},
         year = 2016,
        month = apr,
       volume = {151},
       number = {4},
          eid = {96},
        pages = {96},
          doi = {10.3847/0004-6256/151/4/96},
archivePrefix = {arXiv},
       eprint = {1601.06960},
 primaryClass = {astro-ph.GA},
       adsurl = {https://ui.adsabs.harvard.edu/abs/2016AJ....151...96M}
}

@ARTICLE{MartinNavarro2019,
       author = {{Mart{\'\i}n-Navarro}, Ignacio and {Romanowsky}, Aaron J. and {Brodie}, Jean P. and {Ferr{\'e}-Mateu}, Anna and {Alabi}, Adebusola and {Forbes}, Duncan A. and {Sharina}, Margarita and {Villaume}, Alexa and {Pandya}, Viraj and {Martinez-Delgado}, David},
        title = "{Extreme chemical abundance ratio suggesting an exotic origin for an ultradiffuse galaxy}",
      journal = {MNRAS},
         year = 2019,
        month = apr,
       volume = {484},
       number = {3},
        pages = {3425-3433},
          doi = {10.1093/mnras/stz252},
archivePrefix = {arXiv},
       eprint = {1901.08068},
 primaryClass = {astro-ph.GA},
       adsurl = {https://ui.adsabs.harvard.edu/abs/2019MNRAS.484.3425M}
}

@ARTICLE{Beasley2016,
       author = {{Beasley}, Michael A. and {Romanowsky}, Aaron J. and {Pota}, Vincenzo and {Navarro}, Ignacio Martin and {Martinez Delgado}, David and {Neyer}, Fabian and {Deich}, Aaron L.},
        title = "{An Overmassive Dark Halo around an Ultra-diffuse Galaxy in the Virgo Cluster}",
      journal = {ApJL},
         year = 2016,
        month = mar,
       volume = {819},
       number = {2},
          eid = {L20},
        pages = {L20},
          doi = {10.3847/2041-8205/819/2/L20},
archivePrefix = {arXiv},
       eprint = {1602.04002},
 primaryClass = {astro-ph.GA},
       adsurl = {https://ui.adsabs.harvard.edu/abs/2016ApJ...819L..20B}
}

@ARTICLE{Collins2020,
       author = {{Collins}, Michelle L.~M. and {Tollerud}, Erik J. and {Rich}, R. Michael and {Ibata}, Rodrigo A. and {Martin}, Nicolas F. and {Chapman}, Scott C. and {Gilbert}, Karoline M. and {Preston}, Janet},
        title = "{A detailed study of Andromeda XIX, an extreme local analogue of ultradiffuse galaxies}",
      journal = {MNRAS},
         year = 2020,
        month = jan,
       volume = {491},
       number = {3},
        pages = {3496-3514},
          doi = {10.1093/mnras/stz3252},
archivePrefix = {arXiv},
       eprint = {1910.12879},
 primaryClass = {astro-ph.GA},
       adsurl = {https://ui.adsabs.harvard.edu/abs/2020MNRAS.491.3496C}
}

@ARTICLE{Forbes2019,
       author = {{Forbes}, Duncan A. and {Gannon}, Jonah and {Couch}, Warrick J. and {Iodice}, Enrichetta and {Spavone}, Marilena and {Cantiello}, Michele and {Napolitano}, Nicola and {Schipani}, Pietro},
        title = "{An ultra diffuse galaxy in the NGC 5846 group from the VEGAS survey}",
      journal = {A\&A},
         year = 2019,
        month = jun,
       volume = {626},
          eid = {A66},
        pages = {A66},
          doi = {10.1051/0004-6361/201935499},
archivePrefix = {arXiv},
       eprint = {1905.06415},
 primaryClass = {astro-ph.GA},
       adsurl = {https://ui.adsabs.harvard.edu/abs/2019A&A...626A..66F}
}

@ARTICLE{Danieli2022,
       author = {{Danieli}, Shany and {van Dokkum}, Pieter and {Trujillo-Gomez}, Sebastian and {Kruijssen}, J.~M. Diederik and {Romanowsky}, Aaron J. and {Carlsten}, Scott and {Shen}, Zili and {Li}, Jiaxuan and {Abraham}, Roberto and {Brodie}, Jean and {Conroy}, Charlie and {Gannon}, Jonah S. and {Greco}, Johnny},
        title = "{NGC 5846-UDG1: A Galaxy Formed Mostly by Star Formation in Massive, Extremely Dense Clumps of Gas}",
      journal = {ApJL},
         year = 2022,
        month = mar,
       volume = {927},
       number = {2},
          eid = {L28},
        pages = {L28},
          doi = {10.3847/2041-8213/ac590a},
archivePrefix = {arXiv},
       eprint = {2111.14851},
 primaryClass = {astro-ph.GA},
       adsurl = {https://ui.adsabs.harvard.edu/abs/2022ApJ...927L..28D}
}

@ARTICLE{Chilingarian2019,
       author = {{Chilingarian}, Igor V. and {Afanasiev}, Anton V. and {Grishin}, Kirill A. and {Fabricant}, Daniel and {Moran}, Sean},
        title = "{Internal Dynamics and Stellar Content of Nine Ultra-diffuse Galaxies in the Coma Cluster Prove Their Evolutionary Link with Dwarf Early-type Galaxies}",
      journal = {ApJ},
         year = 2019,
        month = oct,
       volume = {884},
       number = {1},
          eid = {79},
        pages = {79},
          doi = {10.3847/1538-4357/ab4205},
archivePrefix = {arXiv},
       eprint = {1901.05489},
 primaryClass = {astro-ph.GA},
       adsurl = {https://ui.adsabs.harvard.edu/abs/2019ApJ...884...79C}
}

@ARTICLE{Lim2020,
       author = {{Lim}, Sungsoon and {C{\^o}t{\'e}}, Patrick and {Peng}, Eric W. and {Ferrarese}, Laura and {Roediger}, Joel C. and {Durrell}, Patrick R. and {Mihos}, J. Christopher and {Wang}, Kaixiang and {Gwyn}, S.~D.~J. and {Cuillandre}, Jean-Charles and {Liu}, Chengze and {S{\'a}nchez-Janssen}, Rub{\'e}n and {Toloba}, Elisa and {Sales}, Laura V. and {Guhathakurta}, Puragra and {Lan{\c{c}}on}, Ariane and {Puzia}, Thomas H.},
        title = "{The Next Generation Virgo Cluster Survey (NGVS). XXX. Ultra-diffuse Galaxies and Their Globular Cluster Systems}",
      journal = {ApJ},
         year = 2020,
        month = aug,
       volume = {899},
       number = {1},
          eid = {69},
        pages = {69},
          doi = {10.3847/1538-4357/aba433},
archivePrefix = {arXiv},
       eprint = {2007.10565},
 primaryClass = {astro-ph.GA},
       adsurl = {https://ui.adsabs.harvard.edu/abs/2020ApJ...899...69L}
}

@ARTICLE{Karachentsev2017,
       author = {{Karachentsev}, I.~D. and {Makarova}, L.~N. and {Sharina}, M.~E. and {Karachentseva}, V.~E.},
        title = "{KDG218, a nearby ultra-diffuse galaxy}",
      journal = {Astrophysical Bulletin},
         year = 2017,
        month = oct,
       volume = {72},
       number = {4},
        pages = {376-383},
          doi = {10.1134/S1990341317040022},
archivePrefix = {arXiv},
       eprint = {1711.06074},
 primaryClass = {astro-ph.GA},
       adsurl = {https://ui.adsabs.harvard.edu/abs/2017AstBu..72..376K}
}

@ARTICLE{Webb2022,
       author = {{Webb}, Kristi A. and {Villaume}, Alexa and {Laine}, Seppo and {Romanowsky}, Aaron J. and {Balogh}, Michael and {van Dokkum}, Pieter and {Forbes}, Duncan A. and {Brodie}, Jean and {Martin}, Christopher and {Matuszewski}, Matt},
        title = "{Still at odds with conventional galaxy evolution: the star formation history of ultradiffuse galaxy Dragonfly 44}",
      journal = {MNRAS},
         year = 2022,
        month = nov,
       volume = {516},
       number = {3},
        pages = {3318-3341},
          doi = {10.1093/mnras/stac2417},
archivePrefix = {arXiv},
       eprint = {2208.11038},
 primaryClass = {astro-ph.GA},
       adsurl = {https://ui.adsabs.harvard.edu/abs/2022MNRAS.516.3318W}
}

@ARTICLE{Gu2018,
       author = {{Gu}, Meng and {Conroy}, Charlie and {Law}, David and {van Dokkum}, Pieter and {Yan}, Renbin and {Wake}, David and {Bundy}, Kevin and {Merritt}, Allison and {Abraham}, Roberto and {Zhang}, Jielai and {Bershady}, Matthew and {Bizyaev}, Dmitry and {Brinkmann}, Jonathan and {Drory}, Niv and {Grabowski}, Kathleen and {Masters}, Karen and {Pan}, Kaike and {Parejko}, John and {Weijmans}, Anne-Marie and {Zhang}, Kai},
        title = "{Low Metallicities and Old Ages for Three Ultra-diffuse Galaxies in the Coma Cluster}",
      journal = {ApJ},
         year = 2018,
        month = may,
       volume = {859},
       number = {1},
          eid = {37},
        pages = {37},
          doi = {10.3847/1538-4357/aabbae},
archivePrefix = {arXiv},
       eprint = {1709.07003},
 primaryClass = {astro-ph.GA},
       adsurl = {https://ui.adsabs.harvard.edu/abs/2018ApJ...859...37G}
}

@ARTICLE{Fensch2019,
       author = {{Fensch}, J{\'e}r{\'e}my and {van der Burg}, Remco F.~J. and {Je{\v{r}}{\'a}bkov{\'a}}, Tereza and {Emsellem}, Eric and {Zanella}, Anita and {Agnello}, Adriano and {Hilker}, Michael and {M{\"u}ller}, Oliver and {Rejkuba}, Marina and {Duc}, Pierre-Alain and {Durrell}, Patrick and {Habas}, Rebecca and {Lim}, Sungsoon and {Marleau}, Francine R. and {Peng}, Eric W. and {S{\'a}nchez Janssen}, Rub{\'e}n},
        title = "{The ultra-diffuse galaxy NGC 1052-DF2 with MUSE. II. The population of DF2: stars, clusters, and planetary nebulae}",
      journal = {A\&A},
         year = 2019,
        month = may,
       volume = {625},
          eid = {A77},
        pages = {A77},
          doi = {10.1051/0004-6361/201834911},
archivePrefix = {arXiv},
       eprint = {1812.07346},
 primaryClass = {astro-ph.GA},
       adsurl = {https://ui.adsabs.harvard.edu/abs/2019A&A...625A..77F}
}

@ARTICLE{RuizLara2018,
       author = {{Ruiz-Lara}, T. and {Beasley}, M.~A. and {Falc{\'o}n-Barroso}, J. and {Rom{\'a}n}, J. and {Pinna}, F. and {Brook}, C. and {Di Cintio}, A. and {Mart{\'\i}n-Navarro}, I. and {Trujillo}, I. and {Vazdekis}, A.},
        title = "{Spectroscopic characterization of the stellar content of ultra-diffuse galaxies}",
      journal = {MNRAS},
         year = 2018,
        month = aug,
       volume = {478},
       number = {2},
        pages = {2034-2045},
          doi = {10.1093/mnras/sty1112},
archivePrefix = {arXiv},
       eprint = {1803.06298},
 primaryClass = {astro-ph.GA},
       adsurl = {https://ui.adsabs.harvard.edu/abs/2018MNRAS.478.2034R}
}

@ARTICLE{Shen2021,
       author = {{Shen}, Zili and {Danieli}, Shany and {van Dokkum}, Pieter and {Abraham}, Roberto and {Brodie}, Jean P. and {Conroy}, Charlie and {Dolphin}, Andrew E. and {Romanowsky}, Aaron J. and {Kruijssen}, J.~M. Diederik and {Dutta Chowdhury}, Dhruba},
        title = "{A Tip of the Red Giant Branch Distance of 22.1 {\ensuremath{\pm}} 1.2 Mpc to the Dark Matter Deficient Galaxy NGC 1052-DF2 from 40 Orbits of Hubble Space Telescope Imaging}",
      journal = {ApJL},
         year = 2021,
        month = jun,
       volume = {914},
       number = {1},
          eid = {L12},
        pages = {L12},
          doi = {10.3847/2041-8213/ac0335},
archivePrefix = {arXiv},
       eprint = {2104.03319},
 primaryClass = {astro-ph.GA},
       adsurl = {https://ui.adsabs.harvard.edu/abs/2021ApJ...914L..12S}
}

@ARTICLE{Danieli2019,
       author = {{Danieli}, Shany and {van Dokkum}, Pieter and {Conroy}, Charlie and {Abraham}, Roberto and {Romanowsky}, Aaron J.},
        title = "{Still Missing Dark Matter: KCWI High-resolution Stellar Kinematics of NGC1052-DF2}",
      journal = {ApJL},
         year = 2019,
        month = apr,
       volume = {874},
       number = {2},
          eid = {L12},
        pages = {L12},
          doi = {10.3847/2041-8213/ab0e8c},
archivePrefix = {arXiv},
       eprint = {1901.03711},
 primaryClass = {astro-ph.GA},
       adsurl = {https://ui.adsabs.harvard.edu/abs/2019ApJ...874L..12D}
}

@ARTICLE{vanDokkum2019b,
       author = {{van Dokkum}, Pieter and {Wasserman}, Asher and {Danieli}, Shany and {Abraham}, Roberto and {Brodie}, Jean and {Conroy}, Charlie and {Forbes}, Duncan A. and {Martin}, Christopher and {Matuszewski}, Matt and {Romanowsky}, Aaron J. and {Villaume}, Alexa},
        title = "{Spatially Resolved Stellar Kinematics of the Ultra-diffuse Galaxy Dragonfly 44. I. Observations, Kinematics, and Cold Dark Matter Halo Fits}",
      journal = {ApJ},
         year = 2019,
        month = aug,
       volume = {880},
       number = {2},
          eid = {91},
        pages = {91},
          doi = {10.3847/1538-4357/ab2914},
archivePrefix = {arXiv},
       eprint = {1904.04838},
 primaryClass = {astro-ph.GA},
       adsurl = {https://ui.adsabs.harvard.edu/abs/2019ApJ...880...91V}
}

@ARTICLE{vanDokkum2016,
       author = {{van Dokkum}, Pieter and {Abraham}, Roberto and {Brodie}, Jean and {Conroy}, Charlie and {Danieli}, Shany and {Merritt}, Allison and {Mowla}, Lamiya and {Romanowsky}, Aaron and {Zhang}, Jielai},
        title = "{A High Stellar Velocity Dispersion and {\ensuremath{\sim}}100 Globular Clusters for the Ultra-diffuse Galaxy Dragonfly 44}",
      journal = {ApJL},
         year = 2016,
        month = sep,
       volume = {828},
       number = {1},
          eid = {L6},
        pages = {L6},
          doi = {10.3847/2041-8205/828/1/L6},
archivePrefix = {arXiv},
       eprint = {1606.06291},
 primaryClass = {astro-ph.GA},
       adsurl = {https://ui.adsabs.harvard.edu/abs/2016ApJ...828L...6V}
}

@ARTICLE{Peng2016,
       author = {{Peng}, Eric W. and {Lim}, Sungsoon},
        title = "{A Rich Globular Cluster System in Dragonfly 17: Are Ultra-diffuse Galaxies Pure Stellar Halos?}",
      journal = {\apjl},
         year = 2016,
        month = may,
       volume = {822},
       number = {2},
          eid = {L31},
        pages = {L31},
          doi = {10.3847/2041-8205/822/2/L31},
archivePrefix = {arXiv},
       eprint = {1604.07496},
 primaryClass = {astro-ph.GA},
       adsurl = {https://ui.adsabs.harvard.edu/abs/2016ApJ...822L..31P}
}

@ARTICLE{Rong2020,
       author = {{Rong}, Yu and {Zhu}, Kai and {Johnston}, Evelyn J. and {Zhang}, Hong-Xin and {Cao}, Tianwen and {Puzia}, Thomas H. and {Galaz}, Gaspar},
        title = "{Lessons on Star-forming Ultra-diffuse Galaxies from the Stacked Spectra of the Sloan Digital Sky Survey}",
      journal = {\apjl},
         year = 2020,
        month = aug,
       volume = {899},
       number = {1},
          eid = {L12},
        pages = {L12},
          doi = {10.3847/2041-8213/aba8aa},
       adsurl = {https://ui.adsabs.harvard.edu/abs/2020ApJ...899L..12R}
}

@ARTICLE{vanDokkum2017,
       author = {{van Dokkum}, Pieter and {Abraham}, Roberto and {Romanowsky}, Aaron J. and {Brodie}, Jean and {Conroy}, Charlie and {Danieli}, Shany and {Lokhorst}, Deborah and {Merritt}, Allison and {Mowla}, Lamiya and {Zhang}, Jielai},
        title = "{Extensive Globular Cluster Systems Associated with Ultra Diffuse Galaxies in the Coma Cluster}",
      journal = {ApJL},
         year = 2017,
        month = jul,
       volume = {844},
       number = {1},
          eid = {L11},
        pages = {L11},
          doi = {10.3847/2041-8213/aa7ca2},
       adsurl = {https://ui.adsabs.harvard.edu/abs/2017ApJ...844L..11V}
}

@ARTICLE{Mihos2022,
       author = {{Mihos}, J. Christopher and {Durrell}, Patrick R. and {Toloba}, Elisa and {C{\^o}t{\'e}}, Patrick and {Ferrarese}, Laura and {Guhathakurta}, Puragra and {Lim}, Sungsoon and {Peng}, Eric W. and {Sales}, Laura V.},
        title = "{The Distance and Dynamical History of the Virgo Cluster Ultradiffuse Galaxy VCC 615}",
      journal = {ApJ},
         year = 2022,
        month = jan,
       volume = {924},
       number = {2},
          eid = {87},
        pages = {87},
          doi = {10.3847/1538-4357/ac35d9},
       adsurl = {https://ui.adsabs.harvard.edu/abs/2022ApJ...924...87M}
}

@ARTICLE{Forbes2018,
       author = {{Forbes}, Duncan A. and {Read}, Justin I. and {Gieles}, Mark and {Collins}, Michelle L.~M.},
        title = "{Extending the globular cluster system-halo mass relation to the lowest galaxy masses}",
      journal = {MNRAS},
         year = 2018,
        month = dec,
       volume = {481},
       number = {4},
        pages = {5592-5605},
          doi = {10.1093/mnras/sty2584},
       adsurl = {https://ui.adsabs.harvard.edu/abs/2018MNRAS.481.5592F}
}

@ARTICLE{mcconnachie2012,
       author = {{McConnachie}, Alan W.},
        title = "{The Observed Properties of Dwarf Galaxies in and around the Local Group}",
      journal = {AJ},
         year = 2012,
        month = jul,
       volume = {144},
       number = {1},
          eid = {4},
        pages = {4},
          doi = {10.1088/0004-6256/144/1/4},
       adsurl = {https://ui.adsabs.harvard.edu/abs/2012AJ....144....4M}
}

@ARTICLE{Martin2016,
       author = {{Martin}, Nicolas F. and {Ibata}, Rodrigo A. and {Lewis}, Geraint F. and {McConnachie}, Alan and {Babul}, Arif and {Bate}, Nicholas F. and {Bernard}, Edouard and {Chapman}, Scott C. and {Collins}, Michelle M.~L. and {Conn}, Anthony R. and {Crnojevi{\'c}}, Denija and {Fardal}, Mark A. and {Ferguson}, Annette M.~N. and {Irwin}, Michael and {Mackey}, A. Dougal and {McMonigal}, Brendan and {Navarro}, Julio F. and {Rich}, R. Michael},
        title = "{The PAndAS View of the Andromeda Satellite System. II. Detailed Properties of 23 M31 Dwarf Spheroidal Galaxies}",
      journal = {ApJ},
         year = 2016,
        month = dec,
       volume = {833},
       number = {2},
          eid = {167},
        pages = {167},
          doi = {10.3847/1538-4357/833/2/167},
       adsurl = {https://ui.adsabs.harvard.edu/abs/2016ApJ...833..167M}
}

@ARTICLE{torrealba2019,
       author = {{Torrealba}, G. and {Belokurov}, V. and {Koposov}, S.~E. and {Li}, T.~S. and {Walker}, M.~G. and {Sanders}, J.~L. and {Geringer-Sameth}, A. and {Zucker}, D.~B. and {Kuehn}, K. and {Evans}, N.~W. and {Dehnen}, W.},
        title = "{The hidden giant: discovery of an enormous Galactic dwarf satellite in Gaia DR2}",
      journal = {MNRAS},
         year = 2019,
        month = sep,
       volume = {488},
       number = {2},
        pages = {2743-2766},
          doi = {10.1093/mnras/stz1624},
       adsurl = {https://ui.adsabs.harvard.edu/abs/2019MNRAS.488.2743T}
}

@ARTICLE{Saifollahi2022,
       author = {{Saifollahi}, Teymoor and {Zaritsky}, Dennis and {Trujillo}, Ignacio and {Peletier}, Reynier F. and {Knapen}, Johan H. and {Amorisco}, Nicola and {Beasley}, Michael A. and {Donnerstein}, Richard},
        title = "{Implications for galaxy formation models from observations of Globular Clusters around Ultra-Diffuse Galaxies}",
      journal = {MNRAS},
         year = 2022,
        month = feb,
          doi = {10.1093/mnras/stac328},
       adsurl = {https://ui.adsabs.harvard.edu/abs/2022MNRAS.tmp..336S}
}

@ARTICLE{vanDokkum2018,
       author = {{van Dokkum}, Pieter and {Danieli}, Shany and {Cohen}, Yotam and {Merritt}, Allison and {Romanowsky}, Aaron J. and {Abraham}, Roberto and {Brodie}, Jean and {Conroy}, Charlie and {Lokhorst}, Deborah and {Mowla}, Lamiya and {O'Sullivan}, Ewan and {Zhang}, Jielai},
        title = "{A galaxy lacking dark matter}",
      journal = {Nature},
         year = 2018,
        month = mar,
       volume = {555},
       number = {7698},
        pages = {629-632},
          doi = {10.1038/nature25767},
       adsurl = {https://ui.adsabs.harvard.edu/abs/2018Natur.555..629V}
}

@ARTICLE{Tang2025,
       author = {{Tang}, Yimeng and {Romanowsky}, Aaron J. and {Huang}, Song and {Okabe}, Nobuhiro and {Brodie}, Jean P. and {Bundy}, Kevin A. and {Buzzo}, Maria Luisa and {Carleton}, Timothy and {Ferr{\'e}-Mateu}, Anna and {Forbes}, Duncan A. and {Gannon}, Jonah S. and {Janssens}, Steven R. and {Levitskiy}, Arsen and {Musick}, Alexi M.},
        title = "{Connection between Dwarf Galaxies and Globular Clusters: Insights from the Perseus Cluster Using Subaru Imaging and Keck Spectroscopy}",
      journal = {\apj},
         year = 2026,
        month = feb,
       volume = {998},
       number = {2},
          eid = {254},
        pages = {254},
          doi = {10.3847/1538-4357/ae33c3},
       adsurl = {https://ui.adsabs.harvard.edu/abs/2026ApJ...998..254T}
}

@ARTICLE{Morrissey2018,
       author = {{Morrissey}, Patrick and {Matuszewski}, Matuesz and {Martin}, D. Christopher and {Neill}, James D. and {Epps}, Harland and {Fucik}, Jason and {Weber}, Bob and {Darvish}, Behnam and {Adkins}, Sean and {Allen}, Steve and {Bartos}, Randy and {Belicki}, Justin and {Cabak}, Jerry and {Callahan}, Shawn and {Cowley}, Dave and {Crabill}, Marty and {Deich}, Willian and {Delecroix}, Alex and {Doppman}, Greg and {Hilyard}, David and {James}, Ean and {Kaye}, Steve and {Kokorowski}, Michael and {Kwok}, Shui and {Lanclos}, Kyle and {Milner}, Steve and {Moore}, Anna and {O'Sullivan}, Donal and {Parihar}, Prachi and {Park}, Sam and {Phillips}, Andrew and {Rizzi}, Luca and {Rockosi}, Constance and {Rodriguez}, Hector and {Salaun}, Yves and {Seaman}, Kirk and {Sheikh}, David and {Weiss}, Jason and {Zarzaca}, Ray},
        title = "{The Keck Cosmic Web Imager Integral Field Spectrograph}",
      journal = {\apj},
         year = 2018,
        month = sep,
       volume = {864},
       number = {1},
          eid = {93},
        pages = {93},
          doi = {10.3847/1538-4357/aad597},
       adsurl = {https://ui.adsabs.harvard.edu/abs/2018ApJ...864...93M}
}

@ARTICLE{Knowles2023,
       author = {{Knowles}, Adam T. and {Sansom}, A.~E. and {Vazdekis}, A. and {Allende Prieto}, C.},
        title = "{sMILES SSPs: a library of semi-empirical MILES stellar population models with variable [{\ensuremath{\alpha}}/Fe] abundances}",
      journal = {\mnras},
         year = 2023,
        month = aug,
       volume = {523},
       number = {3},
        pages = {3450-3470},
          doi = {10.1093/mnras/stad1647},
       adsurl = {https://ui.adsabs.harvard.edu/abs/2023MNRAS.523.3450K}
}

@ARTICLE{Simon2019,
       author = {{Simon}, Joshua D.},
        title = "{The Faintest Dwarf Galaxies}",
      journal = {\araa},
         year = 2019,
        month = aug,
       volume = {57},
        pages = {375-415},
          doi = {10.1146/annurev-astro-091918-104453},
       adsurl = {https://ui.adsabs.harvard.edu/abs/2019ARA&A..57..375S}
}

@software{Cardiel2010,
       author = {{Cardiel}, Nicolas},
        title = "{indexf: Line-strength Indices in Fully Calibrated FITS Spectra}",
 howpublished = {Astrophysics Source Code Library, record ascl:1010.046},
         year = 2010,
        month = oct,
          eid = {ascl:1010.046},
archivePrefix = {ascl},
       eprint = {1010.046},
       adsurl = {https://ui.adsabs.harvard.edu/abs/2010ascl.soft10046C}
}

@ARTICLE{Ma2016,
       author = {{Ma}, Xiangcheng and {Hopkins}, Philip F. and {Faucher-Gigu{\`e}re}, Claude-Andr{\'e} and {Zolman}, Nick and {Muratov}, Alexander L. and {Kere{\v{s}}}, Du{\v{s}}an and {Quataert}, Eliot},
        title = "{The origin and evolution of the galaxy mass-metallicity relation}",
      journal = {\mnras},
         year = 2016,
        month = feb,
       volume = {456},
       number = {2},
        pages = {2140-2156},
          doi = {10.1093/mnras/stv2659},
       adsurl = {https://ui.adsabs.harvard.edu/abs/2016MNRAS.456.2140M}
}

@ARTICLE{Rodriguez2023,
       author = {{Rodriguez}, Carl L. and {Hafen}, Zachary and {Grudi{\'c}}, Michael Y. and {Lamberts}, Astrid and {Sharma}, Kuldeep and {Faucher-Gigu{\`e}re}, Claude-Andr{\'e} and {Wetzel}, Andrew},
        title = "{Great balls of FIRE II: The evolution and destruction of star clusters across cosmic time in a Milky Way-mass galaxy}",
      journal = {\mnras},
         year = 2023,
        month = may,
       volume = {521},
       number = {1},
        pages = {124-147},
          doi = {10.1093/mnras/stad578},
       adsurl = {https://ui.adsabs.harvard.edu/abs/2023MNRAS.521..124R}
}

@ARTICLE{Meng2022,
       author = {{Meng}, Xi and {Gnedin}, Oleg Y.},
        title = "{Tidal disruption of star clusters in galaxy formation simulations}",
      journal = {\mnras},
         year = 2022,
        month = sep,
       volume = {515},
       number = {1},
        pages = {1065-1077},
          doi = {10.1093/mnras/stac1751},
       adsurl = {https://ui.adsabs.harvard.edu/abs/2022MNRAS.515.1065M}
}

@ARTICLE{Doll2025,
       author = {{Doll}, Goran and {Buttitta}, Chiara and {Iodice}, Enrichetta and {Ferr{\'e}-Mateu}, Anna and {Falc{\'o}n-Barroso}, Jesus and {Mart{\'\i}n-Navarro}, Ignacio and {Paolillo}, Maurizio and {Rossi}, Luca and {Forbes}, Duncan A. and {Spiniello}, Chiara and {Hartke}, Johanna and {Gullieuszik}, Marco and {Arnaboldi}, Magda and {Cantiello}, Michele and {Corsini}, Enrico Maria and {D'Ago}, Giuseppe and {Hilker}, Michael and {La Marca}, Antonio and {Mieske}, Steffen and {Mirabile}, Marco and {Rejkuba}, Marina and {Spavone}, Marilena},
        title = "{Looking into the faintEst WIth MUSE (LEWIS): Exploring the nature of ultra-diffuse galaxies in the Hydra-I cluster: V. Integrated stellar population properties}",
      journal = {\aap},
         year = 2026,
        month = feb,
       volume = {707},
          eid = {A88},
        pages = {A88},
          doi = {10.1051/0004-6361/202556736},
       adsurl = {https://ui.adsabs.harvard.edu/abs/2026A&A...707A..88D}
}

@ARTICLE{Bidaran2022,
       author = {{Bidaran}, Bahar and {La Barbera}, Francesco and {Pasquali}, Anna and {Peletier}, Reynier and {van de Ven}, Glenn and {Grebel}, Eva K. and {Falc{\'o}n-Barroso}, Jesus and {Sybilska}, Agnieszka and {Gadotti}, Dimitri A. and {Coccato}, Lodovico},
        title = "{On the accretion of a new group of galaxies on to Virgo - II. The effect of pre-processing on the stellar population content of dEs}",
      journal = {\mnras},
         year = 2022,
        month = sep,
       volume = {515},
       number = {3},
        pages = {4622-4638},
          doi = {10.1093/mnras/stac2005},
       adsurl = {https://ui.adsabs.harvard.edu/abs/2022MNRAS.515.4622B}
}

@ARTICLE{Li2025,
       author = {{Li}, Dayi (David) and {Eadie}, Gwendolyn M. and {Brown}, Patrick E. and {Harris}, William E. and {Abraham}, Roberto G. and {van Dokkum}, Pieter and {Janssens}, Steven R. and {Berek}, Samantha C. and {Danieli}, Shany and {Romanowsky}, Aaron J. and {Speagle}, Joshua S.},
        title = "{Discovery of Two Ultra-diffuse Galaxies with Unusually Bright Globular Cluster Luminosity Functions via a Mark-dependently Thinned Point Process (MATHPOP)}",
      journal = {\apj},
         year = 2025,
        month = may,
       volume = {984},
       number = {2},
          eid = {147},
        pages = {147},
          doi = {10.3847/1538-4357/adc71f},
       adsurl = {https://ui.adsabs.harvard.edu/abs/2025ApJ...984..147L}
}

@ARTICLE{Tau2025,
       author = {{Tau}, Elisa A. and {Monachesi}, Antonela and {G{\'o}mez}, Facundo A. and {Grand}, Robert J.~J. and {Pakmor}, R{\"u}diger and {van de Voort}, Freeke and {Marinacci}, Federico and {Bieri}, Rebekka},
        title = "{Age and metallicity of low-mass galaxies: from their centres to their stellar halos}",
      journal = {arXiv e-prints},
         year = 2025,
        month = nov,
          eid = {arXiv:2511.20806},
        pages = {arXiv:2511.20806},
          doi = {10.48550/arXiv.2511.20806},
archivePrefix = {arXiv},
       eprint = {2511.20806},
 primaryClass = {astro-ph.GA},
       adsurl = {https://ui.adsabs.harvard.edu/abs/2025arXiv251120806T}
}

@ARTICLE{Spolaor2009,
       author = {{Spolaor}, Max and {Proctor}, Robert N. and {Forbes}, Duncan A. and {Couch}, Warrick J.},
        title = "{The Mass-Metallicity Gradient Relation of Early-Type Galaxies}",
      journal = {\apjl},
         year = 2009,
        month = feb,
       volume = {691},
       number = {2},
        pages = {L138-L141},
          doi = {10.1088/0004-637X/691/2/L138},
       adsurl = {https://ui.adsabs.harvard.edu/abs/2009ApJ...691L.138S}
}

@ARTICLE{Mercado2021,
       author = {{Mercado}, Francisco J. and {Bullock}, James S. and {Boylan-Kolchin}, Michael and {Moreno}, Jorge and {Wetzel}, Andrew and {El-Badry}, Kareem and {Graus}, Andrew S. and {Fitts}, Alex and {Hopkins}, Philip F. and {Faucher-Gigu{\`e}re}, Claude-Andr{\'e} and {Gurvich}, Alexander B.},
        title = "{A relationship between stellar metallicity gradients and galaxy age in dwarf galaxies}",
      journal = {\mnras},
         year = 2021,
        month = mar,
       volume = {501},
       number = {4},
        pages = {5121-5134},
          doi = {10.1093/mnras/staa3958},
       adsurl = {https://ui.adsabs.harvard.edu/abs/2021MNRAS.501.5121M}
}

@ARTICLE{Bidaran2023,
       author = {{Bidaran}, Bahar and {La Barbera}, Francesco and {Pasquali}, Anna and {van de Ven}, Glenn and {Peletier}, Reynier and {Falc{\'o}n-Barroso}, Jesus and {Gadotti}, Dimitri A. and {Sybilska}, Agnieszka and {Grebel}, Eva K.},
        title = "{On the accretion of a new group of galaxies onto Virgo - III. The stellar population radial gradients of dEs}",
      journal = {\mnras},
         year = 2023,
        month = nov,
       volume = {525},
       number = {3},
        pages = {4329-4346},
          doi = {10.1093/mnras/stad2546},
       adsurl = {https://ui.adsabs.harvard.edu/abs/2023MNRAS.525.4329B}
}

@ARTICLE{Forbes2022,
       author = {{Forbes}, Duncan A. and {Ferr{\'e}-Mateu}, Anna and {Gannon}, Jonah S. and {Romanowsky}, Aaron J. and {Carlin}, Jeffrey L. and {Brodie}, Jean P. and {Day}, Jacob},
        title = "{Low-metallicity globular clusters in the low-mass isolated spiral galaxy NGC 2403}",
      journal = {\mnras},
         year = 2022,
        month = may,
       volume = {512},
       number = {1},
        pages = {802-810},
          doi = {10.1093/mnras/stac503},
       adsurl = {https://ui.adsabs.harvard.edu/abs/2022MNRAS.512..802F}
}

@ARTICLE{Peng2008,
       author = {{Peng}, Eric W. and {Jord{\'a}n}, Andr{\'e}s and {C{\^o}t{\'e}}, Patrick and {Takamiya}, Marianne and {West}, Michael J. and {Blakeslee}, John P. and {Chen}, Chin-Wei and {Ferrarese}, Laura and {Mei}, Simona and {Tonry}, John L. and {West}, Andrew A.},
        title = "{The ACS Virgo Cluster Survey. XV. The Formation Efficiencies of Globular Clusters in Early-Type Galaxies: The Effects of Mass and Environment}",
      journal = {\apj},
         year = 2008,
        month = jul,
       volume = {681},
       number = {1},
        pages = {197-224},
          doi = {10.1086/587951},
       adsurl = {https://ui.adsabs.harvard.edu/abs/2008ApJ...681..197P}
}

@ARTICLE{Liu2019,
       author = {{Liu}, Yiqing and {Peng}, Eric W. and {Jord{\'a}n}, Andr{\'e}s and {Blakeslee}, John P. and {C{\^o}t{\'e}}, Patrick and {Ferrarese}, Laura and {Puzia}, Thomas H.},
        title = "{The ACS Fornax Cluster Survey. III. Globular Cluster Specific Frequencies of Early-type Galaxies}",
      journal = {\apj},
         year = 2019,
        month = apr,
       volume = {875},
       number = {2},
          eid = {156},
        pages = {156},
          doi = {10.3847/1538-4357/ab12d9},
       adsurl = {https://ui.adsabs.harvard.edu/abs/2019ApJ...875..156L}
}

@ARTICLE{Worthey1992,
       author = {{Worthey}, Guy and {Faber}, S.~M. and {Gonzalez}, J.~J.},
        title = "{MG and Fe Absorption Features in Elliptical Galaxies}",
      journal = {\apj},
         year = 1992,
        month = oct,
       volume = {398},
        pages = {69},
          doi = {10.1086/171836},
       adsurl = {https://ui.adsabs.harvard.edu/abs/1992ApJ...398...69W}
}

@ARTICLE{Liu2016,
       author = {{Liu}, Yiqing and {Peng}, Eric W. and {Blakeslee}, John and {C{\^o}t{\'e}}, Patrick and {Ferrarese}, Laura and {Jord{\'a}n}, Andr{\'e}s and {Puzia}, Thomas H. and {Toloba}, Elisa and {Zhang}, Hong-Xin},
        title = "{Evidence for the Rapid Formation of Low-mass Early-type Galaxies in Dense Environments}",
      journal = {\apj},
         year = 2016,
        month = feb,
       volume = {818},
       number = {2},
          eid = {179},
        pages = {179},
          doi = {10.3847/0004-637X/818/2/179},
       adsurl = {https://ui.adsabs.harvard.edu/abs/2016ApJ...818..179L}
}

@ARTICLE{Pasquali2019,
       author = {{Pasquali}, A. and {Smith}, R. and {Gallazzi}, A. and {De Lucia}, G. and {Zibetti}, S. and {Hirschmann}, M. and {Yi}, S.~K.},
        title = "{Physical properties of SDSS satellite galaxies in projected phase space}",
      journal = {\mnras},
         year = 2019,
        month = apr,
       volume = {484},
       number = {2},
        pages = {1702-1723},
          doi = {10.1093/mnras/sty3530},
       adsurl = {https://ui.adsabs.harvard.edu/abs/2019MNRAS.484.1702P}
}

@ARTICLE{Gallazzi2021,
       author = {{Gallazzi}, Anna R. and {Pasquali}, A. and {Zibetti}, S. and {Barbera}, F. La},
        title = "{Galaxy evolution across environments as probed by the ages, stellar metallicities, and [{\ensuremath{\alpha}} /Fe] of central and satellite galaxies}",
      journal = {\mnras},
         year = 2021,
        month = apr,
       volume = {502},
       number = {3},
        pages = {4457-4478},
          doi = {10.1093/mnras/stab265},
       adsurl = {https://ui.adsabs.harvard.edu/abs/2021MNRAS.502.4457G}
}

@ARTICLE{Wittmann2017,
       author = {{Wittmann}, Carolin and {Lisker}, Thorsten and {Ambachew Tilahun}, Liyualem and {Grebel}, Eva K. and {Conselice}, Christopher J. and {Penny}, Samantha and {Janz}, Joachim and {Gallagher}, John S. and {Kotulla}, Ralf and {McCormac}, James},
        title = "{A population of faint low surface brightness galaxies in the Perseus cluster core}",
      journal = {\mnras},
         year = 2017,
        month = sep,
       volume = {470},
       number = {2},
        pages = {1512-1525},
          doi = {10.1093/mnras/stx1229},
       adsurl = {https://ui.adsabs.harvard.edu/abs/2017MNRAS.470.1512W}
}

@ARTICLE{LaMarca2022,
       author = {{La Marca}, Antonio and {Iodice}, Enrichetta and {Cantiello}, Michele and {Forbes}, Duncan A. and {Rejkuba}, Marina and {Hilker}, Michael and {Arnaboldi}, Magda and {Greggio}, Laura and {Spiniello}, Chiara and {Mieske}, Steffen and {Venhola}, Aku and {Spavone}, Marilena and {D'Ago}, Giuseppe and {Raj}, Maria Angela and {Ragusa}, Rossella and {Mirabile}, Marco and {Rampazzo}, Roberto and {Peletier}, Reynier and {Paolillo}, Maurizio and {Challapa}, Nelvy Choque and {Schipani}, Pietro},
        title = "{Galaxy populations in the Hydra I cluster from the VEGAS survey. II. The ultra-diffuse galaxy population}",
      journal = {\aap},
         year = 2022,
        month = sep,
       volume = {665},
          eid = {A105},
        pages = {A105},
          doi = {10.1051/0004-6361/202142367},
       adsurl = {https://ui.adsabs.harvard.edu/abs/2022A&A...665A.105L}
}

@ARTICLE{Janowiecki2019,
       author = {{Janowiecki}, Steven and {Jones}, Michael G. and {Leisman}, Lukas and {Webb}, Andrew},
        title = "{The environment of H I-bearing ultra-diffuse galaxies in the ALFALFA survey}",
      journal = {\mnras},
         year = 2019,
        month = nov,
       volume = {490},
       number = {1},
        pages = {566-577},
          doi = {10.1093/mnras/stz1868},
       adsurl = {https://ui.adsabs.harvard.edu/abs/2019MNRAS.490..566J}
}

@INPROCEEDINGS{Jones2023,
       author = {{Jones}, Michael and {Karunakaran}, Ananthan and {Bennet}, Paul and {Sand}, David and {Spekkens}, Kristine and {Mutlu Pakdil}, Burcin and {Crnojevic}, Denija and {Janowiecki}, Steven and {Leisman}, Lukas and {Fielder}, Catherine},
        title = "{Gas-rich, field, ultra-diffuse galaxies host few globular clusters}",
    booktitle = {American Astronomical Society Meeting Abstracts},
         year = 2023,
       series = {American Astronomical Society Meeting Abstracts},
       volume = {241},
        month = jan,
          eid = {426.01},
        pages = {426.01},
       adsurl = {https://ui.adsabs.harvard.edu/abs/2023AAS...24142601J}
}

@ARTICLE{Yozin2015,
       author = {{Yozin}, C. and {Bekki}, K.},
        title = "{The quenching and survival of ultra diffuse galaxies in the Coma cluster}",
      journal = {\mnras},
         year = 2015,
        month = sep,
       volume = {452},
       number = {1},
        pages = {937-943},
          doi = {10.1093/mnras/stv1073},
       adsurl = {https://ui.adsabs.harvard.edu/abs/2015MNRAS.452..937Y}
}

@ARTICLE{Mistani2016,
       author = {{Mistani}, Pouria A. and {Sales}, Laura V. and {Pillepich}, Annalisa and {Sanchez-Janssen}, Rub{\'e}n and {Vogelsberger}, Mark and {Nelson}, Dylan and {Rodriguez-Gomez}, Vicente and {Torrey}, Paul and {Hernquist}, Lars},
        title = "{On the assembly of dwarf galaxies in clusters and their efficient formation of globular clusters}",
      journal = {\mnras},
         year = 2016,
        month = jan,
       volume = {455},
       number = {3},
        pages = {2323-2336},
          doi = {10.1093/mnras/stv2435},
       adsurl = {https://ui.adsabs.harvard.edu/abs/2016MNRAS.455.2323M}
}

@ARTICLE{Buzzo2024,
       author = {{Buzzo}, Maria Luisa and {Forbes}, Duncan A. and {Jarrett}, Thomas H. and {Marleau}, Francine R. and {Duc}, Pierre-Alain and {Brodie}, Jean P. and {Romanowsky}, Aaron J. and {Ferr{\'e}-Mateu}, Anna and {Hilker}, Michael and {Gannon}, Jonah S. and {Pfeffer}, Joel and {Haacke}, Lydia},
        title = "{The multiple classes of ultra-diffuse galaxies: can we tell them apart?<SUP></SUP>}",
      journal = {\mnras},
         year = 2025,
        month = jan,
       volume = {536},
       number = {3},
        pages = {2536-2557},
          doi = {10.1093/mnras/stae2700},
       adsurl = {https://ui.adsabs.harvard.edu/abs/2025MNRAS.536.2536B}
}

@ARTICLE{Forbes2025,
       author = {{Forbes}, Duncan A. and {Buzzo}, Maria Luisa and {Ferre-Mateu}, Anna and {Romanowsky}, Aaron J. and {Gannon}, Jonah and {Brodie}, Jean P. and {Collins}, Michelle L.~M.},
        title = "{Why do some ultra diffuse Galaxies have rich globular cluster systems?}",
      journal = {\mnras},
         year = 2025,
        month = jan,
       volume = {536},
       number = {2},
        pages = {1217-1225},
          doi = {10.1093/mnras/stae2675},
       adsurl = {https://ui.adsabs.harvard.edu/abs/2025MNRAS.536.1217F}
}

@ARTICLE{Grishin2021,
       author = {{Grishin}, Kirill A. and {Chilingarian}, Igor V. and {Afanasiev}, Anton V. and {Fabricant}, Daniel and {Katkov}, Ivan Yu. and {Moran}, Sean and {Yagi}, Masafumi},
        title = "{Transforming gas-rich low-mass disky galaxies into ultra-diffuse galaxies by ram pressure}",
      journal = {Nature Astronomy},
         year = 2021,
        month = dec,
       volume = {5},
        pages = {1308-1318},
          doi = {10.1038/s41550-021-01470-5},
       adsurl = {https://ui.adsabs.harvard.edu/abs/2021NatAs...5.1308G}
}

@ARTICLE{Juanis2022,
       author = {{Junais} and {Boissier}, S. and {Boselli}, A. and {Ferrarese}, L. and {C{\^o}t{\'e}}, P. and {Gwyn}, S. and {Roediger}, J. and {Lim}, S. and {Peng}, E.~W. and {Cuillandre}, J.-C. and {Longobardi}, A. and {Fossati}, M. and {Hensler}, G. and {Koda}, J. and {Bautista}, J. and {Boquien}, M. and {Ma{\l}ek}, K. and {Amram}, P. and {Roehlly}, Y.},
        title = "{A Virgo Environmental Survey Tracing Ionised Gas Emission (VESTIGE). XIII. The role of ram-pressure stripping in transforming the diffuse and ultra-diffuse galaxies in the Virgo cluster}",
      journal = {\aap},
         year = 2022,
        month = nov,
       volume = {667},
          eid = {A76},
        pages = {A76},
          doi = {10.1051/0004-6361/202244237},
       adsurl = {https://ui.adsabs.harvard.edu/abs/2022A&A...667A..76J}
}

@software{Bradley2024,
  author       = {Larry Bradley and
                  Brigitta Sip{\H o}cz and
                  Thomas Robitaille and
                  Erik Tollerud and
                  Z\`e Vin{\'{\i}}cius and
                  Christoph Deil and
                  Kyle Barbary and
                  Tom J Wilson and
                  Ivo Busko and
                  Axel Donath and
                  Hans Moritz G{\"u}nther and
                  Mihai Cara and
                  P. L. Lim and
                  Sebastian Me{\ss}linger and
                  Simon Conseil and
                  Zach Burnett and
                  Azalee Bostroem and
                  Michael Droettboom and
                  E. M. Bray and
                  Lars Andersen Bratholm and
                  Adam Ginsburg and
                  William Jamieson and
                  Geert Barentsen and
                  Matt Craig and
                  Brett M. Morris and
                  Marshall Perrin and
                  Shivangee Rathi and
                  Sergio Pascual and
                  Iskren Y. Georgiev},
  title        = {astropy/photutils: 2.0.2},
  month        = oct,
  year         = 2024,
  publisher    = {Zenodo},
  version      = {2.0.2},
  doi          = {10.5281/zenodo.13989456},
  url          = {https://doi.org/10.5281/zenodo.13989456},
}

@ARTICLE{Revaz2018,
       author = {{Revaz}, Yves and {Jablonka}, Pascale},
        title = "{Pushing back the limits: detailed properties of dwarf galaxies in a {\ensuremath{\Lambda}}CDM universe}",
      journal = {\aap},
         year = 2018,
        month = aug,
       volume = {616},
          eid = {A96},
        pages = {A96},
          doi = {10.1051/0004-6361/201832669},
       adsurl = {https://ui.adsabs.harvard.edu/abs/2018A&A...616A..96R}
}

@ARTICLE{Benitez2013,
       author = {{Ben{\'\i}tez-Llambay}, Alejandro and {Navarro}, Julio F. and {Abadi}, Mario G. and {Gottl{\"o}ber}, Stefan and {Yepes}, Gustavo and {Hoffman}, Yehuda and {Steinmetz}, Matthias},
        title = "{Dwarf Galaxies and the Cosmic Web}",
      journal = {\apjl},
         year = 2013,
        month = feb,
       volume = {763},
       number = {2},
          eid = {L41},
        pages = {L41},
          doi = {10.1088/2041-8205/763/2/L41},
       adsurl = {https://ui.adsabs.harvard.edu/abs/2013ApJ...763L..41B}
}

@ARTICLE{Pipino2008,
       author = {{Pipino}, A. and {D'Ercole}, A. and {Matteucci}, F.},
        title = "{Formation of [{\ensuremath{\alpha}}/Fe] radial gradients in the stars of elliptical galaxies}",
      journal = {\aap},
         year = 2008,
        month = jun,
       volume = {484},
       number = {3},
        pages = {679-691},
          doi = {10.1051/0004-6361:20078121},
       adsurl = {https://ui.adsabs.harvard.edu/abs/2008A&A...484..679P}
}

@ARTICLE{Forbes2023,
       author = {{Forbes}, Duncan A. and {Gannon}, Jonah and {Iodice}, Enrichetta and {Hilker}, Michael and {Doll}, Goran and {Buttitta}, Chiara and {Marca}, Antonio La and {Arnaboldi}, Magda and {Cantiello}, Michele and {D'Ago}, G. and {Falcon Barroso}, Jesus and {Greggio}, Laura and {Gullieuszik}, Marco and {Hartke}, Johanna and {Mieske}, Steffen and {Mirabile}, Marco and {Rampazzo}, Roberto and {Rejkuba}, Marina and {Spavone}, Marilena and {Spiniello}, Chiara and {Capasso}, Giulio},
        title = "{Ultra diffuse galaxies in the Hydra I cluster from the LEWISProject: Phase-Space distribution and globular cluster richness}",
      journal = {\mnras},
         year = 2023,
        month = oct,
       volume = {525},
       number = {1},
        pages = {L93-L97},
          doi = {10.1093/mnrasl/slad101},
       adsurl = {https://ui.adsabs.harvard.edu/abs/2023MNRAS.525L..93F}
}

@ARTICLE{Mori1997,
       author = {{Mori}, Masao and {Yoshii}, Yuzuru and {Tsujimoto}, Takuji and {Nomoto}, Ken'ichi},
        title = "{The Evolution of Dwarf Galaxies with Star Formation in an Outward-propagating Supershell}",
      journal = {\apjl},
         year = 1997,
        month = mar,
       volume = {478},
       number = {1},
        pages = {L21-L24},
          doi = {10.1086/310547},
       adsurl = {https://ui.adsabs.harvard.edu/abs/1997ApJ...478L..21M}
}

@ARTICLE{Gannon2026,
       author = {{Gannon}, Jonah S. and {Ferr{\'e}-Mateu}, Anna and {Forbes}, Duncan A.},
        title = "{The Dawes review 14: A decade of ultra-diffuse galaxies}",
      journal = {\pasa},
         year = 2026,
        month = mar,
       volume = {43},
          eid = {e031},
        pages = {e031},
          doi = {10.1017/pasa.2026.10169},
       adsurl = {https://ui.adsabs.harvard.edu/abs/2026PASA...43...31G}
}

@ARTICLE{Pfeffer2024,
       author = {{Pfeffer}, Joel and {Janssens}, Steven R. and {Buzzo}, Maria Luisa and {Gannon}, Jonah S. and {Bastian}, Nate and {Bekki}, Kenji and {Brodie}, Jean P. and {Couch}, Warrick J. and {Crain}, Robert A. and {Forbes}, Duncan A. and {Kruijssen}, J.~M. Diederik and {Romanowsky}, Aaron J.},
        title = "{Origin of the correlation between stellar kinematics and globular cluster system richness in ultradiffuse galaxies}",
      journal = {\mnras},
         year = 2024,
        month = apr,
       volume = {529},
       number = {4},
        pages = {4914-4928},
          doi = {10.1093/mnras/stae850},
       adsurl = {https://ui.adsabs.harvard.edu/abs/2024MNRAS.529.4914P}
}

@ARTICLE{Haines2012,
       author = {{Haines}, C.~P. and {Pereira}, M.~J. and {Sanderson}, A.~J.~R. and {Smith}, G.~P. and {Egami}, E. and {Babul}, A. and {Edge}, A.~C. and {Finoguenov}, A. and {Moran}, S.~M. and {Okabe}, N.},
        title = "{LoCuSS: A Dynamical Analysis of X-Ray Active Galactic Nuclei in Local Clusters}",
      journal = {\apj},
         year = 2012,
        month = aug,
       volume = {754},
       number = {2},
          eid = {97},
        pages = {97},
          doi = {10.1088/0004-637X/754/2/97},
       adsurl = {https://ui.adsabs.harvard.edu/abs/2012ApJ...754...97H}
}

@ARTICLE{Schiavon2017,
       author = {{Schiavon}, Ricardo P. and {Zamora}, Olga and {Carrera}, Ricardo and {Lucatello}, Sara and {Robin}, A.~C. and {Ness}, Melissa and {Martell}, Sarah L. and {Smith}, Verne V. and {Garc{\'\i}a-Hern{\'a}ndez}, D.~A. and {Manchado}, Arturo and {Sch{\"o}nrich}, Ralph and {Bastian}, Nate and {Chiappini}, Cristina and {Shetrone}, Matthew and {Mackereth}, J. Ted and {Williams}, Rob A. and {M{\'e}sz{\'a}ros}, Szabolcs and {Allende Prieto}, Carlos and {Anders}, Friedrich and {Bizyaev}, Dmitry and {Beers}, Timothy C. and {Chojnowski}, S. Drew and {Cunha}, Katia and {Epstein}, Courtney and {Frinchaboy}, Peter M. and {Garc{\'\i}a P{\'e}rez}, Ana E. and {Hearty}, Fred R. and {Holtzman}, Jon A. and {Johnson}, Jennifer A. and {Kinemuchi}, Karen and {Majewski}, Steven R. and {Muna}, Demitri and {Nidever}, David L. and {Nguyen}, Duy Cuong and {O'Connell}, Robert W. and {Oravetz}, Daniel and {Pan}, Kaike and {Pinsonneault}, Marc and {Schneider}, Donald P. and {Schultheis}, Matthias and {Simmons}, Audrey and {Skrutskie}, Michael F. and {Sobeck}, Jennifer and {Wilson}, John C. and {Zasowski}, Gail},
        title = "{Chemical tagging with APOGEE: discovery of a large population of N-rich stars in the inner Galaxy}",
      journal = {\mnras},
         year = 2017,
        month = feb,
       volume = {465},
       number = {1},
        pages = {501-524},
          doi = {10.1093/mnras/stw2162},
       adsurl = {https://ui.adsabs.harvard.edu/abs/2017MNRAS.465..501S}
}

@ARTICLE{Kruijssen2014,
       author = {{Kruijssen}, J.~M. Diederik},
        title = "{Globular cluster formation in the context of galaxy formation and evolution}",
      journal = {Classical and Quantum Gravity},
         year = 2014,
        month = dec,
       volume = {31},
       number = {24},
          eid = {244006},
        pages = {244006},
          doi = {10.1088/0264-9381/31/24/244006},
       adsurl = {https://ui.adsabs.harvard.edu/abs/2014CQGra..31x4006K}
}

@ARTICLE{Sandoval2025,
       author = {{Sandoval Ascencio}, Loraine and {Cooper}, M.~C. and {Zaritsky}, Dennis and {Donnerstein}, Richard and {Khim}, Donghyeon J. and {Baxter}, Devontae C.},
        title = "{Caught in the Act of Quenching? ─ A Population of Post-Starburst Ultra-Diffuse Galaxies}",
      journal = {The Open Journal of Astrophysics},
         year = 2025,
        month = aug,
       volume = {8},
          eid = {110},
        pages = {110},
          doi = {10.33232/001c.142946},
archivePrefix = {arXiv},
       eprint = {2502.00117},
 primaryClass = {astro-ph.GA},
       adsurl = {https://ui.adsabs.harvard.edu/abs/2025OJAp....8E.110S}
}

@ARTICLE{More2015,
       author = {{More}, Surhud and {Diemer}, Benedikt and {Kravtsov}, Andrey V.},
        title = "{The Splashback Radius as a Physical Halo Boundary and the Growth of Halo Mass}",
      journal = {\apj},
         year = 2015,
        month = sep,
       volume = {810},
       number = {1},
          eid = {36},
        pages = {36},
          doi = {10.1088/0004-637X/810/1/36},
archivePrefix = {arXiv},
       eprint = {1504.05591},
 primaryClass = {astro-ph.CO},
       adsurl = {https://ui.adsabs.harvard.edu/abs/2015ApJ...810...36M}
}

@ARTICLE{Williamson2016,
       author = {{Williamson}, David and {Martel}, Hugo and {Romeo}, Alessandro B.},
        title = "{Chemodynamic Evolution of Dwarf Galaxies in Tidal Fields}",
      journal = {\apj},
         year = 2016,
        month = nov,
       volume = {831},
       number = {1},
          eid = {1},
        pages = {1},
          doi = {10.3847/0004-637X/831/1/1},
archivePrefix = {arXiv},
       eprint = {1608.06849},
 primaryClass = {astro-ph.GA},
       adsurl = {https://ui.adsabs.harvard.edu/abs/2016ApJ...831....1W}
}

%\begin{thebibliography}{99}
%\end{thebibliography}

%%%%%%%%%%%%%%%%%%%%%%%%%%%%%%%%%%%%%%%%%%%%%%%%%%

%%%%%%%%%%%%%%%%% APPENDICES %%%%%%%%%%%%%%%%%%%%%

%\appendix



%%%%%%%%%%%%%%%%%%%%%%%%%%%%%%%%%%%%%%%%%%%%%%%%%%

% Don't change these lines
\bsp	% typesetting comment
\label{lastpage}
\end{document}